\documentclass[]{aa}
\usepackage{color}
\usepackage{colortbl}
\usepackage{graphicx}
\usepackage{epstopdf}
\usepackage{natbib}
\bibpunct{(}{)}{;}{a}{}{,}
\graphicspath{{figures/}}
\usepackage{subfigure}
\usepackage{hyperref}
\usepackage{algorithm,algorithmicx}
\begin{document}
%
%
%
%
\title{A fast multi-dimensional magnetohydrodynamic formulation of the transition region adaptive conduction (TRAC) method}
  \author{C. D. Johnston\inst{1,2,3}
  \and A. W. Hood\inst{1} 
  \and I. De Moortel\inst{1,4}
  \and P. Pagano\inst{5,6}
  \and T. A. Howson\inst{1} 
  }
  \institute{School of Mathematics and Statistics, University 
  of St Andrews, St Andrews, Fife, KY16 9SS, UK.
  \and
  Department of Physics and Astronomy, George Mason University, 
  Fairfax, VA 22030, USA. 
  \and
  NASA Goddard Space Flight Center, Greenbelt, MD 20771, USA.
  \and
  Rosseland Centre for Solar Physics, University of Oslo, PO 
  Box 1029  Blindern, NO-0315 Oslo, Norway.
  \and
  Dipartimento di Fisica \& Chimica, Universit\`{a} di Palermo, Piazza del 
  Parlamento 1, I-90134 Palermo, Italy.
  \and
  INAF-Osservatorio Astronomico di Palermo, Piazza del Parlamento 1, I-90134 
  Palermo, Italy.
  \\
  \email{craig.d.johnston@nasa.gov}
  }
%
%
%
%
\abstract
  {
  We have demonstrated that
  the Transition Region Adaptive Conduction (TRAC) 
  method permits fast and accurate numerical solutions of the field-aligned 
  hydrodynamic equations, successfully removing the influence of 
  numerical resolution on the coronal density response to impulsive heating. 
  This is achieved by adjusting the parallel thermal 
  conductivity, radiative loss, and heating rates  to broaden the 
  transition region (TR), below a global cutoff temperature, so that the 
  steep gradients are spatially resolved even when using coarse
  numerical grids.    
  Implementing the original 1D formulation of TRAC in   
  multi-dimensional magnetohydrodynamic (MHD) models 
  would require tracing a large number of magnetic field lines at every 
  time step in order to 
  prescribe a global cutoff temperature to each field line.
  In this paper, we present 
  a highly efficient formulation of the TRAC method for 
  use in multi-dimensional MHD simulations, 
  which does not rely on field line tracing. 
  In the TR, 
  adaptive local cutoff 
  temperatures are used instead of
  global cutoff temperatures
  to broaden any unresolved parts of the atmosphere.
  These local cutoff temperatures are calculated using only local grid cell 
  quantities, enabling the MHD extension of TRAC to efficiently account for 
  the magnetic field evolution, without tracing field lines.
  Consistent with analytical predictions, we show that this approach
  successfully preserves the 
  properties of the original TRAC method.
  In particular, the total radiative losses and heating remain conserved
  under the MHD formulation.
  Results from 2D MHD simulations of 
  impulsive heating in unsheared and sheared arcades of coronal loops are 
  also presented. 
  These simulations benchmark the MHD TRAC method against a series of 1D 
  models and demonstrate the versatility and robustness of the method in  
  multi-dimensional magnetic fields.
  We show, for the first time, that 
  pressure differences, generated
  during the evaporation phase of impulsive heating events,
  can produce current layers that are significantly narrower than the
  transverse energy deposition.
  }
  \titlerunning{A fast multi-dimensional MHD formulation of the TRAC method}
  \maketitle
  %
  %
%
%
\section{Introduction}
  \label{Sect:Intro}
  \indent
  By using multi-dimensional magnetohydrodynamic (MHD) models to study the 
  physics of magnetically closed loops in the solar atmosphere, we 
  have learned a great deal about the storage and release of energy in the 
  corona   
  \citep[see e.g.][]{paper:Reale2014,paper:Pontin&Hornig2020}.  
  Simulating the plasma 
  response to the heating
  in such models requires a physical connection between the corona, 
  transition region (TR), and chromosphere in order to account for 
  field-aligned thermal conduction,
  optically thin radiation, and chromospheric evaporation. These 
  processes 
  control the evolution of the temperature and density of the confined 
  plasma, which determine the brightness of the emission from
  the coronal loops.
  \\
  \indent
  One of the main difficulties encountered 
  when including such additional physics
  in MHD models
  is the need to implement a grid that fully resolves the 
  steep gradients in the TR, which are associated with thermal 
  conduction between the corona and chromosphere
  \citep[e.g.][]{paper:Antiochos&Sturrock1978,paper:Veseckyetal1979}.
  Resolving these gradients in numerical simulations
  requires very small grid cell widths, typically less than 1~km
  \citep{paper:Bradshaw&Cargill2013}, 
  which, in turn, acts as a major constraint on the time step, as required 
  for numerical stability. 
  Obtaining this spatial resolution in 
  active-region-sized 3D MHD
  models poses a serious challenge 
  \citep[e.g.][]{paper:Reidetal2018,paper:Reidetal2020,
  paper:Knizhniketal2019,
  paper:Kohutova2020},  for
  simulations to be run in a realistic time.
  \\
  \indent
  As pointed out by
  \cite{paper:Bradshaw&Cargill2013},
  the main consequence of not properly resolving the
  TR 
  when using the standard
  \citet[][hereafter SH]{paper:Spitzer1953} conduction method is that the 
  resulting coronal density~($n$) is artificially low. 
  This happens because the 
  downward heat flux is forced to jump
  across an under-resolved TR to the 
  chromosphere, where the incoming energy is then strongly 
  radiated. 
  Since the emission measure scales with $n^2$, 
  such underestimations in the density can then potentially lead to
  inaccurate conclusions 
  when numerical predictions are compared with real observational data.
  \\
  \indent
  Furthermore, for the case of steady footpoint heating, 
  \cite{paper:Johnstonetal2019} 
  demonstrated that inadequate TR resolution 
  can result in the suppression of the thermal non-equilibrium cycles
  \citep{paper:Fromentetal2018,paper:Winebargeretal2018,
  paper:Klimchuk&Luna2019} 
  that are present when the TR is properly resolved.
  Similar results were also reported by
  \cite{paper:Zhouetal2021} for the formation of prominences
  \citep[e.g.][]{paper:Antiochosetal1999,paper:Xiaetal2012}. 
  Both are examples where the predicted observational signatures are 
  significantly different 
  depending on the size of grid cells used in the
  TR.
  \\
  \indent
  In a recent paper, 
  \citet[][hereafter JB19]{paper:Johnston&Bradshaw2019}
  demonstrated that modelling the Transition Region 
  with the use of an Adaptive Conduction (TRAC) method
  successfully removes this influence of numerical resolution 
  on the coronal density response to heating while
  maintaining high levels of agreement with fully resolved 
  hydrodynamic (HD) models.
  When employed with the coarser spatial resolutions, 
  typically achieved in multi-dimensional MHD codes, the
  TRAC simulations gave peak density 
  errors of less than $5\%$, whereas without 
  TRAC, in the equivalent coarse resolution simulations,
  the errors can be as high as 75\%
  \citep[see e.g.][JB19]{paper:Johnstonetal2017a,paper:Johnstonetal2020}.
  This is achieved by enforcing adjustments to
  the parallel thermal conductivity ($\kappa_\parallel(T)$),
  radiative loss ($\Lambda(T)$), and
  heating ($Q$) rates that
  are due, in their original form, to
  \cite{paper:Lionelloetal2009} and \cite{paper:Mikicetal2013}
  and were subsequently extended by
  \cite{paper:Johnstonetal2020}.
  These conditions act to broaden any unresolved parts of the
  TR,
  below an adaptive
  cutoff temperature ($T_c$),
  while ensuring that $\kappa_\parallel(T)\Lambda(T)$
  and $\kappa_\parallel(T)Q(T)$
  give the same function of temperature as for
  $T \geq T_c$.
  \cite{paper:Johnstonetal2020} showed that 
  modifications of this form 
  allow the TR to be modelled
  in HD simulations with 
  tractable grid sizes, of order 50~km, 
  because they preserve the energy balance in the TR 
  and conserve the total amount of energy that
  is delivered to the chromosphere, consistent 
  with fully resolved models. 
  \\
  \indent
  The natural extension of the original TRAC method,
  from 1D HD to multi-dimensional MHD,
  requires the 
  tracing of magnetic field lines at each time step
  in order to
  identify the global
  cutoff temperature that is associated with each field line.
  Recently, \cite{paper:Zhouetal2021} proposed such an approach,
  applying the 1D TRAC method in 2D MHD 
  simulations of prominence formation  
  by using two different field line tracing techniques.
  However, multi-dimensional implementations of TRAC that employ
  the original 1D formulation suffer from
  limitations that are associated with 
  the need to prescribe a global cutoff temperature to individual
  field lines.
  In particular, tracing a sufficient number of 
  magnetic field lines at every time step  
  is computationally very time consuming
  because of
  the global communication that is required between all of the grid cells
  in the numerical domain.
  The outcome is that field line tracing implementations of the
  TRAC method are unlikely to be practical in 3D MHD simulations of
  coronal heating, where the
  energy release is generated 
  self-consistently through the
  build-up of magnetic energy in the coronal field 
  and subsequent dissipation through magnetic reconnection events
  \citep[e.g.][]{paper:Hoodetal2016,
  paper:Realeetal2016,
  paper:Reidetal2018,paper:Reidetal2020}.
  \\
  \indent
  In this paper, we address these shortcomings 
  by presenting an
  extension
  of the TRAC method for use in multi-dimensional MHD simulations, 
  without the need to trace magnetic field lines.
  This is achieved by 
  prescribing an adaptive cutoff temperature local to each
  grid cell, using only local grid cell quantities.
  The full details of the MHD TRAC method are described in
  Sect. \ref{Sect:TRAC},
  where we also show that moving from a global to a local 
  cutoff temperature preserves the properties of the original
  TRAC method.
  Section \ref{Sect:Model_and_experiments} outlines the numerical 
  experiments, and in Sect. \ref{Sect:2D_Results} we present
  the results from
  2D MHD simulations that model the thermodynamic response to
  impulsive heating events in unsheared and sheared arcades
  of coronal loops.
  The unsheared arcade simulation is used to
  benchmark the MHD TRAC method against a series of 1D models,
  while the sheared arcade model
  demonstrates the performance of the MHD extension of TRAC
  in a multi-dimensional
  magnetic field configuration.
  We conclude with a discussion of the MHD TRAC method in Sect. 
  \ref{Sect:Discussion} and present supplementary material
  in Appendix \ref{App:IoNR}.
  %
  %
%
%
\section{The TRAC method
  \label{Sect:TRAC}}
  \indent
  In \cite{paper:Johnstonetal2020}, 
  we presented an extensive description of the TRAC method 
  for the highly efficient numerical integration of the field-aligned 
  HD equations through the computationally demanding TR. 
  The extension of the method to multi-dimensional MHD
  simulations will be presented in the following subsections.
  %
  %
%
%
\subsection{MHD model
  \label{Sect:MHD_model}}
  \indent  
  To model the magnetic field evolution and plasma response to heating,
  we considered the following set of MHD equations, 
  which include gravitational stratification and an energy 
  equation that incorporates the effects of thermal conduction and optically 
  thin radiation, 
  \begin{align}
    &
    \frac{\partial\rho}{\partial t}  
    + \nabla \cdot (\rho {\bf v})
    = 0; 
    \label{Eqn:mhd_continuity}
    \\[1.5mm]
    &
    \rho \frac{D{\bf v}}{Dt}
    = - 
    \nabla P - \rho {\bf g} + {\bf j \times B} + 
    {\bf F}_{\textrm{visc.}};
    \label{Eqn:mhd_motion}
    \\[1.5mm]
    &
    \rho \frac{D \epsilon}{Dt}
    = -P\nabla \cdot{\bf v} - 
    	 \nabla \cdot {\bf q}  + Q_{\textrm{visc.}} + Q
    - \! n^2 \Lambda(T) + \frac{|\textbf{j}|^2}{\sigma};
    \label{Eqn:mhd_ee}
    \\[1.5mm]
    &
    \frac{\partial {\bf B}}{\partial t}
    = 
    \nabla \times ({\bf v \times B})
    - \nabla \times ({\bf \eta \nabla \times B});
    \label{Eqn:mhd_induct}
    \\[1.5mm]
    & 
    P = 2 \, k_B n T.
    \label{Eqn:mhd_gas_law}
  \end{align} 
  Here, 
  $\rho$ is the mass density,  
  ${\bf v}$ is the velocity, 
  $P$ is the gas pressure, 
  ${\bf g}$ is the gravitational acceleration,
  ${\bf j}$ is the electric current density,
  ${\bf B}$ is the magnetic field,
  ${\bf F}_{\textrm{visc.}}$ represents the viscous force,
  $\epsilon=P/(\gamma-1) \rho $ is the specific 
  internal energy density
  (where $\gamma=5/3$ is the ratio of specific heats),
  ${\bf q}$ is the heat flux vector, 
  $Q_{\textrm{visc.}}$ represents the viscous heating,
  $Q$ is a heating function that includes uniform background heating
  and a time-dependent component that can be dependent on position,
  $n$ is the number density ($n=\rho/1.2m_p$, where $m_p$ is the
  proton mass),
  $\Lambda(T)$ is the 
  radiative loss function in an optically thin plasma, 
  which we approximated using the 
  piecewise continuous function defined in
  \cite{paper:Klimchuketal2008},
  $\sigma$ is the electrical conductivity,
  $\eta$ is the resistivity,
  $k_B$ is the Boltzmann 
  constant and
  $T$ is 
  the temperature.
  \\
  \indent
  We solved the MHD equations 
  \eqref{Eqn:mhd_continuity}--\eqref{Eqn:mhd_gas_law}   
  using the Lagrangian Remap (Lare) code described in 
  \citep{paper:Arber2001}.
  Two small shock viscosity terms were included to ensure numerical 
  stability together with a small background viscosity
  \citep{paper:Reidetal2020}. 
  These contribute a force, ${\bf F}_{\textrm{visc.}}$, on the 
  right-hand side of the equation of motion \eqref{Eqn:mhd_motion} and a
  heating term, $Q_{\textrm{visc.}}$,
  to the energy equation \eqref{Eqn:mhd_ee}.
  The thermal conduction model is based on the
  \cite{paper:Braginskii1965} heat flux in the presence of a 
  magnetic field,
  where the heat flux vector,
  \begin{align}
    {\bf q} = -
    \dfrac{\kappa_\parallel(T)}{B^2+b^2_{\rm min}} 
	\left( 
	({\bf B \cdot \nabla}T){\bf B}
	+
	b^2_{\rm min} \nabla T
	\right),
	\label{Eqn:mhd_q_vector}
  \end{align}  
  recovers the anisotropic SH parallel thermal 
  conductivity \citep{paper:Spitzer1953} in the limit 
  $B^2 \gg b^2_{\rm min}$.
  Here, $\kappa_\parallel(T) = \kappa_0 T^{5/2}$ 
  is the SH parallel
  coefficient of thermal conduction
  with $\kappa_0=10^{-11}$~Jm$^{-1}$K$^{-7/2}$s$^{-1}$
  and the perpendicular conductivity is given by
  \begin{align}
    \kappa_\perp (T) =
    \dfrac{\kappa_\parallel(T)}
    {1+ B^2/b^2_{\rm min}},
	\label{Eqn:mhd_kappa_perp}
  \end{align}  
  where $B^2/b^2_{\rm min}$ is used to approximate   
  the square of the
  product between the electron gyrofrequency and electron collision time
  $(\omega_{ce} \tau_e)^2$,
  with $b_{\rm min}=0.1$~G used throughout this paper.
  In the strong field limit, 
  $\kappa_\perp (T)$
  is proportional to 
  $\kappa_\parallel(T)/B^2$
  and
  when $B^2 \ll b^2_{\rm min}$, we note that the 
  thermal conductivity reduces to 
  isotropic.
  \\
  \indent
  Time-splitting methods 
  are used to update thermal conduction
  and optically thin radiation 
  separately from the advection terms, 
  as discussed in Appendix A
  of \cite{paper:Johnstonetal2017a}.
  Furthermore, to treat thermal conduction, we use super time stepping
  methods, as described in 
  \cite{paper:Meyeretal2012,paper:Meyeretal2014} and discussed
  in Appendix B of \cite{paper:Johnstonetal2017a}. 
  This time integration strategy is a computational
  efficient way of dealing with the potentially
  large difference between the 
  advection ($dt_{\textrm{adv}}$) and conduction
  ($dt_{\textrm{cond}}$)
  time step 
  restrictions that are required for numerical 
  stability in an explicit numerical scheme, where
  $dt_{\textrm{cond}} \ll dt_{\textrm{adv}}$
  is typical for
  coronal plasma evolution.
  %
  %
%
%
\subsection{The TRAC method: Extension to MHD models
  \label{Sect:TRAC_MHD}}
  \indent
  The extension of the TRAC method to 
  multi-dimensional MHD models requires a more
  sophisticated treatment than the field-aligned 
  HD implementation. The main challenge that needs to be 
  addressed is how the magnetic field evolution modifies the 
  prescription of the adaptive cutoff temperature along a field line. 
  As pointed out by \cite{paper:Ruanetal2020},
  continuing with the same approach as JB19 and
  \cite{paper:Johnstonetal2020} 
  requires the 
  tracking of magnetic field lines and 
  identification of a 
  cutoff temperature, associated with each field line,
  at each time step of the numerical simulation.
  However, this approach, which was subsequently
  pursued by \cite{paper:Zhouetal2021},
  is computationally expensive 
  and non-trivial to parallelise with a strong scaling
  due to the substantial communication required between all of the 
  grid cells in the numerical domain.
  \\
  \indent
  Therefore, it is desirable to develop an
  optimised extension
  of the TRAC method, for use in MHD simulations, 
  that prescribes an adaptive cutoff temperature local to each
  grid cell, using only local grid cell quantities.
  The full details of such an implementation are described next, 
  starting first with the field-aligned HD formulation, 
  followed by the 
  generalisation to multi-dimensional MHD.
  %
  %
%
%
\subsubsection{Hydrodynamic implementation
  \label{Sect:TRAC_HD}}
  \indent
  To formulate
  the extension of the TRAC method,
  we begin with a steady state version of the energy
  equation that approximates the SH temperature gradient
  as given by Eq.~(8) in
  \cite{paper:Johnstonetal2020},
  \begin{align}
    \dfrac{T}{L_T}
    \!
    =
    \!
    \!
    \dfrac{ 
    \!
    5k_B J 
    \!
    \!
    \pm
    \!
    \!
    \sqrt{
    \!
    25k_B^2J^2
    \!
    \!
    +
    \!
    4\dfrac{\!\kappa_\parallel(T)}{T}
    \!
    \!
    \left[
    \!
    \!
    \left(
    \!
    \dfrac{\!P\!}{\!2k_BT\!}
    \!
    \right)^{\!\!2}
    \!
    \!
    \Lambda(T)
    \!
    - 
    \!
    Q
    \!
    \right]  
    \!
    \!
    }
    }
    {2\dfrac{\!\kappa_\parallel(T)}{T}},
    \!
    \!
    \label{Eqn:T_L_T_quadratic_formula}
  \end{align}
  where
  \begin{align}
    L_T(s) = \dfrac{T(s)}{dT(s)/ds}
    \label{Eqn:L_T} 
  \end{align}
  is 
  the temperature length scale, $s$ is the spatial co-ordinate
  along the magnetic field, 
  $J=nv$ is the mass flux, and
  $v$ is the velocity parallel 
  to the magnetic field.
  We note that
  the
  positive (negative) root corresponds to an
  increasing (decreasing) temperature gradient. 
  \\
  \indent
  The aim is to construct a conductivity model that broadens the TR, 
  giving a new local temperature length scale ($L_T(s)$) 
  that satisfies the
  minimum resolution criteria of 
  \cite{paper:Johnstonetal2017a,paper:Johnstonetal2017b}.  
  This requires that
  \begin{align}
    L_T(s) = \dfrac{L_R(s)}{\delta},
    \label{Eqn:L_T_resolution_criteria} 
  \end{align}
  where $L_R(s)=\Delta s$ is the local grid cell
  width and
  $\delta=1/2$ is a parameter that controls the number
  of grid cells used to resolve $L_T(s)$.
  \\
  \indent
  Combining Eq. \eqref{Eqn:T_L_T_quadratic_formula} 
  and \eqref{Eqn:L_T_resolution_criteria}, taking the absolute value of 
  the mass flux term (see below) and considering only the positive root
  (to remove the dependence on sign of the local temperature gradient),
  we obtain
  the following expression for a
  conductivity model
  that is fitted to resolve the local 
  temperature length 
  scale,
  \begin{align}
    \kappa_\parallel^{\!\textsc{trac}}(T)
    \!
    = 
    \!
    \!
    \dfrac{ 
    \!
    5k_B |J| 
    \!
    \!
    +
    \!
    \!
    \sqrt{
    \!
    25k_B^2J^2
    \!
    \!
    +
    \!
    4\dfrac{\!\kappa_\parallel(T)}{T}
    \!
    \!
    \left[
    \!
    \!
    \left(
    \!
    \dfrac{\!P\!}{\!2k_BT\!}
    \!
    \right)^{\!\!2}
    \!
    \!
    \Lambda(T)
    \!
    - 
    \!
    Q
    \!
    \right]  
    \!
    \!
    }
    }
    {\dfrac{2\delta}{L_R}}
    \!
    ,
    \!
    \!
    \!
    \label{Eqn:TRAC_conductivity}
  \end{align}
  which we refer to
  as the TRAC
  parallel thermal conductivity. 
  We note that
  Eq. \eqref{Eqn:TRAC_conductivity} can also be 
  interpreted as the calculation of a local cutoff 
  temperature in each grid cell, using only local grid cell quantities.
  \\
  \indent
  As formulated, 
  the TRAC conductivity
  only exceeds the SH value ($\kappa_\parallel(T)$) 
  in grid cells that would be under-resolved with the
  SH conductivity. That is locations in the TR
  where $L_T(s) < L_R(s)/\delta$.
  On the other hand,
  $\kappa_\parallel^{\textsc{trac}}(T)$ is smaller than
  the SH value
  in properly resolved grid cells (e.g. in the corona
  where $L_T(s) > L_R(s)/\delta$).
  This is the case during the evaporation and peak density phases 
  of an impulsively heated loop
  because
  the approximation of the SH temperature gradient,
  used in the calculation of
  the TRAC conductivity,
  reduces to the simplified expressions presented in 
  \cite{paper:Johnstonetal2020}, for the two limits of
  strong evaporation (neglecting radiation) and peak density (neglecting 
  mass flux terms).
  Thus, the TRAC conductivity is calculated using an accurate approximation
  of the SH temperature gradient, during these first two
  phases.
  \\
  \indent
  We note that it is necessary to take the absolute value
  of the mass flux term in order
  to ensure that the TRAC conductivity remains smooth
  during the evaporation phase, when large upflows in the TR are 
  accompanied by small downflows at the base of the TR
  \citep[see e.g.][]{paper:Johnstonetal2020}.
  The evaporation phase is prioritised because
  this phase has the most severe requirements 
  for resolving the downward heat flux.
  However,
  for the decay phase of an impulsively heated loop the mass
  flux term is negative (downflow). Therefore, the approximation 
  of the temperature gradient used in Eq.
  \eqref{Eqn:TRAC_conductivity} does not recover the radiative 
  cooling limit (neglecting thermal conduction) due to the
  sign enforced by the
  $|J|$ term. 
  The outcome is that the TRAC conductivity can lead to an over-broadening  
  of the TR during the decay phase.
  \\
  \indent
  To mitigate this over-broadening effect, we imposed a limiter
  on the TRAC conductivity, which is derived by neglecting the mass flux 
  terms,
  \begin{align}
    \kappa_\parallel^{\textsc{lim}}(T)
    \!
    = 
    \dfrac{ 
    \!
    \!
    \sqrt{
    \!
    4\dfrac{\!\kappa_\parallel(T)}{T}
    \!
    \!
    \left[
    \!
    \!
    \left(
    \!
    \dfrac{\!P\!}{\!2k_BT\!}
    \!
    \right)^{\!\!2}
    \!
    \!
    \Lambda(T)
    \!
    - 
    \!
    Q
    \!
    \right]  
    \!
    \!
    }
    }
    {\dfrac{2\delta}{L_R}}
    \!
    .
    \!
    \!
    \label{Eqn:TRAC_lim}
  \end{align}
  This limited conductivity is used
  when grid cells would be under-resolved with the SH conductivity
  and over-resolved with the TRAC conductivity given by
  Eq. \eqref{Eqn:TRAC_conductivity}.
  Such grid cells are 
  regarded as being over-resolved
  if the new local temperature length scale, 
  given by Eq. \eqref{Eqn:L_T},
  is broadened beyond
  twice the value intended by
  Eq. \eqref{Eqn:L_T_resolution_criteria}.
  Therefore,
  the threshold
  for transitioning to $\kappa_\parallel^{\textsc{lim}}(T)$
  is taken as $L_T(T) > 2L_R(T)/\delta$.
  \\
  \indent
  Incorporating the TRAC conductivity 
  with the broadening limiter,
  we set the parallel thermal
  conductivity to be of the form
  \begin{align}
    \kappa_\parallel^{\prime}(T)
    \!    
    = 
    \!
    \!
    \left\{
    \begin{array}{l}
      \!
      \!
      \!
      \!
      \textrm{max}
      \{
      \!
      \kappa_\parallel^{\textsc{trac}}(T),
      \,
      \kappa_\parallel(T)
      \!
      \},
      \,\textrm{for}\,
      L_T(T) \leq  \dfrac{2L_R(T)}{\delta}
      \\[3mm]
      \!
      \!
      \!
      \!
      \textrm{max}
      \{
      \!
      \kappa_\parallel^{\textsc{lim}}(T),
      \,
      \kappa_\parallel(T)
      \!
      \},
      \, \, \, \, \textrm{for}\, 
      L_T(T) > \dfrac{2L_R(T)}{\delta}
    \end{array}
    \right.
    \! \!
    \! \!
    \! ,
    \! \!
    \label{Eqn:prime_conductivity}
  \end{align}
  which increases the conductivity in under-resolved grid
  cells and
  reduces to the classical 
  SH value elsewhere,
  as we show in the next subsection.
  This calculation of the 
  $\kappa_\parallel^{\prime}(T)$ conductivity
  model comprises the first part of the TRAC method
  as described in JB19.
  \\
  \indent
  The second part of the method
  is to broaden the steep temperature and   
  density gradients in the TR.
  This is achieved using
  an approach first proposed by 
  \cite{paper:Linkeretal2001},
  \cite{paper:Lionelloetal2009} and 
  \cite{paper:Mikicetal2013}.
  \\
  \indent
  Using the 
  $\kappa_\parallel^{\prime}(T)$ conductivity
  model, 
  we modify the radiative   
  loss rate 
  ($\Lambda(T)$)
  to preserve
  $\kappa_\parallel(T)\Lambda(T)=
  \kappa_\parallel^{\prime}(T)\Lambda^{\prime}(T)$,
  \begin{align}
    \Lambda^{\prime}(T) &= \Lambda(T) 
    \dfrac{\kappa_\parallel(T)}{\kappa_\parallel^{\prime}(T)},
    \label{Eqn:prime_losses} 
  \end{align}
  and the heating rate ($Q(T)$) to preserve
  $\kappa_\parallel(T) Q(T)=
  \kappa_\parallel^{\prime}(T)Q^{\prime}(T)$,
  \begin{align}
    Q^{\prime}(T) &= Q(T) 
    \dfrac{\kappa_\parallel(T)}{\kappa_\parallel^{\prime}(T)}.
    \label{Eqn:prime_heating} 
  \end{align}
  Reducing both $\Lambda (T)$ and $Q(T)$, in this way, ensures that the 
  total radiation and heating, integrated across the TR remains the same
  for both the SH and TRAC methods   
  \citep{paper:Johnstonetal2020}.
  %
  %
%
%
\subsubsection{Comparison between the SH and TRAC models}
  %
  %
  %
  %
  %
  %
\begin{figure*}
  \vspace*{-6mm}
  \hspace*{-0.05\linewidth}
  \subfigure{\includegraphics[width=0.53\linewidth]
  {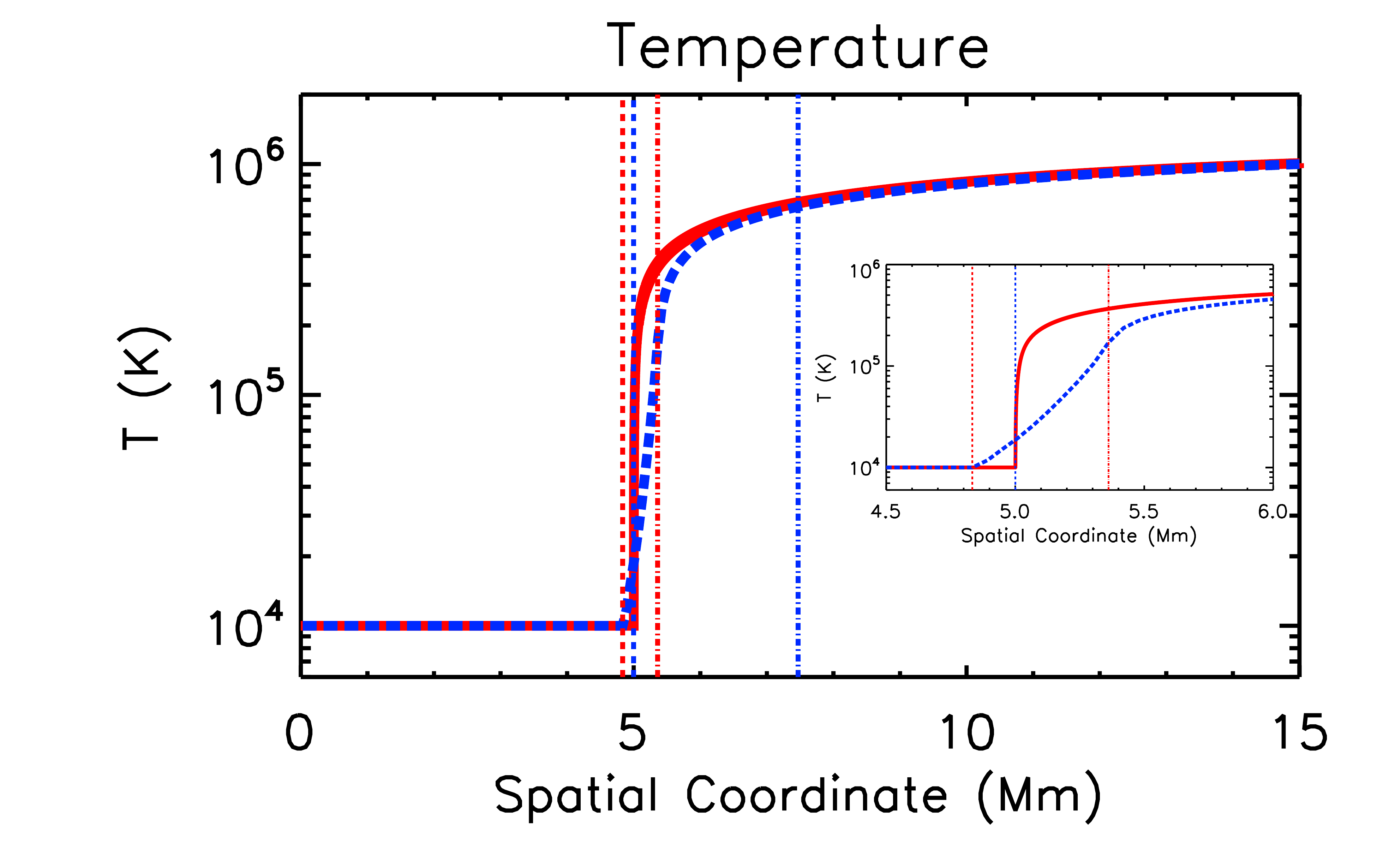}}
  \hspace*{-0.03\linewidth}
  \subfigure{\includegraphics[width=0.53\linewidth]
  {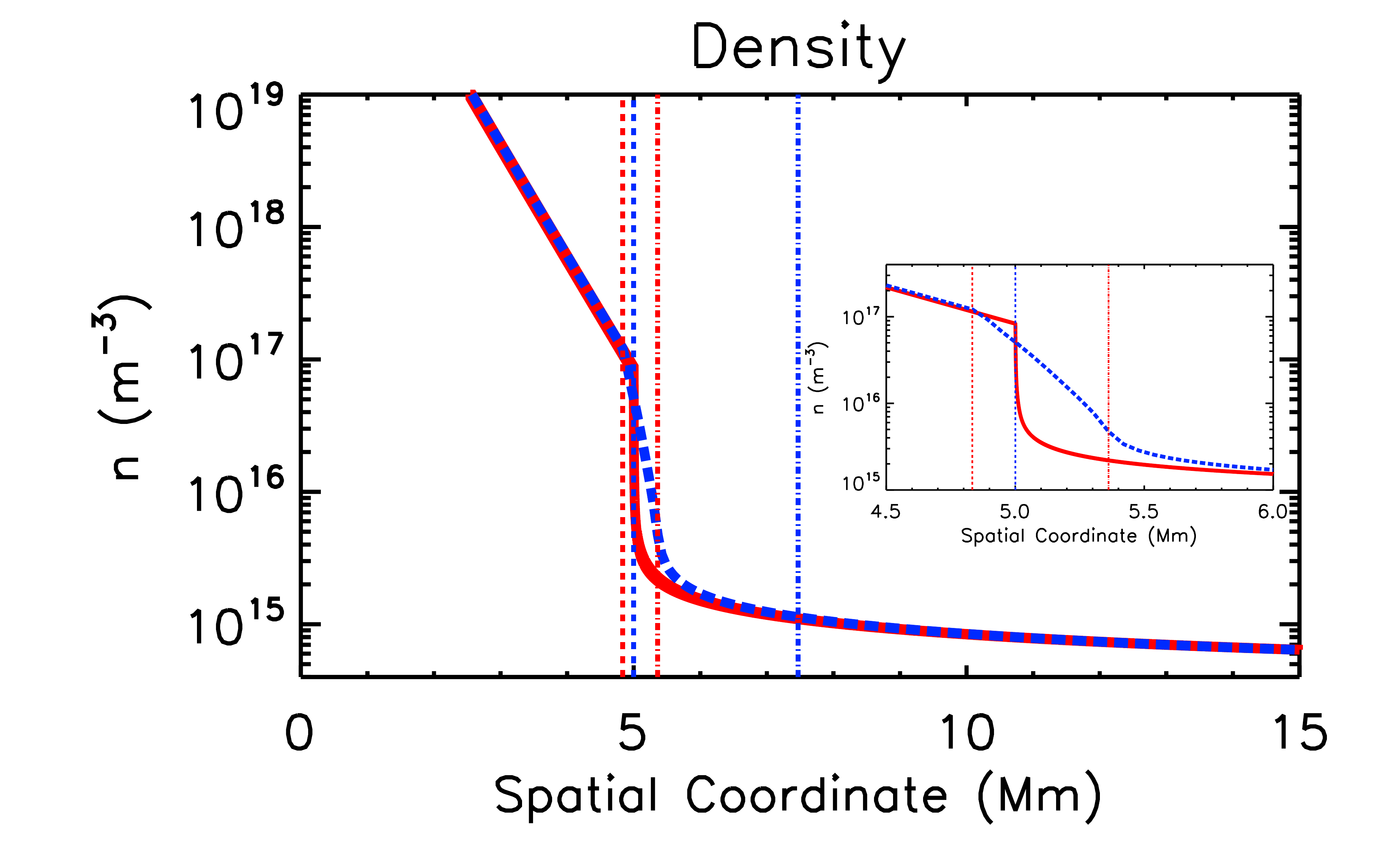}}
  \\[-6mm]
  \hspace*{-0.05\linewidth}
  \subfigure{\includegraphics[width=0.53\linewidth]
  {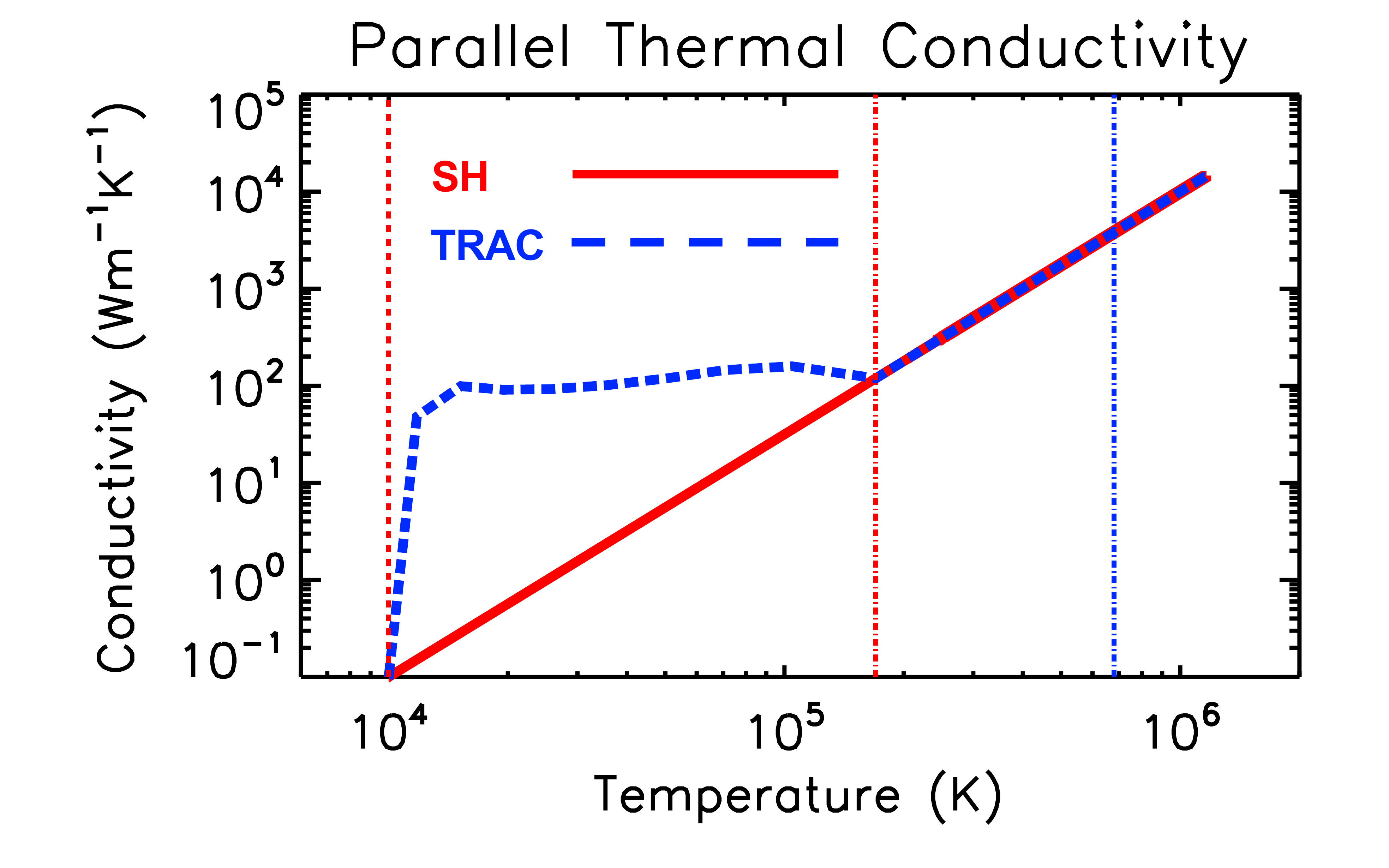}}
  \hspace*{-0.03\linewidth}
  \subfigure{\includegraphics[width=0.53\linewidth]
  {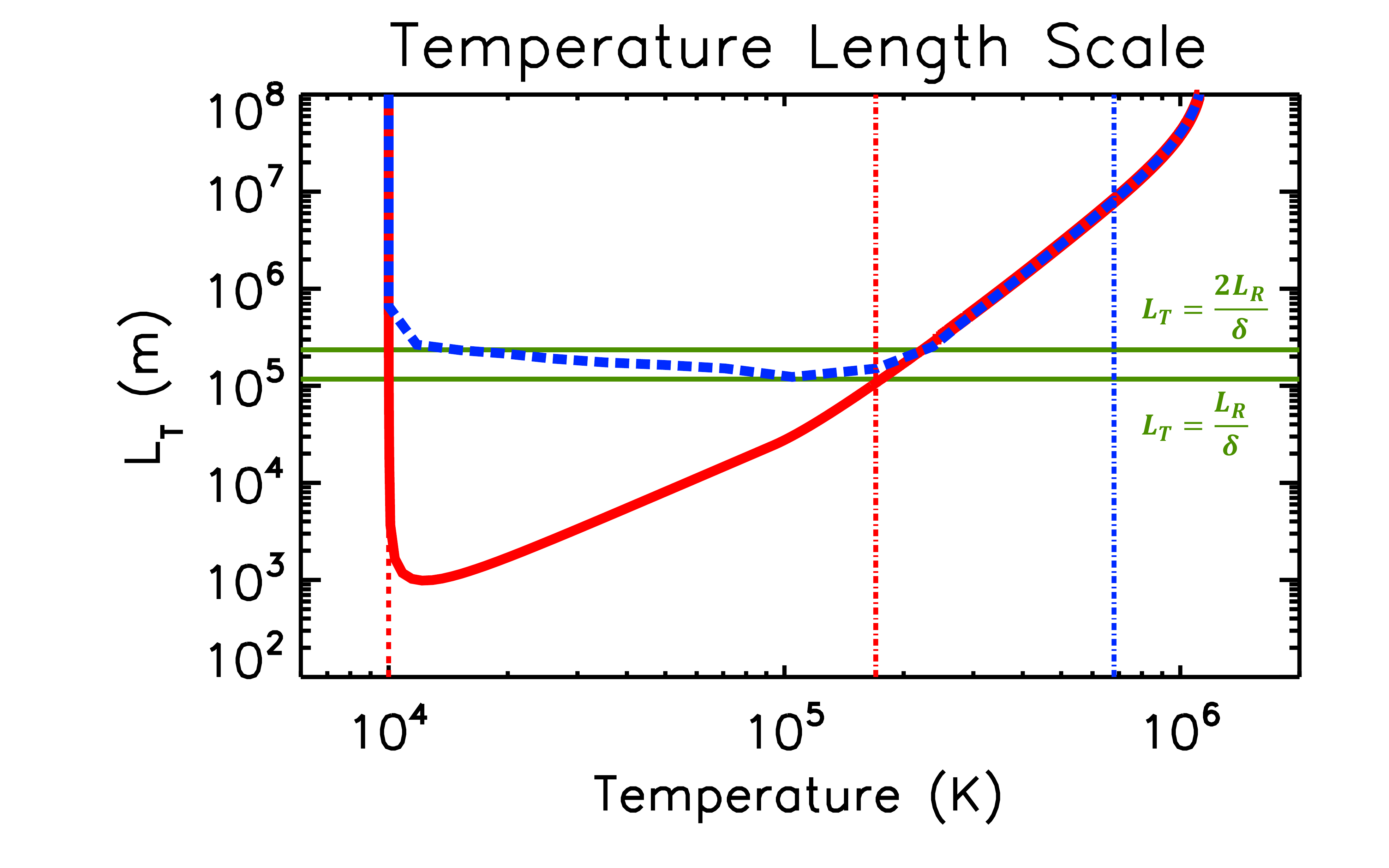}}
  \\[-7mm]
  \hspace*{-0.05\linewidth}
  \subfigure{\includegraphics[width=0.53\linewidth]
  {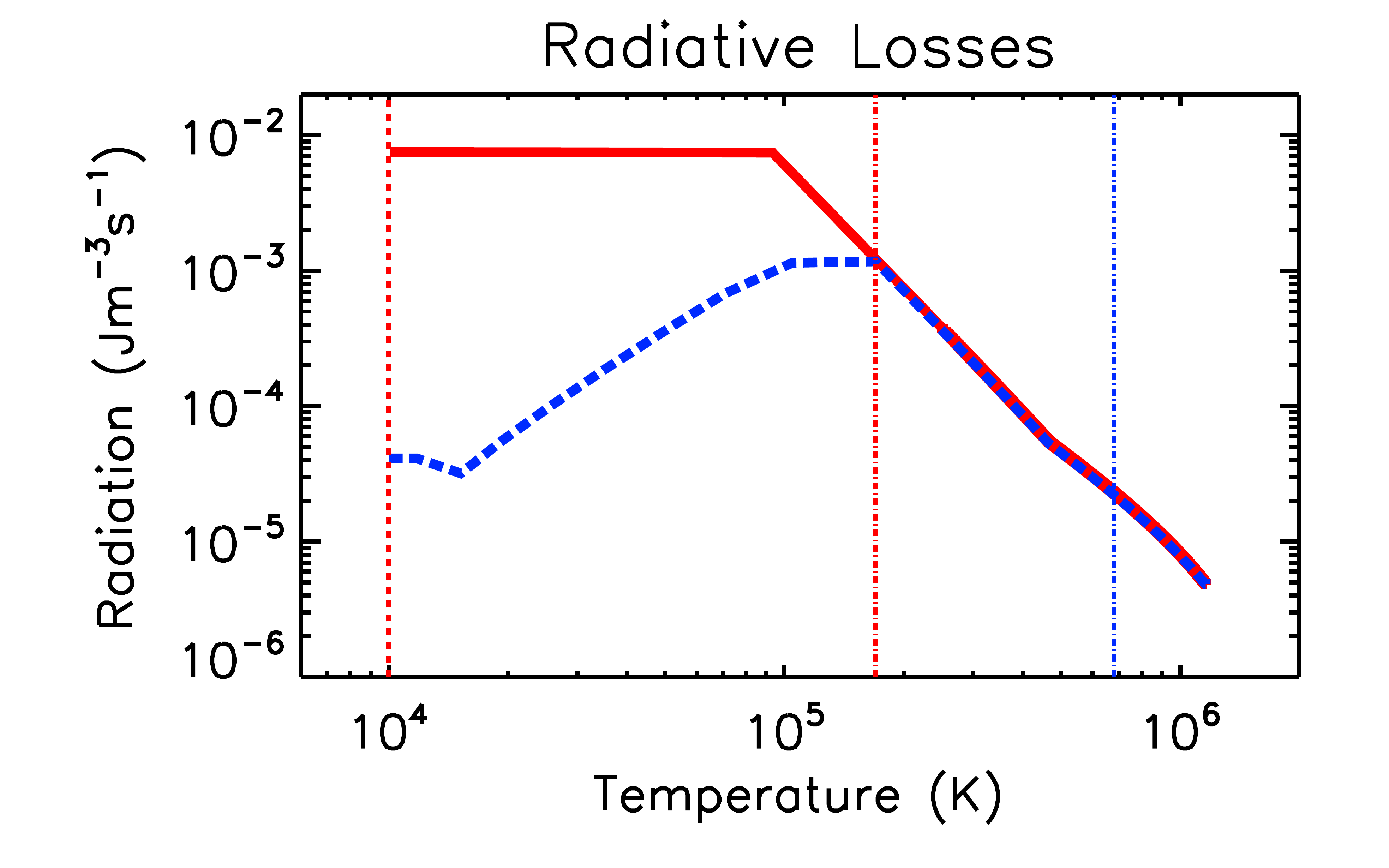}}
  \hspace*{-0.03\linewidth}
  \subfigure{\includegraphics[width=0.53\linewidth]
  {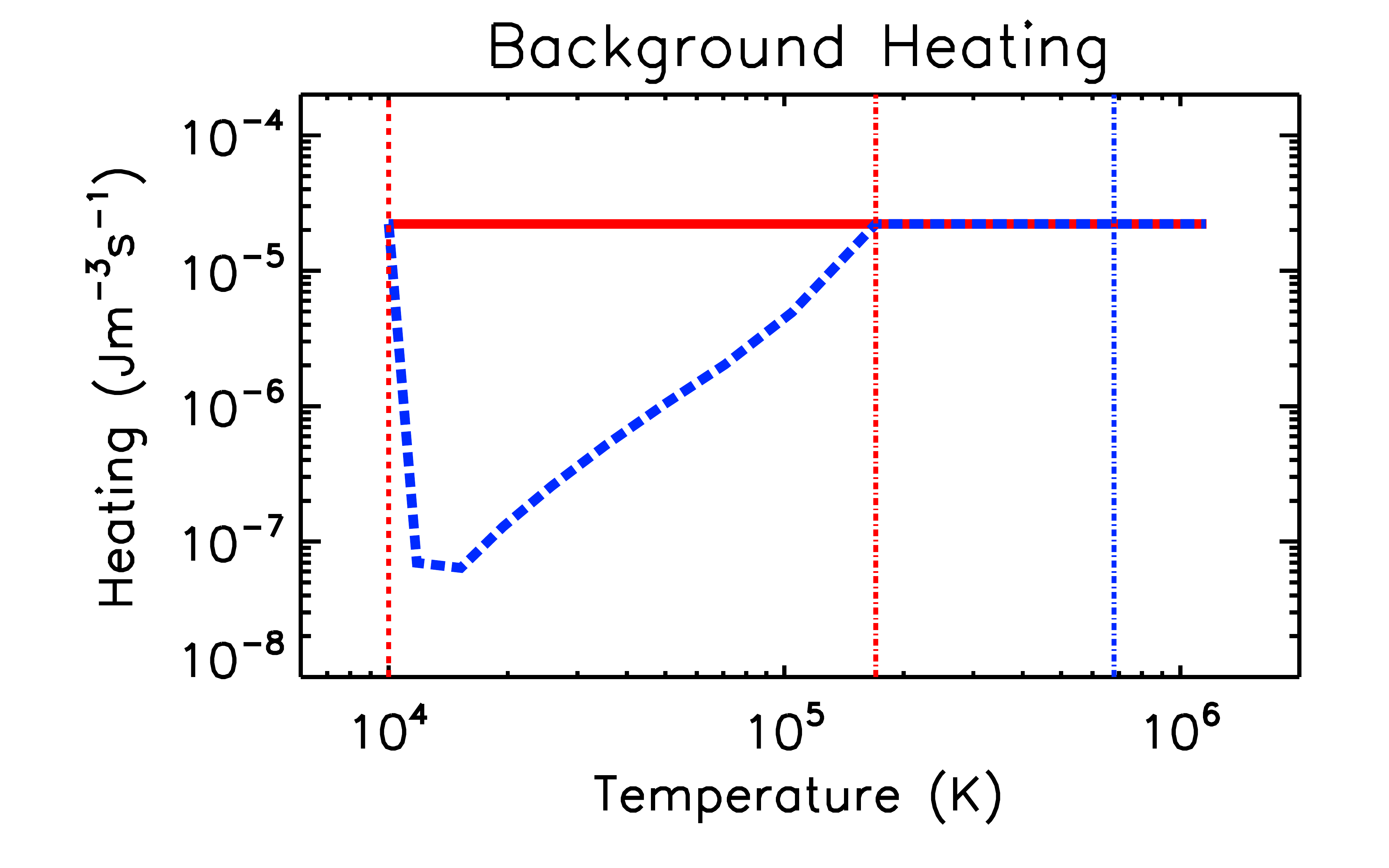}}
  \\[-7mm]
  \hspace*{-0.05\linewidth}
  \subfigure{\includegraphics[width=0.53\linewidth]
  {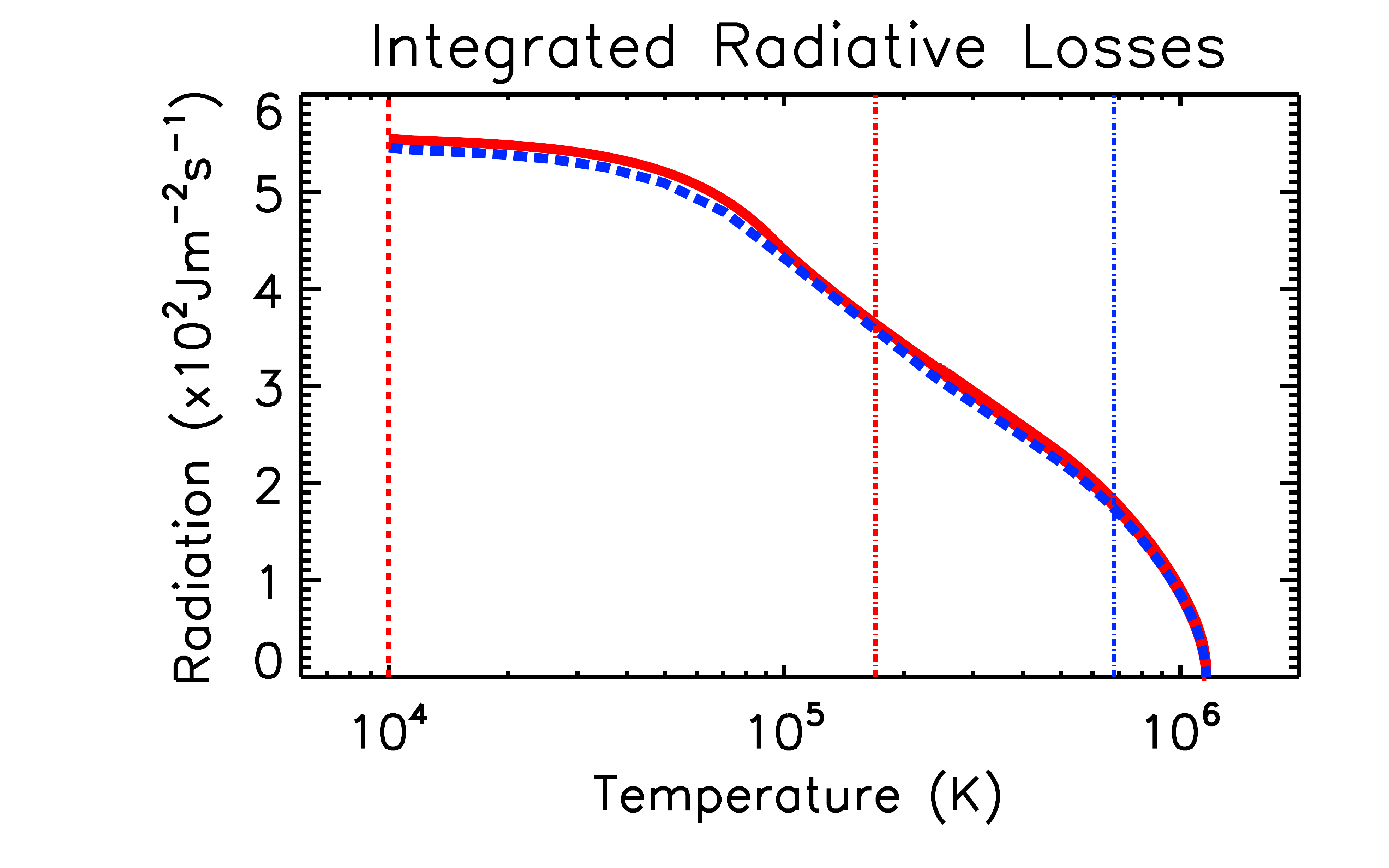}}
  \hspace*{-0.03\linewidth}
  \subfigure{\includegraphics[width=0.53\linewidth]
  {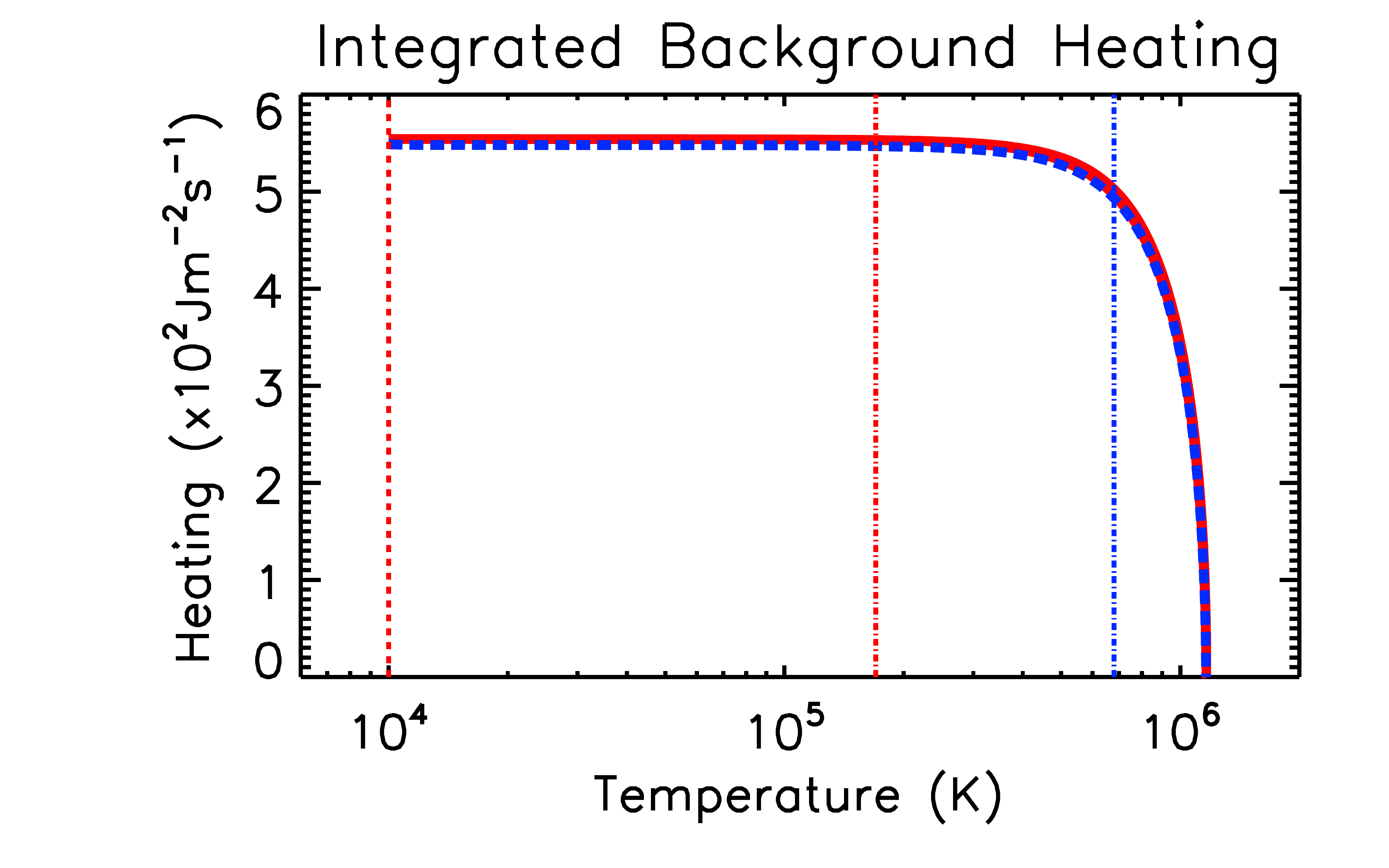}}
  \\[-8mm]
  \caption{Comparison of the
  SH and TRAC conduction methods
  for a one-dimensional loop in hydrostatic
  equilibrium.
  Upper two panels: Temperature and density
  as functions of position along the loop.
  Lower six panels: Parallel thermal conductivity,
  temperature length scale, local radiative losses,
  local background heating rate, 
  integrated radiative losses, and
  integrated background heating
  as functions of temperature.
  The lines are colour-coded in a way that reflects the 
  conduction method used with dashed blue (solid red) representing the TRAC 
  (SH) solution.
  The dashed red (blue) vertical line indicates the base of the TRAC (SH) TR 
  and the dot-dashed red (blue) vertical line the temperature 
  at the top of the TRAC region (the temperature at the top of~the~TR).
  \label{Fig:TRAC_SH_1D_equilibrium_comparison}
  }
\end{figure*}
  \indent
  Figure \ref{Fig:TRAC_SH_1D_equilibrium_comparison}
  shows the outcome of 
  implementing the method outlined above
  for a loop of total length 60~Mm, 
  in hydrostatic equilibrium with an apex
  temperature of 1.16~MK.
  In the upper two panels we focus on the TR,
  showing the temperature and density as a function of position. 
  An enlargement about the broadened region that uses the TRAC 
  conductivity, 
  referred to as the TRAC region, is also shown inset. 
  In the lower six panels, we show the parallel thermal conductivity,
  temperature length scale, local radiative losses, 
  local background heating, integrated radiative losses and
  integrated background heating
  as functions of temperature. 
  The integrated quantities are defined as being from the apex of the loop
  downwards to the base of the TR and are shown on a linear scale.
  In these panels, the solid red and dashed blue lines are the 
  SH and TRAC solutions, respectively.
  The TRAC (SH) solution is calculated using
  a grid size of approximately 60~km (60~m).
  Starting from the left, 
  the first 
  dashed red (blue) vertical line shows the base of 
  the TR for the TRAC (SH) solution, and the next
  dot-dashed red 
  line the top of the TRAC region. 
  The rightmost 
  vertical dot-dashed blue line is the top of the actual TR, 
  defined by 
  where the
  downward conduction changes sign from 
  a loss to a gain 
  \citep[e.g.][]{paper:Veseckyetal1979,
  paper:Klimchuketal2008}. 
  \\
  \indent
  The upper two panels (row 1) show the TR broadening 
  that is associated with the TRAC method,
  on the temperature and density structure in the lower TR.
  In particular, the TRAC region extends the 
  TR both below and above the SH location, as was also shown 
  for static loops by \cite{paper:Lionelloetal2009} and dynamic loops 
  in response to heating by
  \cite{paper:Johnstonetal2020}.
  The third panel demonstrates that the TRAC thermal
  conductivity is increased relative to the SH value
  only in under-resolved grid cells and reduces to the SH model elsewhere.
  This helps to broaden 
  the temperature length scale (fourth panel)
  in grid cells that would be under-resolved with the SH
  conduction method.
  For example,
  the minimum $L_T$ with the TRAC (SH) conduction method
  is of order 100~km (1~km)
  for the loop shown in   
  Fig. \ref{Fig:TRAC_SH_1D_equilibrium_comparison}.
  Thus, the TRAC temperature length scale
  satisfies the minimum resolution criteria 
  presented in Eq.~\eqref{Eqn:L_T_resolution_criteria}, 
  shown as the 
  lower solid green line in the fourth panel, 
  and the extent of the broadening 
  is bounded by the over-resolution limit described above
  in Eq. \eqref{Eqn:prime_conductivity} 
  (shown as the upper solid green line).
  The broadened temperature length scale obtained with TRAC prevents the
  heat flux from jumping across any unresolved regions while 
  maintaining accuracy in the
  properly resolved parts of the atmosphere
  (see e.g. JB19).
  \\
  \indent
  The lower four panels  of
  Fig. \ref{Fig:TRAC_SH_1D_equilibrium_comparison}
  demonstrate that moving from a global to a local cutoff temperature 
  conserves the properties of the original TRAC method.
  In particular, consistent with the analytical predictions of
  \cite{paper:Johnstonetal2020}, 
  the TRAC broadening modifications
  to the local radiative loss and heating rates (row 3), conserve 
  the total radiative losses and  total heating (row 4), 
  when integrated over the loop. 
  This preserves the energy balance in the TR and conserves the total 
  amount of energy 
  that is delivered to the chromosphere.
  The plots  
  also show that
  the top of the TR is at 0.68~MK while the top of the TRAC region is
  at 0.17~MK,
  corresponding to roughly 60\% and 15\% of the maximum loop temperature,
  respectively.
  The thickness of the TRAC region 
  is thus a small fraction of the TR thickness. 
  Therefore, the TRAC method has limited influence on the coronal
  properties of the loop.
  The result is that the SH 
  and TRAC temperature and density profiles converge a short 
  distance above the top of the TRAC region.
  \\
  \indent
  We also tested this implementation
  of TRAC on loops that evolve dynamically in response to heating.
  These simulations show the same fundamental properties as
  the hydrostatic loop and excellent agreement with
  \cite{paper:Johnstonetal2020},
  as discussed further in Appendix \ref{App:IoNR}.
  %
  %
%
%
\subsubsection{Magnetohydrodynamic implementation}
  \indent
  Equations 
  \eqref{Eqn:TRAC_conductivity}--\eqref{Eqn:prime_heating} 
  describe the
  field-aligned
  formulation of the TRAC 
  method that is used for the 
  multi-dimensional implementation. 
  The extension to
  MHD requires the generalisation of
  the
  mass flux ($J$), resolution ($L_R$)
  and temperature length scale ($L_T$) terms
  to account for the 
  magnetic field evolution.
  We define the mass flux parallel to the magnetic
  field as
  \begin{align}
    J =
    \dfrac{n({\bf B \cdot v} 
    + (b^2_{\rm min} |{\bf v}|^2)^{1/2})}
    {(B^2+b^2_{\rm min})^{1/2}},
  \end{align}  
  the field-aligned resolution is given by
  \begin{align}
    L_R =
    \dfrac{{\bf B \cdot L_R}
    + (b^2_{\rm min} |{\bf L_R}|^2)^{1/2})}
    {(B^2+b^2_{\rm min})^{1/2}},
  \end{align}
  where
  $
    {\bf L_R} =
    (\Delta x, \Delta y, \Delta z),
  $
  and the temperature length scale parallel to the magnetic field
  is defined as
  \begin{align}
    L_T = \dfrac{T(B^2+b^2_{\rm min})^{1/2}}
    {({\bf B \cdot \nabla}T 
    + (b^2_{\rm min} |{\bf \nabla}T|^2)^{1/2})}.
  \end{align}  
  We note that,
  analogous to the conduction model described in Eq.
  \eqref{Eqn:mhd_q_vector},
  finite $b_{\rm min}$ is used 
  here to make the TRAC
  conductivity isotropic when ${\bf B}={\bf 0}$.
  Furthermore,
  when TRAC is employed in the MHD model, 
  we calculate $\kappa_\parallel^{\prime}(T)$ in
  every grid cell and then
  solve the full set of MHD 
  equations \eqref{Eqn:mhd_continuity}--\eqref{Eqn:mhd_gas_law}, 
  but with the use of the modified
  $\kappa_\parallel^{\prime}(T)$, $\Lambda^{\prime}(T)$ and 
  $Q^{\prime}(T)$. 
  %
  %
%
%
\section{Numerical model and experiments
  \label{Sect:Model_and_experiments}}
  %
  %
%
%
\subsection{Unsheared and sheared arcades}
  \indent
  To demonstrate the viability of the MHD implementation of TRAC,
  we consider a 2D
  coronal arcade 
  given in cylindrical coordinates $(R, \theta, y)$ by
  \begin{align}
    {\bf B} = \left( 0, B_0\frac{R}{r}, B_1 \right),
  \end{align}  
  where $r$ is the radius and $R$, $B_0$ and $B_1$ are constants.
  If $B_1 = 0$~G, then the arcade is unsheared. 
  \\
  \indent
  In particular, we consider a surface given by $r = R$ and 
  express the
  azimuthal distance along the unsheared field as $x = R\theta$,
  where $L = R\pi$ is the length of a field line.  
  This variable transformation
  enables the derivatives along the magnetic field to be written in the
  form
  \begin{align}
    {\bf B} \cdot \nabla 
    = \left( \frac{B_0}{R}\frac{\partial}{\partial \theta}
    + B_1\frac{\partial}{\partial y}\right) 
    = \left( B_0 \frac{\partial}{\partial x} 
    + B_1 \frac{\partial }{\partial y}\right).
  \end{align}  
  Hence, we can model an arcade of coronal loops
  using a straight field geometry
  with a spatially varying gravity, where $x$ 
  is the spatial coordinate along the magnetic field and $y$ represents the 
  transverse direction.
  The expression for the
  gravitational acceleration along the field is then given by
  \begin{align}
    -\frac{{\bf g} \cdot {\bf B}}{|{\bf B}|} =  -g
    \frac{B_0 \cos (\pi x/L)}{\sqrt{B_0^2+B_1^2}}.
  \end{align}  
  \\
  \indent
  For our unsheared arcade model, we take $B_0=100$~G and consider a
  computational domain of dimensions 
  $(x,y) = 60$~Mm~$\times$~2.4~Mm.
  The numerical grid used to resolve this domain is comprised
  of 1024 grid points in $x$ (field-aligned direction),
  while the influence of using low and high resolution 
  in $y$~(cross-field direction) 
  will be examined in Sections \ref{Sect:1D_Results}
  \& \ref{Sect:2D_Sheared_arcade}.
  We stratify the initial atmosphere  using the broadened,
  field-aligned temperature and density profiles 
  shown in Figure \ref{Fig:TRAC_SH_1D_equilibrium_comparison},
  which are calculated using the TRAC conduction method
  and a grid size of approximately 60~km ($N_x=1024$) along the field.
  It is demonstrated in Appendix \ref{App:IoNR} that this 
  field-aligned resolution is adequate to fully resolve the TR,
  when the TRAC method is employed,
  for all of the simulations presented in this paper.
  Here, each field line in the arcade starts in static equilibrium
  and the total length of each field line ($L=60$~Mm) includes 
  a 5~Mm chromosphere 
  at the base of each TR. 
  Following \cite{paper:Johnstonetal2017a}, 
  the chromosphere is modelled as a
  mass reservoir with
  a constant temperature of $10^4$~K, and 
  gravitationally stratified density, while
  the optically thin radiative losses 
  are reduced to zero to maintain the isothermal temperature
  \citep{paper:Klimchuketal1987,paper:Bradshaw&Cargill2013}.
  \\
  \indent
  To model a sheared arcade we set $B_1=4$~G. 
  This tilts the field so that the field lines are no longer 
  aligned with the numerical grid.
  The initial conditions are adjusted accordingly
  to ensure the initial state is an 
  equilibrium.
  %
  %
%
%
\subsection{Non-uniform coronal heating pulse}
  %
  %
  %
  %
  %
  %
  \begin{figure}
    \subfigure{\includegraphics[width=\linewidth]
    {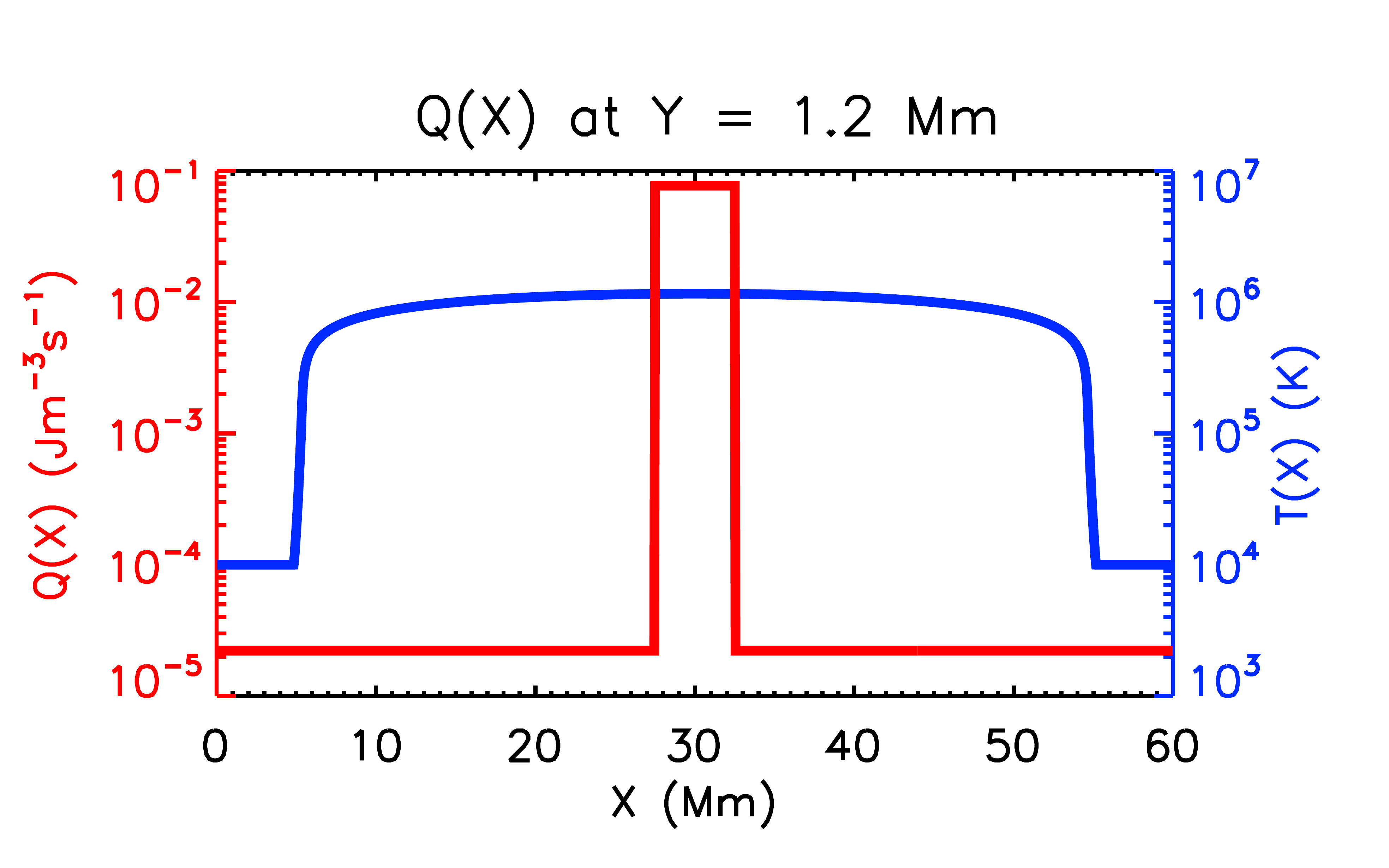}}
    \\[-8mm]
    \subfigure{\includegraphics[width=\linewidth]
    {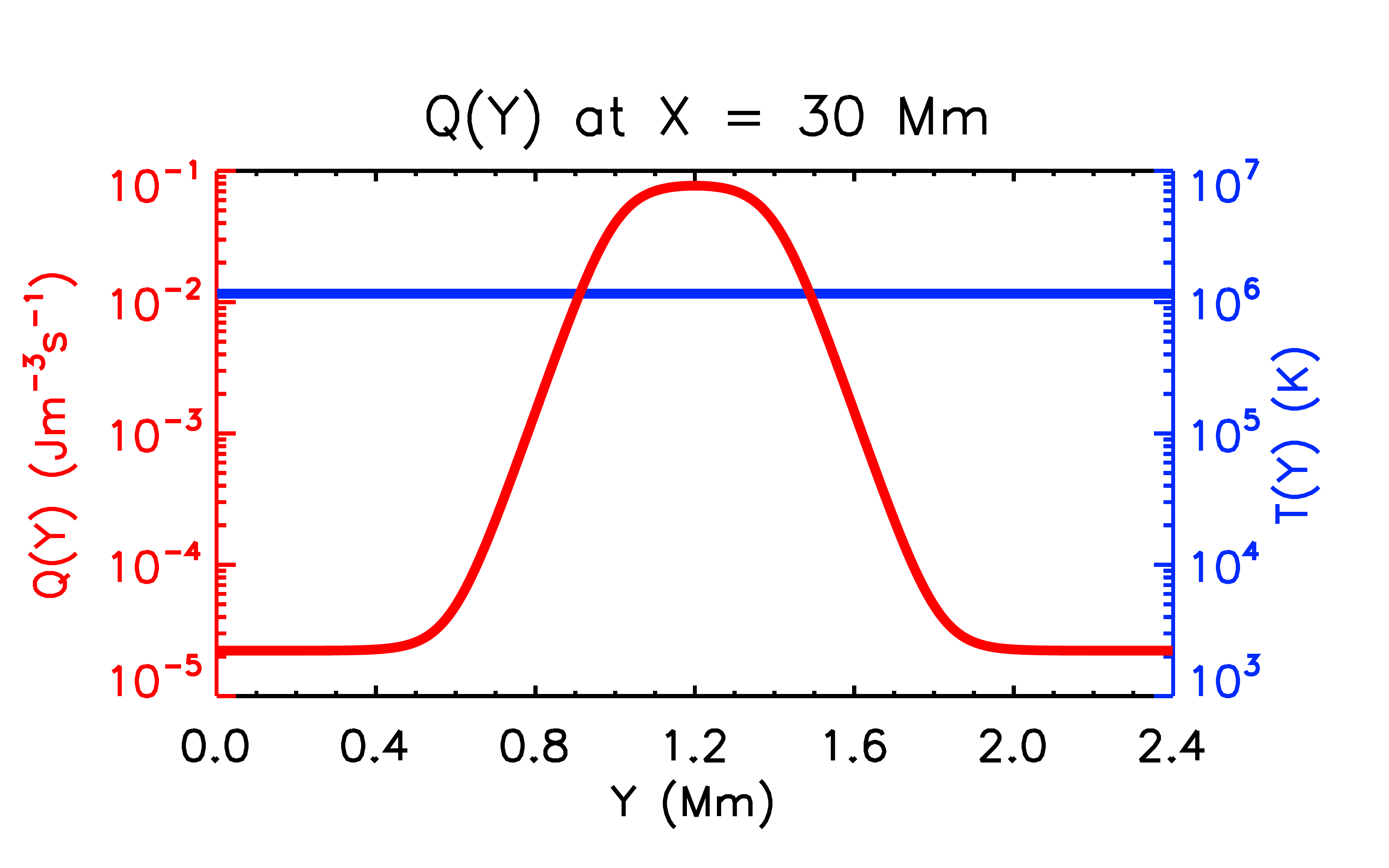}}
    \\[-6mm]
    \caption{
    \label{Fig:Heating_Function}
    Spatially non-uniform heating profile $Q(x,y)$
    (solid red line, left-hand axis) 
    used in Sections \ref{Sect:Model_and_experiments} 
    and \ref{Sect:2D_Results}, 
    imposed on top of the temperature initial condition
    (solid blue line, right-hand axis). 
    The upper (lower) panel shows the variation of the
    heating profile in the  field-aligned (transverse) direction
    at the time of peak heating.
    }
  \end{figure}
  \indent
  The effectiveness of the MHD implementation of TRAC
  is investigated by considering 
  a spatially non-uniform, impulsive coronal heating event,
  comprised of a short pulse 
  that lasts for a total duration of 60 s, where the energy deposition
  is localised at the centre of the computational domain.
  The temporal profile of the heating pulse is triangular
  with a linear ramp up to the peak heating rate
  ($Q_{H_0}$) followed by a 
  linear decrease, while the field-aligned heating profile 
  is square, confining 
  the energy release to the uppermost 5~Mm located at the apex of the loop,
  as shown in Figure \ref{Fig:Heating_Function}
  ($Q(x)$ at $y=1.2$~Mm is displayed in the upper panel as the red curve).
  However, we note that 
  the coronal response to apex heating is similar to that of
  uniform heating 
  because thermal conduction is very efficient at coronal 
  temperatures 
  \citep[see e.g.][]{paper:Johnstonetal2017a,paper:Johnstonetal2017b}.
  Thus,
  the results presented here are not highly sensitive to 
  the form of heating profile that is used in the field-aligned direction,
  given that the same 
  total amount of energy is released at a sufficient height above the
  footpoints of the loop.
  \\
  \indent
  For the transverse direction, the spatial profile of the heating pulse
  is given by
  \begin{align}
    Q_H(y) =
    \dfrac{Q_{H_0}}{2}
    \left(
    \tanh
    \left(
    \dfrac{y - y_L}{y_H}
    \right)
    -
    \tanh
    \left(
    \dfrac{y - y_R}{y_H}
    \right)
    \right),
    \label{Eqn:heating_function} 
  \end{align}
  where $Q_{H_0}$
  is the maximum heating rate, 
  $y_H$ = 100~km is the transverse length scale of heat deposition
  and we take $y_L=1$~Mm and $y_R=1.4$~Mm to give the maximal
  heating value at
  $y=1.2$~Mm.
  This is broadly consistent with the spatial distributions of heating 
  reported by \cite{paper:Reidetal2020}.
  We use a maximum heating rate of
  $Q_{H_0} = 8 \times 10^{-2}$~Jm$^{-3}$s$^{-1}$,
  which 
  corresponds roughly 
  to an active region nanoflare \citep{paper:Bradshaw&Cargill2013}.
  \\
  \indent
  A small spatially uniform background heating term is also
  present so that $Q(x,y) = Q_{bg} + Q_H(x,y)$,
  where $Q_{bg} = 2.2167 \times 10^{-5}$~Jm$^{-3}$s$^{-1}$.
  The initial state of the loop is determined using just $Q_{bg}$, 
  leading to an apex
  temperature of 1.16~MK.
  Defining the transverse heating length scale as
  \begin{align}
    L_{Q_y}=\dfrac{Q}{|\partial Q/ \partial y|},
    \label{Eqn:L_Q_y} 
  \end{align}
  the lower panel of 
  Figure \ref{Fig:Heating_Function} shows that 
  moving outwards from  $y=1.2$~Mm,
  the transverse heating profile decreases smoothly
  from $Q_{H_0}$ to $Q_{bg}$
  with a
  minimum heating length scale of
  $L_{Q_y}=50$~km.
  %
  %
%
%
\subsection{One-dimensional simulations of the unsheared arcade
  \label{Sect:1D_Results}}
  %
  %
  %
  %
  %
  %
\begin{figure*}
  \hspace*{-0.03\linewidth}
  \subfigure{\includegraphics[width=0.53\linewidth]
  {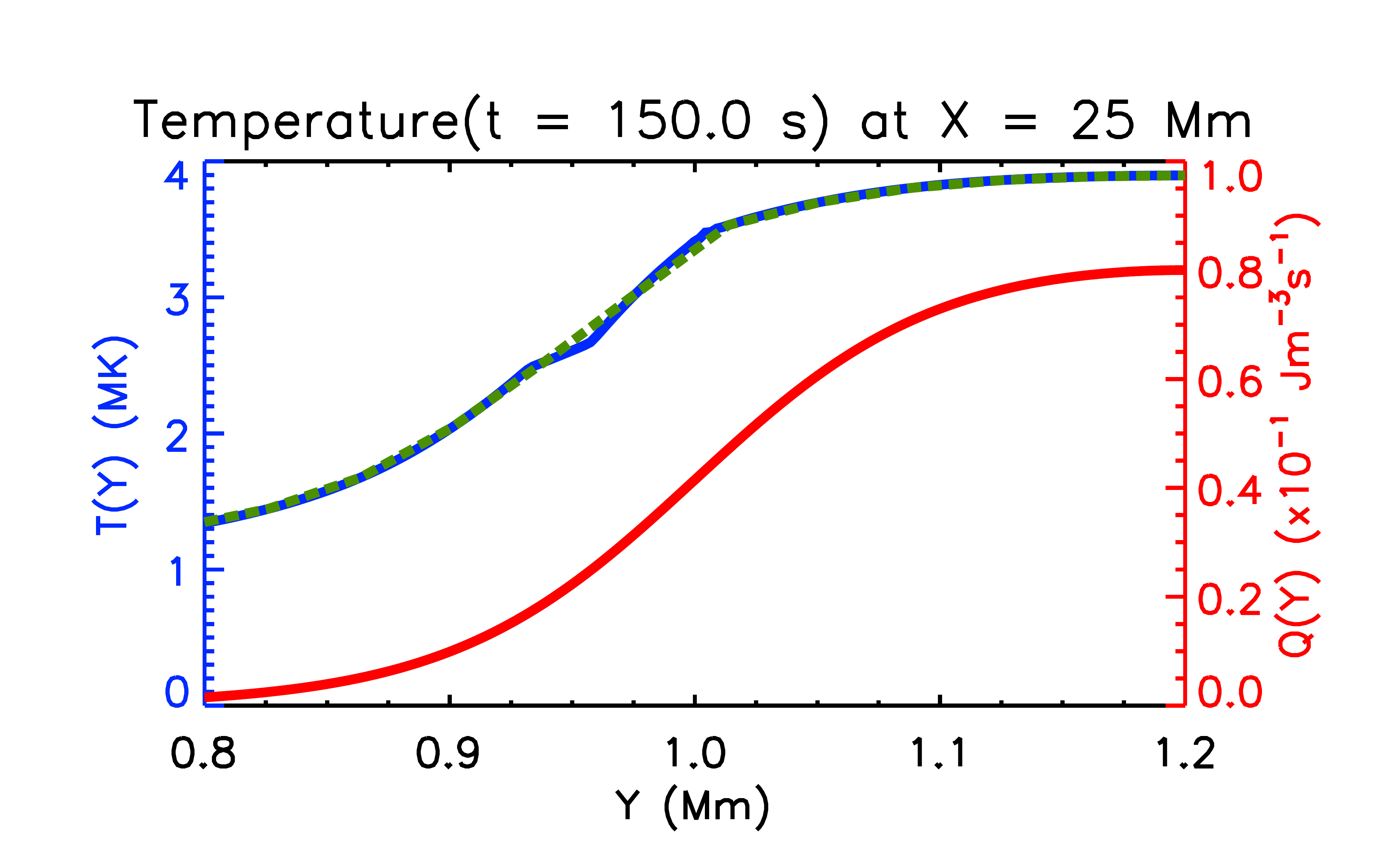}}
  \hspace*{-0.03\linewidth}
  \subfigure{\includegraphics[width=0.53\linewidth]
  {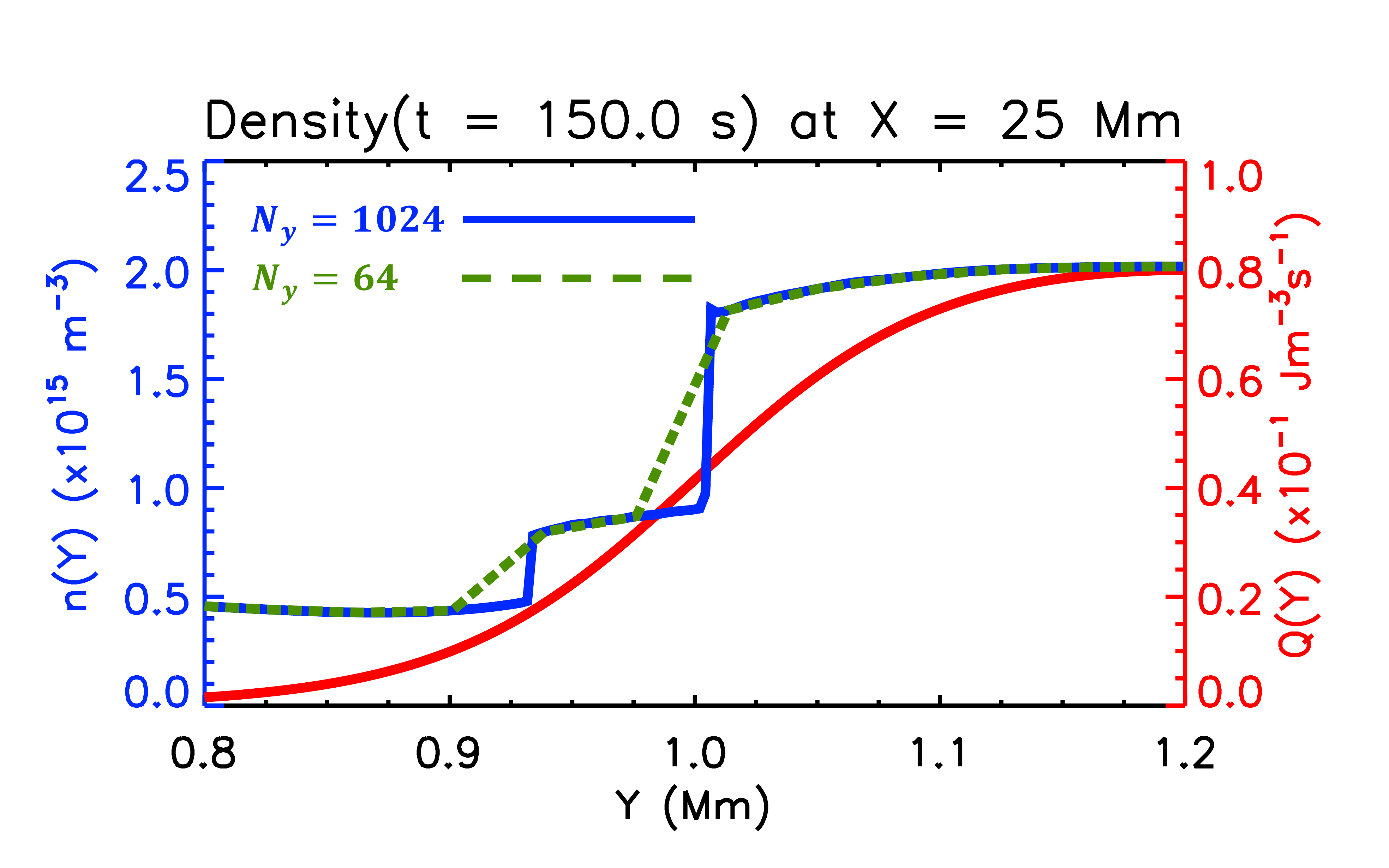}}
  \\[-8mm]
  \hspace*{-0.03\linewidth}
  \subfigure{\includegraphics[width=0.53\linewidth]
  {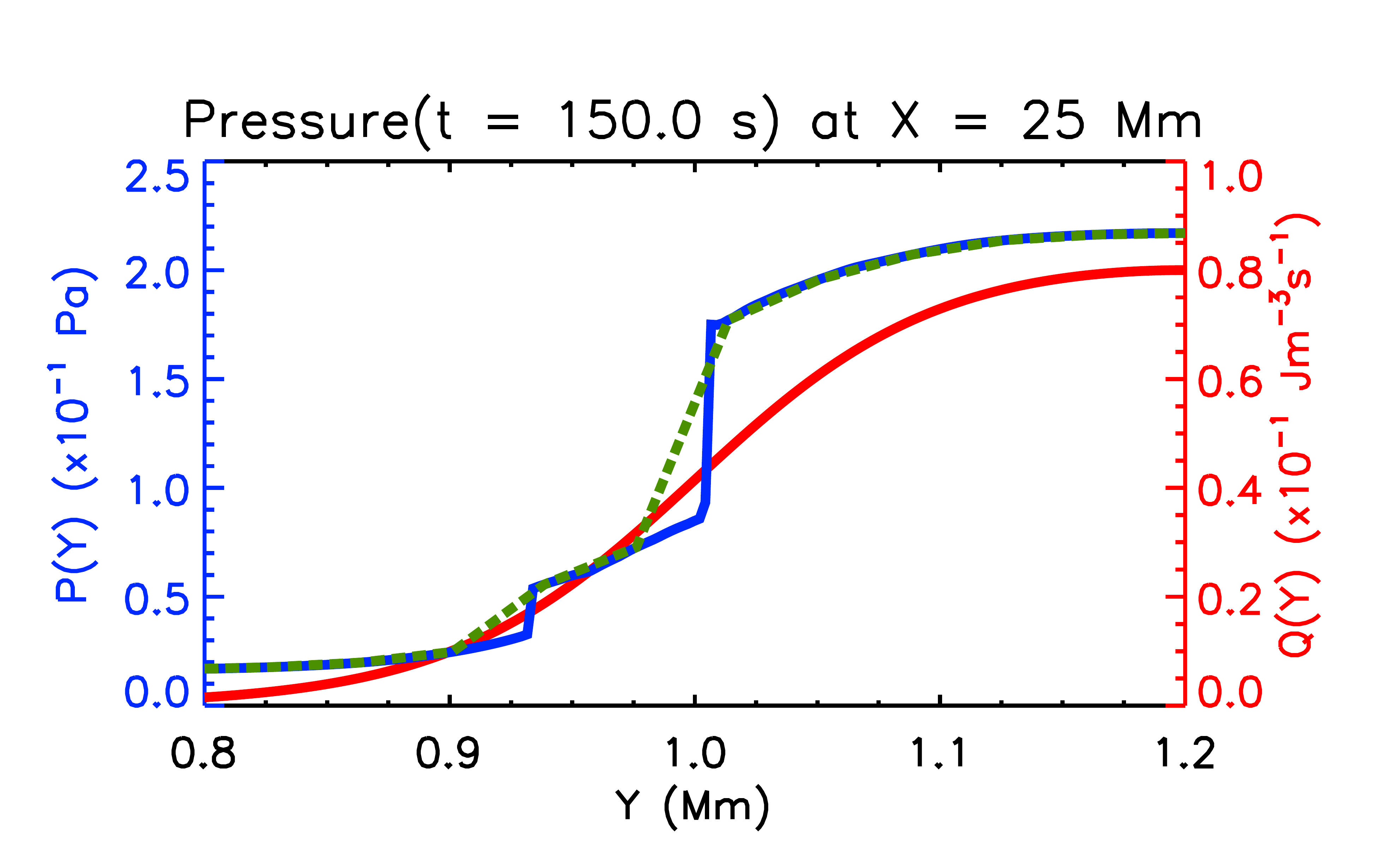}}
  \hspace*{-0.03\linewidth}
  \subfigure{\includegraphics[width=0.53\linewidth]
  {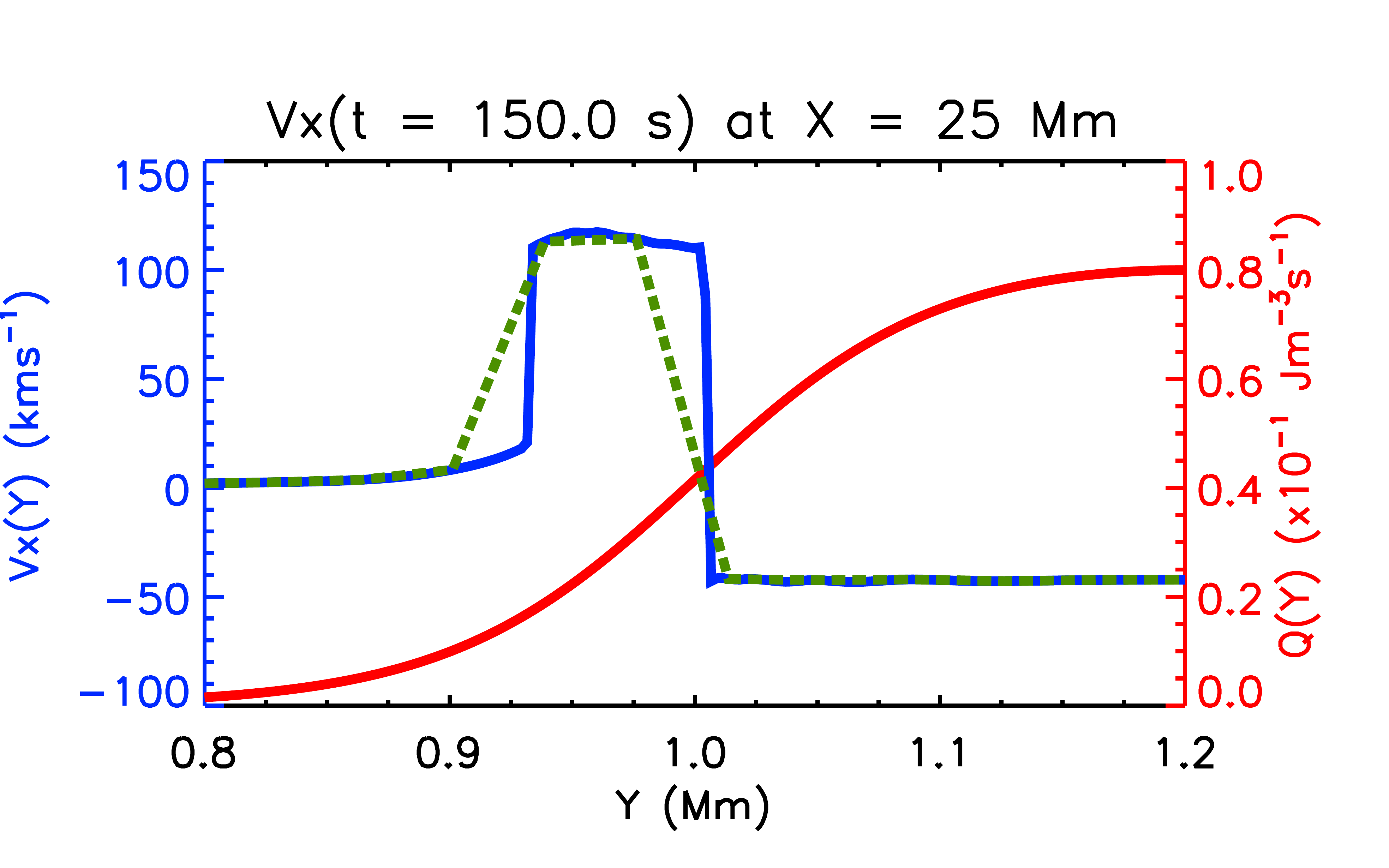}}
  \\[-6mm]
  \caption{
    Results for the reconstruction of the 
    non-uniform coronal heating pulse using one-dimensional 
    HD simulations of the unsheared arcade
    (Sect. \ref{Sect:1D_Results}).
    The panels show the temperature, density, pressure, and 
    field-aligned velocity 
    as functions of position across the arcade~(left-hand axis),
    at a coronal height of $x=25$~Mm,
    at the time of the first density peak $(t=150$~s).
    The lines are colour coded in a way that reflects the
    transverse resolution used with solid blue (dashed green)
    representing the $N_y=1024$ ($N_y=64$) solution,
    which is imposed on top of the transverse heating profile 
    (solid red line, right-hand axis). 
    \label{Fig:1D_Reconstruction_transverse_profiles}
  }
\end{figure*}
  %
  %
  %
  %
  %
  %
  %
  %
\begin{sidewaysfigure*}
  \hspace*{0.02\linewidth}
  \subfigure{\includegraphics[width=0.28\linewidth]
  {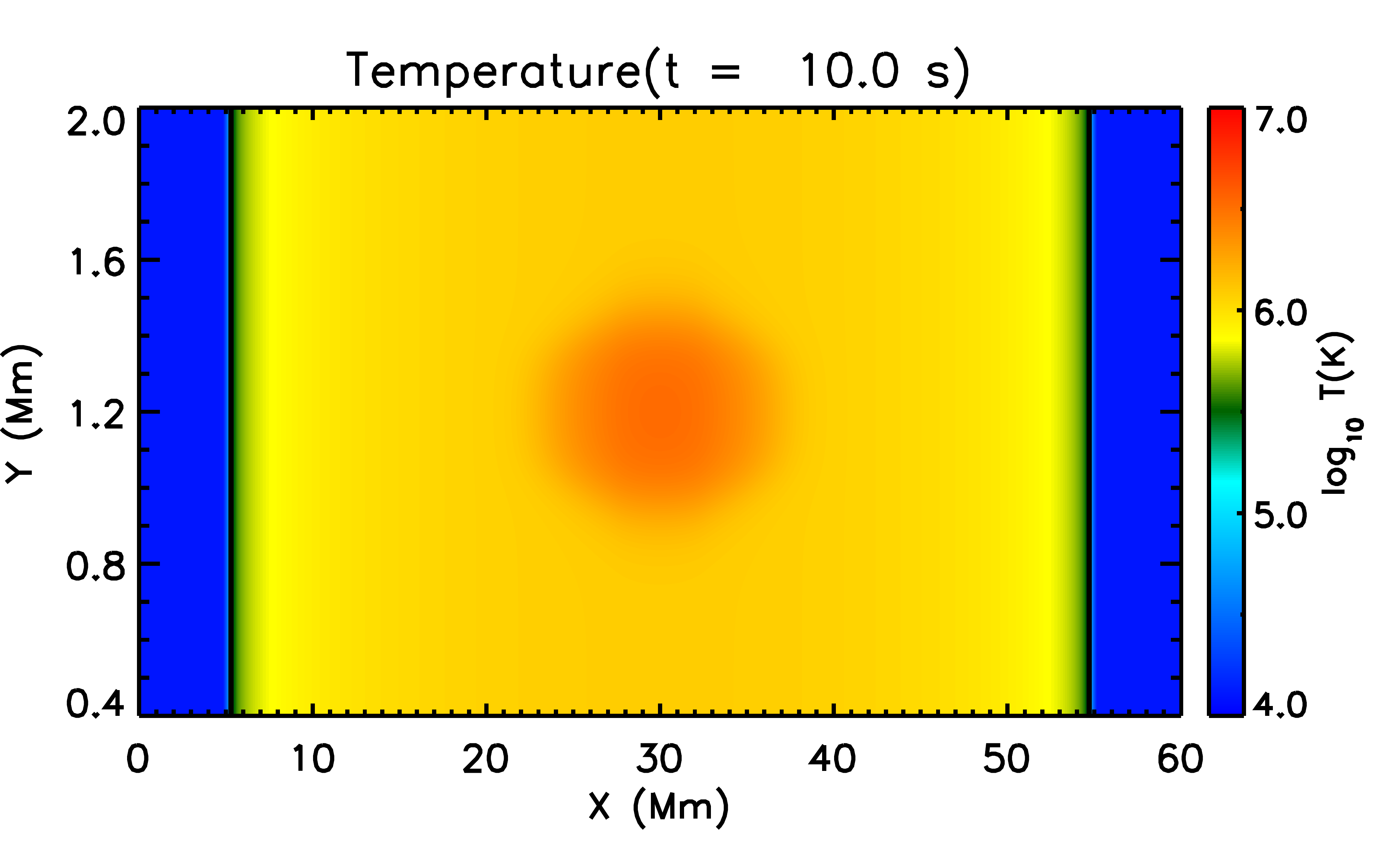}}
  \hspace*{0.04\linewidth}
  \subfigure{\includegraphics[width=0.28\linewidth]
  {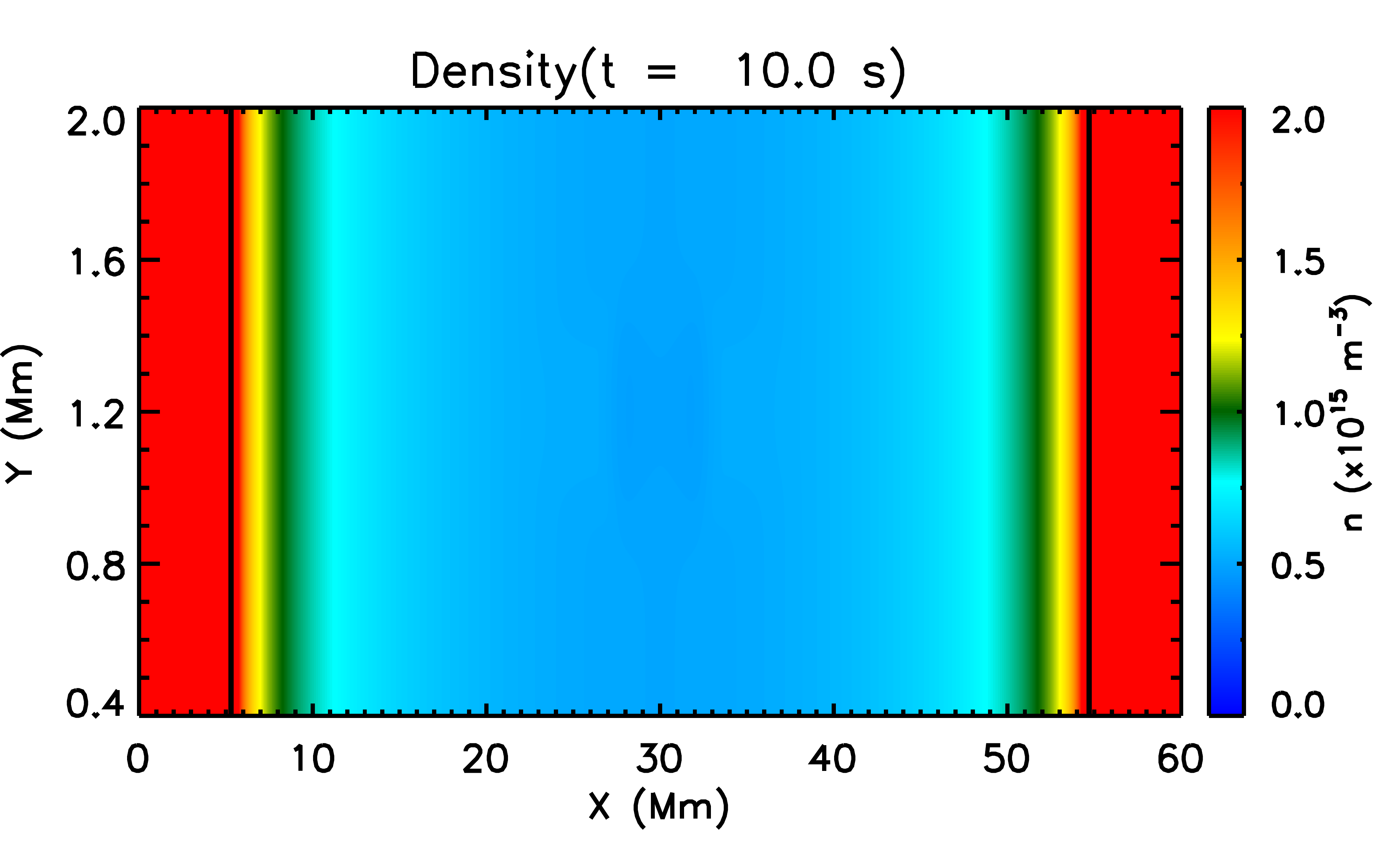}}
  \hspace*{0.04\linewidth}
  \subfigure{\includegraphics[width=0.28\linewidth]
  {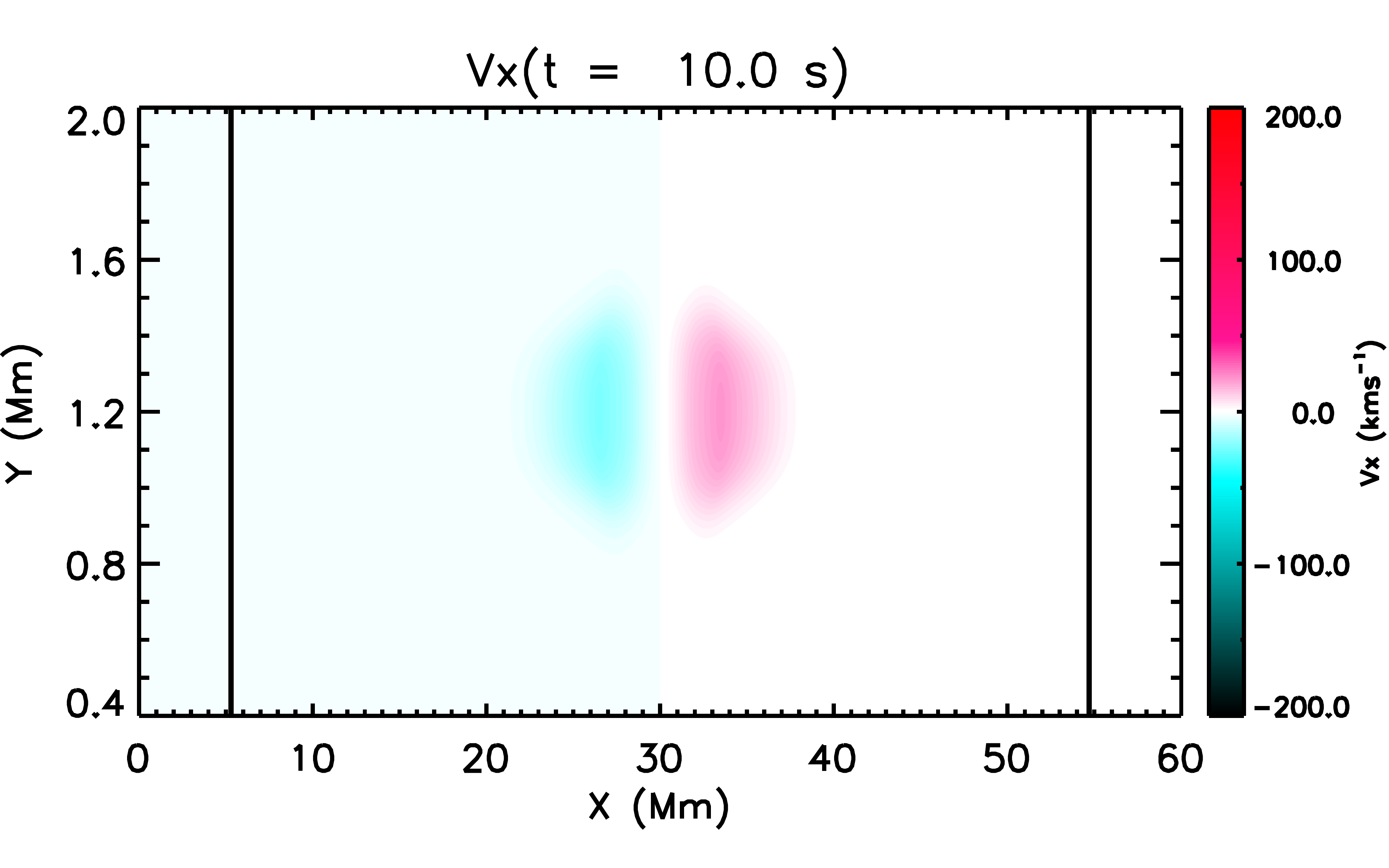}}
  \\[-2mm]
  \hspace*{0.02\linewidth}
  \subfigure{\includegraphics[width=0.28\linewidth]
  {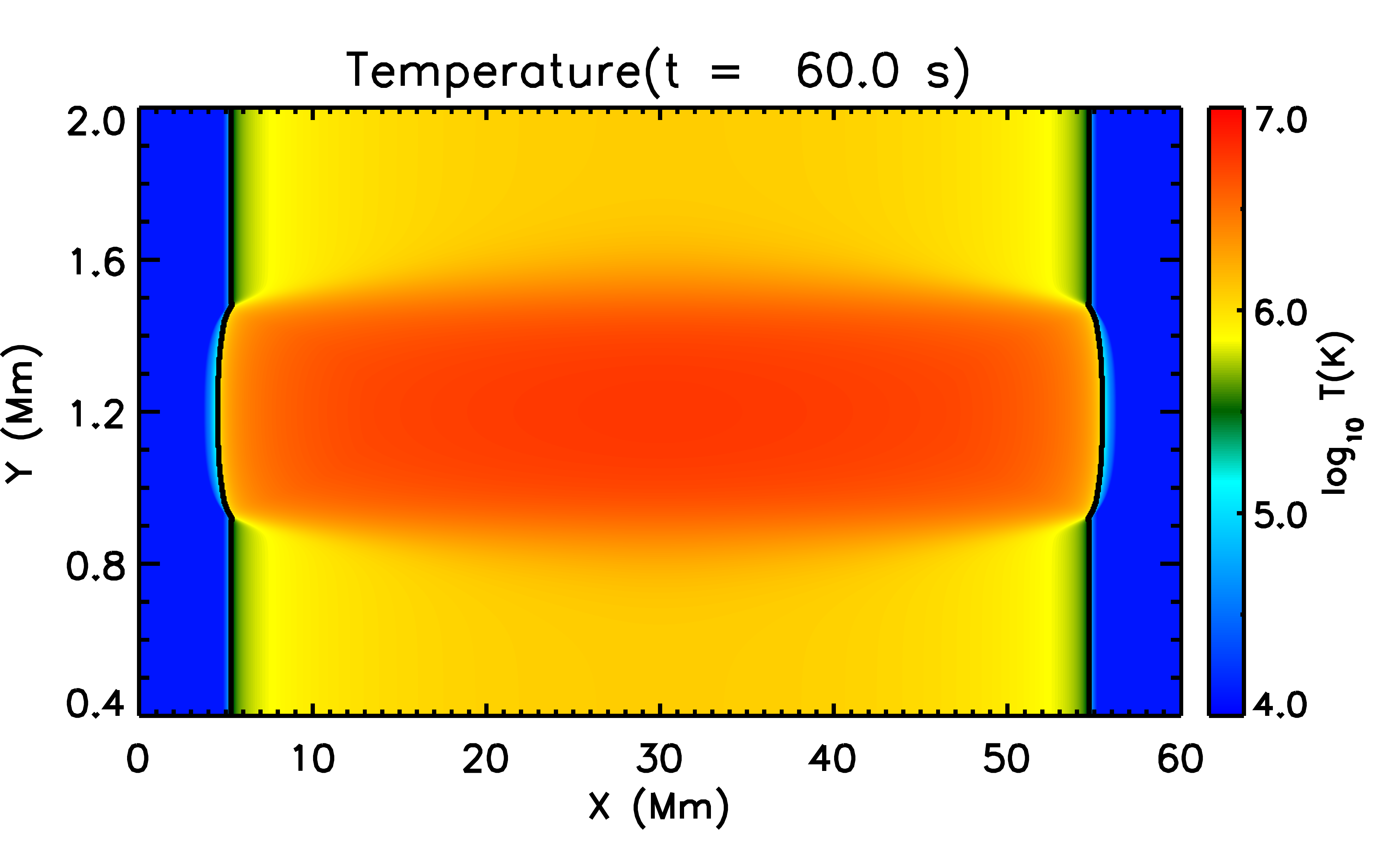}}
  \hspace*{0.04\linewidth}
  \subfigure{\includegraphics[width=0.28\linewidth]
  {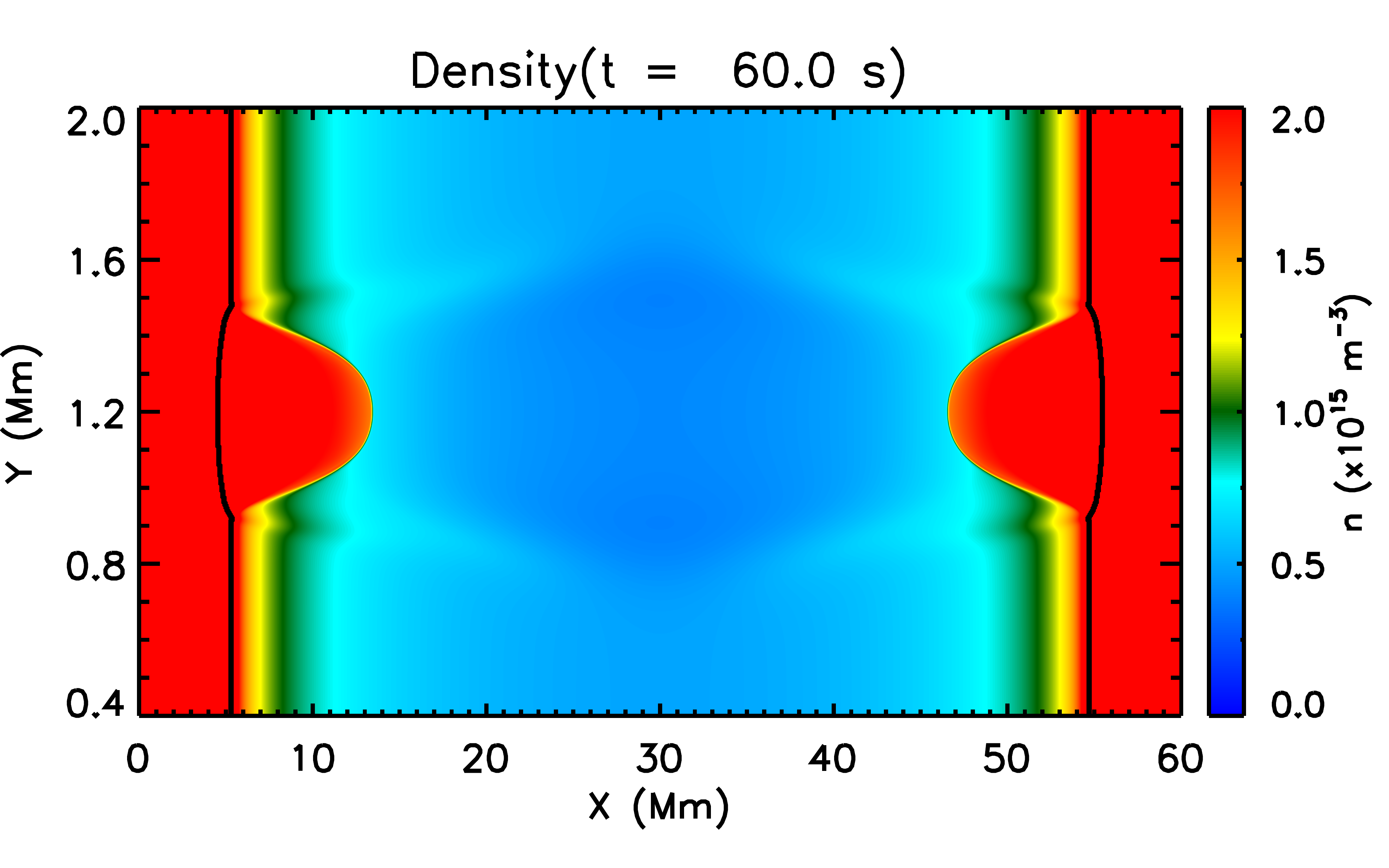}}
  \hspace*{0.04\linewidth}
  \subfigure{\includegraphics[width=0.28\linewidth]
  {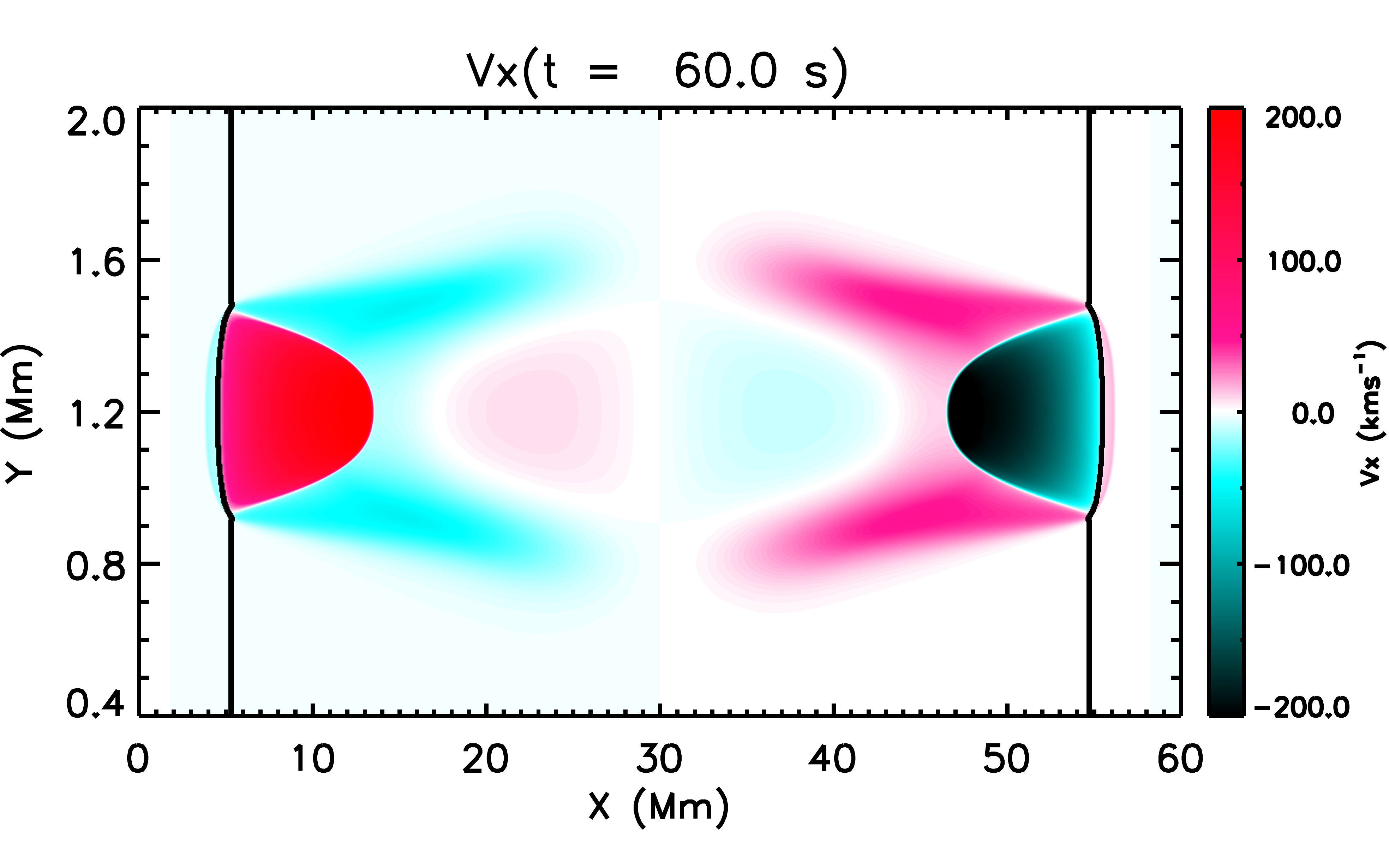}}
  \\[-2mm]
  \hspace*{0.02\linewidth}
  \subfigure{\includegraphics[width=0.28\linewidth]
  {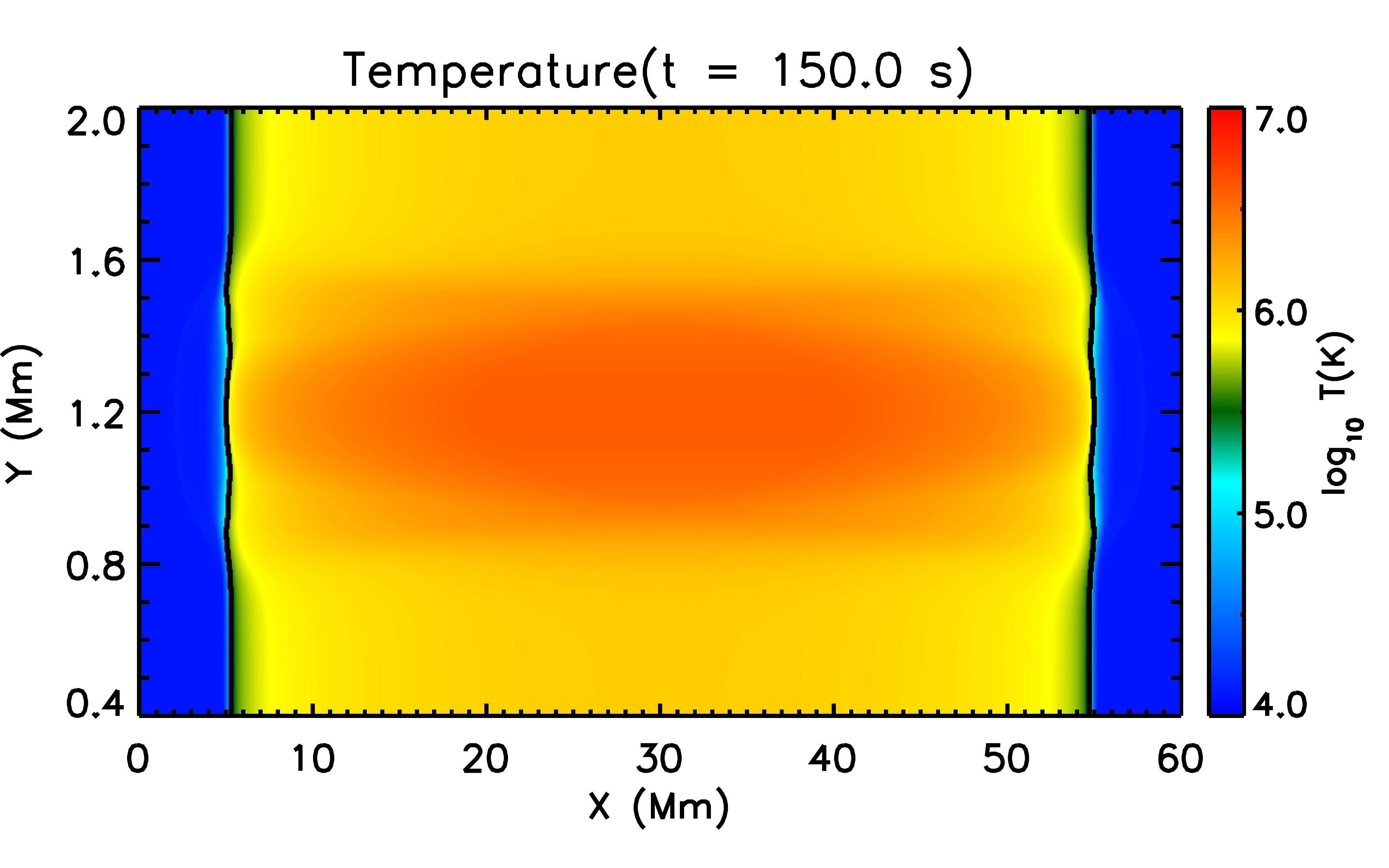}}
  \hspace*{0.04\linewidth}
  \subfigure{\includegraphics[width=0.28\linewidth]
  {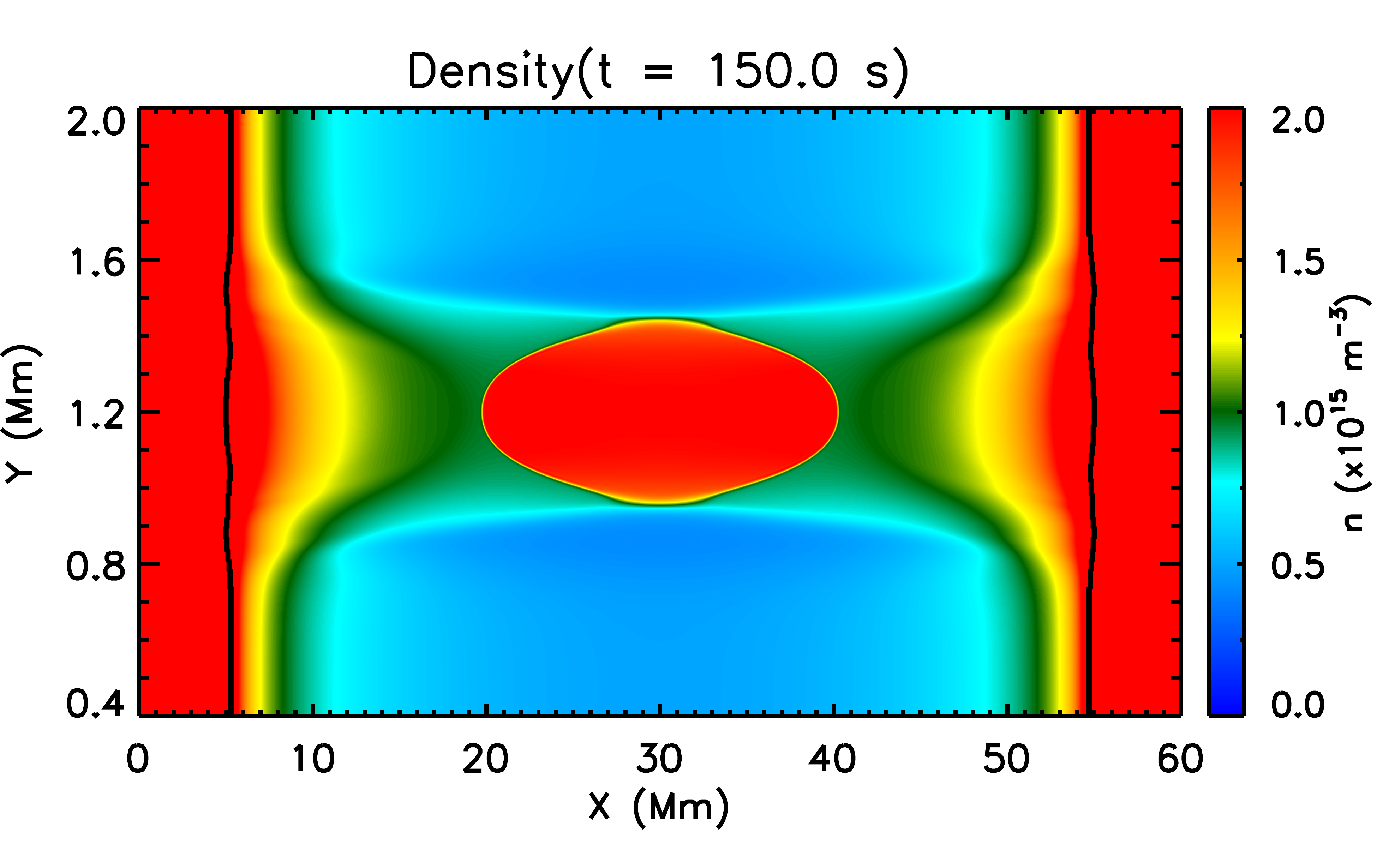}}
  \hspace*{0.04\linewidth}
  \subfigure{\includegraphics[width=0.28\linewidth]
  {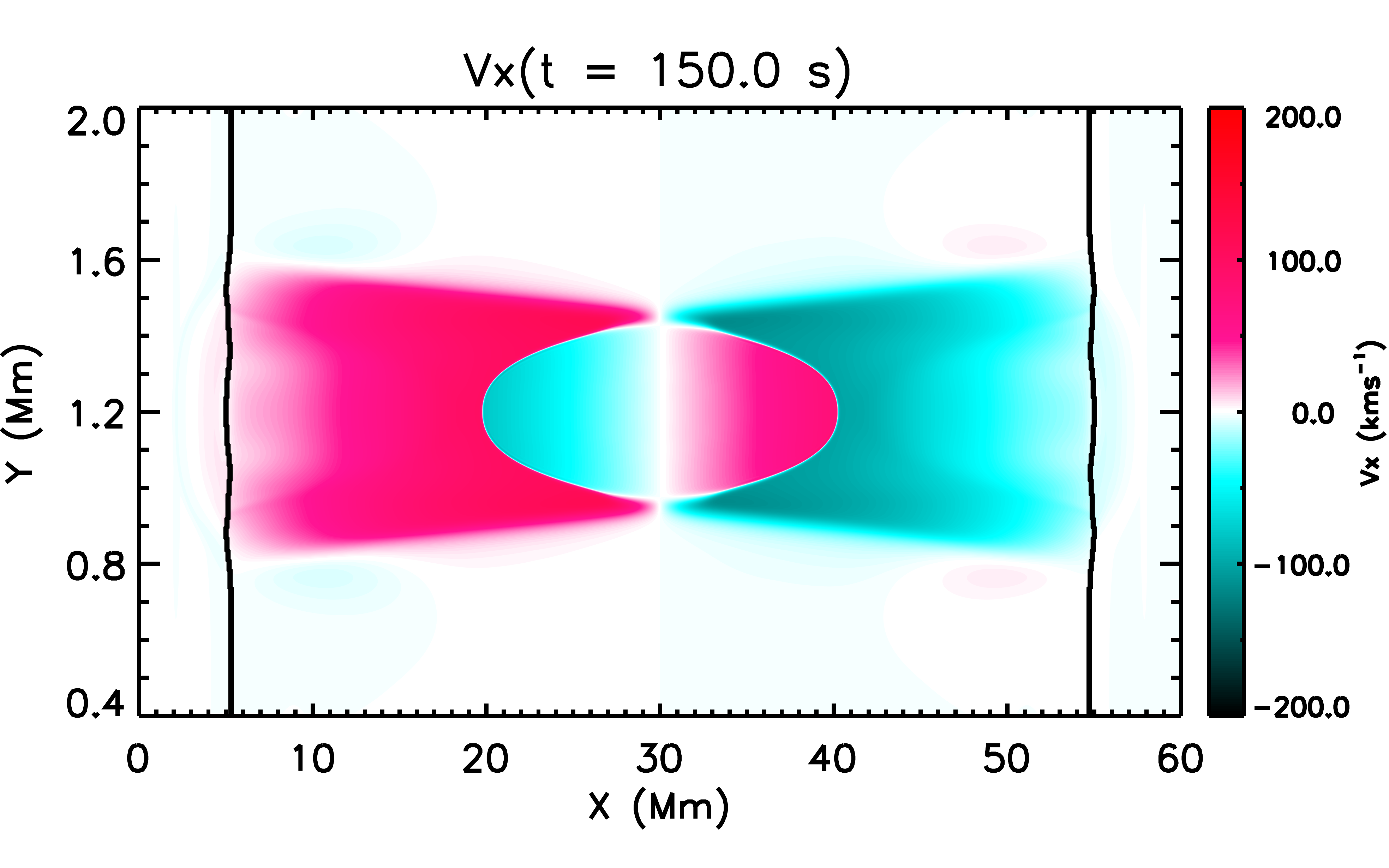}}
  \\[-2mm]
  \hspace*{0.02\linewidth}
  \subfigure{\includegraphics[width=0.28\linewidth]
  {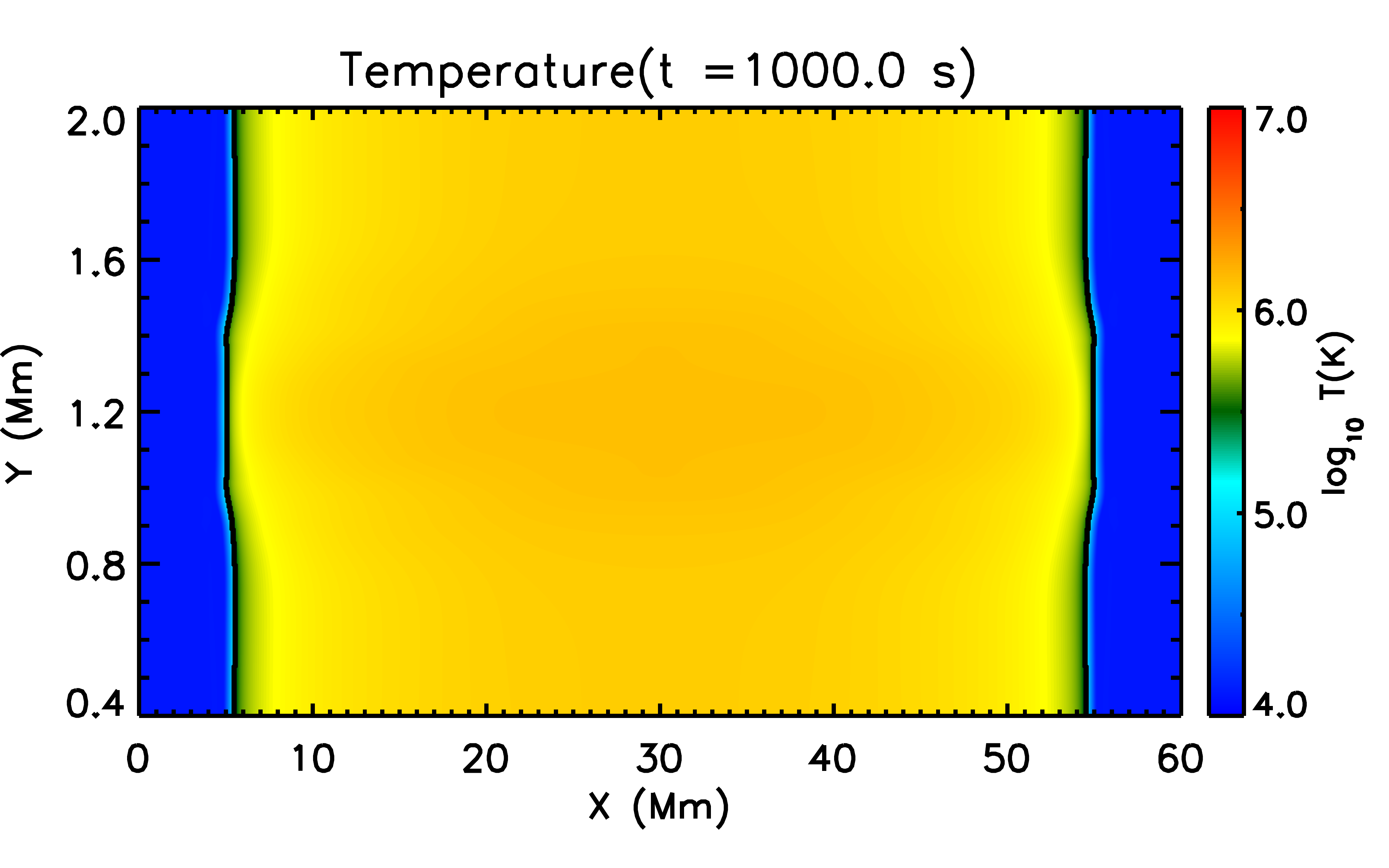}}
  \hspace*{0.04\linewidth}
  \subfigure{\includegraphics[width=0.28\linewidth]
  {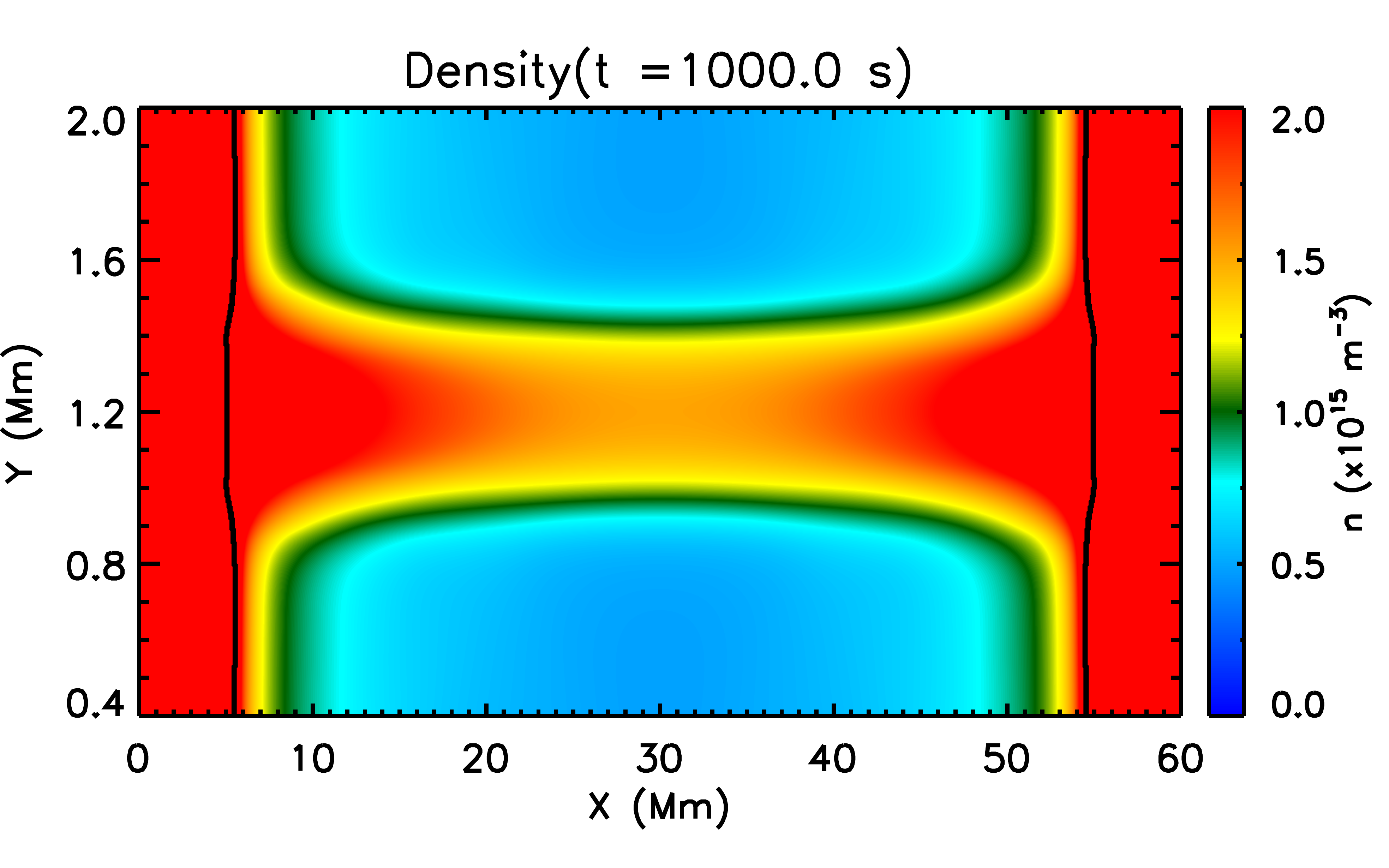}}
  \hspace*{0.04\linewidth}
  \subfigure{\includegraphics[width=0.28\linewidth]
  {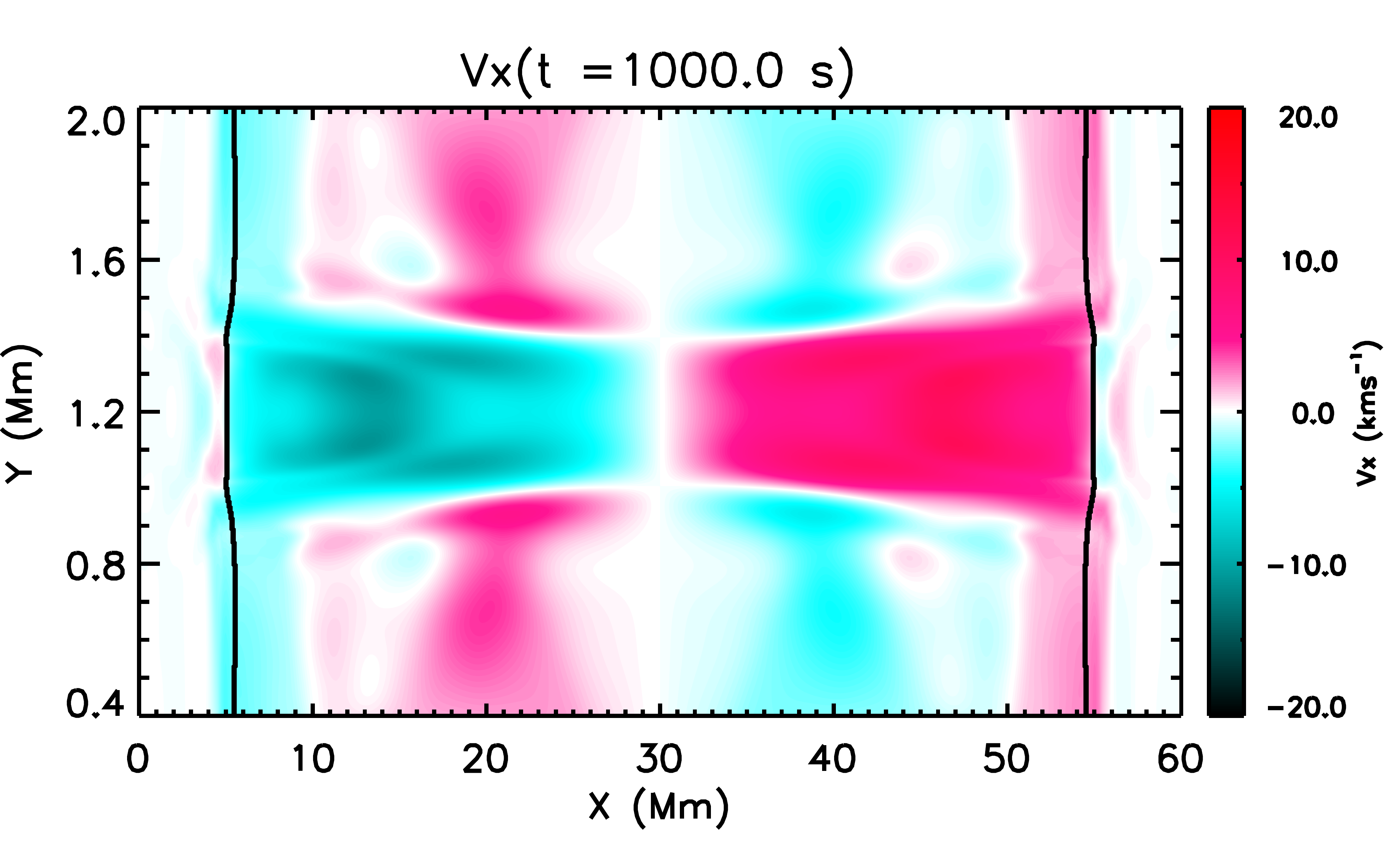}}
  \\[-6mm]
  \caption{
    Results for the 
    non-uniform coronal heating pulse using a two-dimensional
    MHD simulation of the unsheared arcade
    (Sect. \ref{Sect:2D_Unsheared_arcade}).
    Starting from the left, the columns show time ordered snapshots of the
    temperature ($T$), density ($n$) and field-aligned velocity ($v_x$)
    for times during the heating, evaporation and decay phases. 
    The contours are drawn according to the scales shown in the 
    colour tables.
    The black curves indicate the top of the TRAC region
    at each of the footpoints of the unsheared arcade.    
    Movies of the full time evolution of the $T$, $n$ and $v_x$
    contour plots can be viewed in the online version of this article.
    \label{Fig:2D_Simulation_UA_time_evolution_contours}
  }
\end{sidewaysfigure*}
  %
  %
  %
  %
  %
  %
  %
  %
\begin{figure*}
  \hspace*{-0.05\linewidth}
  \subfigure{\includegraphics[width=0.53\linewidth]
  {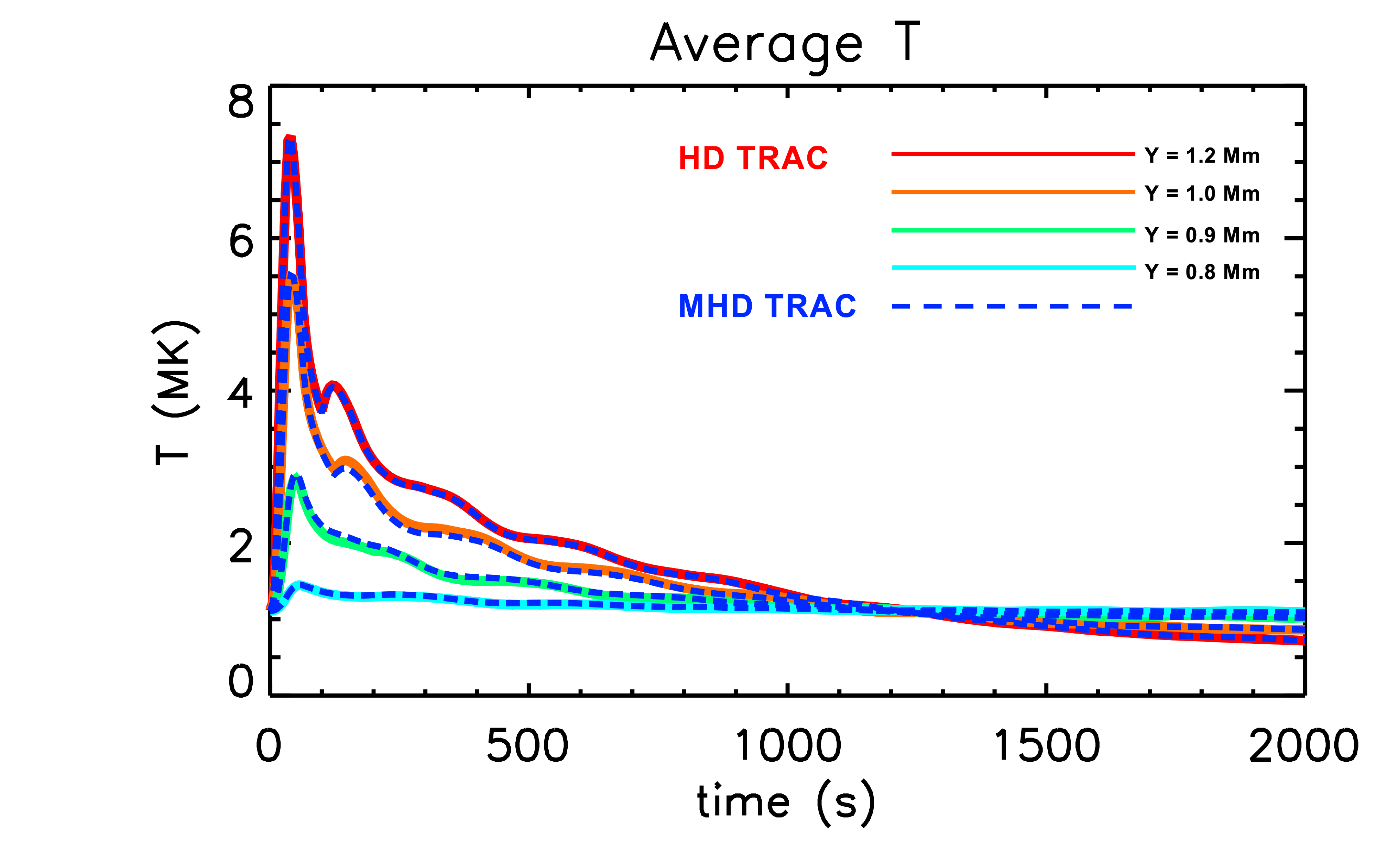}}
  \hspace*{-0.03\linewidth}
  \subfigure{\includegraphics[width=0.53\linewidth]
  {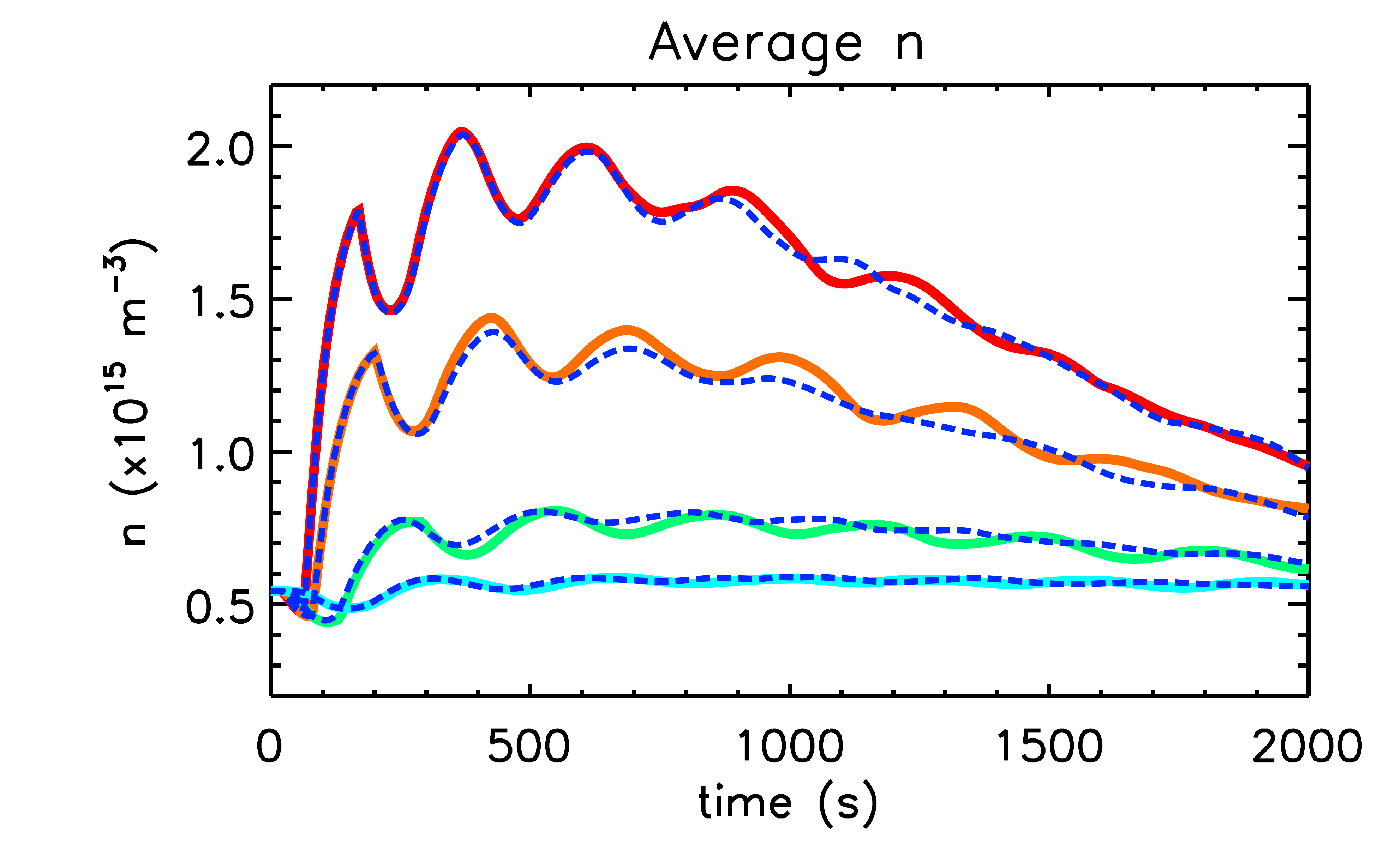}}
  \\[-8mm]
  \caption{
    Comparison of the coronal evolution
    obtained from the two different models of the unsheared arcade
    (Sect. \ref{Sect:2D_Unsheared_arcade}), 
    in response to the 
    non-uniform coronal heating pulse.
    The panels show the coronal averaged temperature and density as 
    functions of time, at four different
    positions across the magnetic field.
    The various solid curves represent the HD TRAC solution at
    these different 
    transverse locations
    (with the lines colour coded in a way that reflects 
    the distance across the field
    from the centre of the arcade) 
    and the dashed blue curves correspond to the 
    MHD TRAC solution at these locations.
    \label{Fig:Unsheared_arcade_coronal_averages}
  }
\end{figure*}
  %
  %
  %
  %
  %
  %
  %
  %
\begin{figure*}
  \hspace*{-0.05\linewidth}
  \subfigure{\includegraphics[width=0.53\linewidth]
  {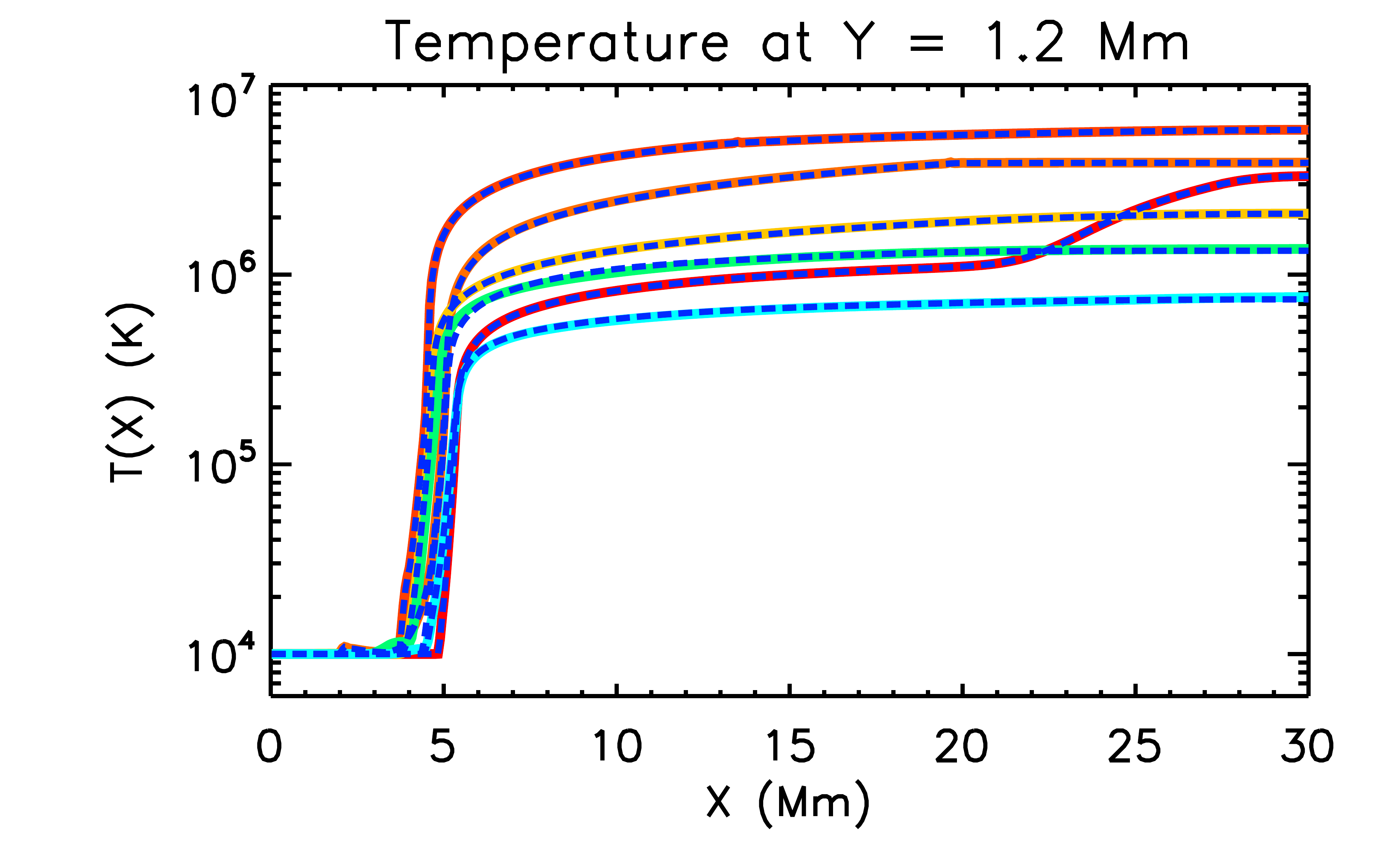}}
  \hspace*{-0.03\linewidth}
  \subfigure{\includegraphics[width=0.53\linewidth]
  {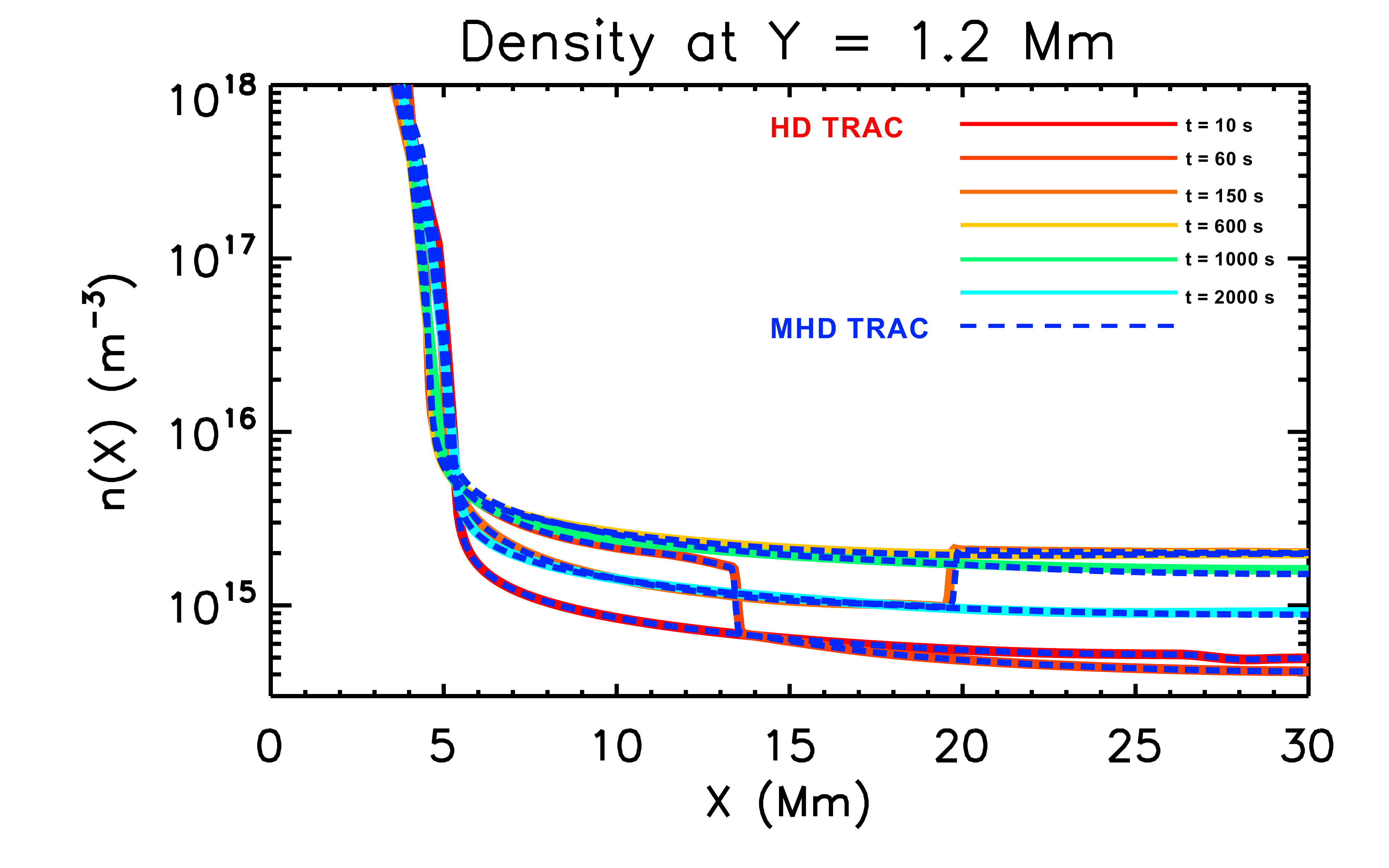}}
  \\[-5mm]
  \hspace*{-0.05\linewidth}
  \subfigure{\includegraphics[width=0.53\linewidth]
  {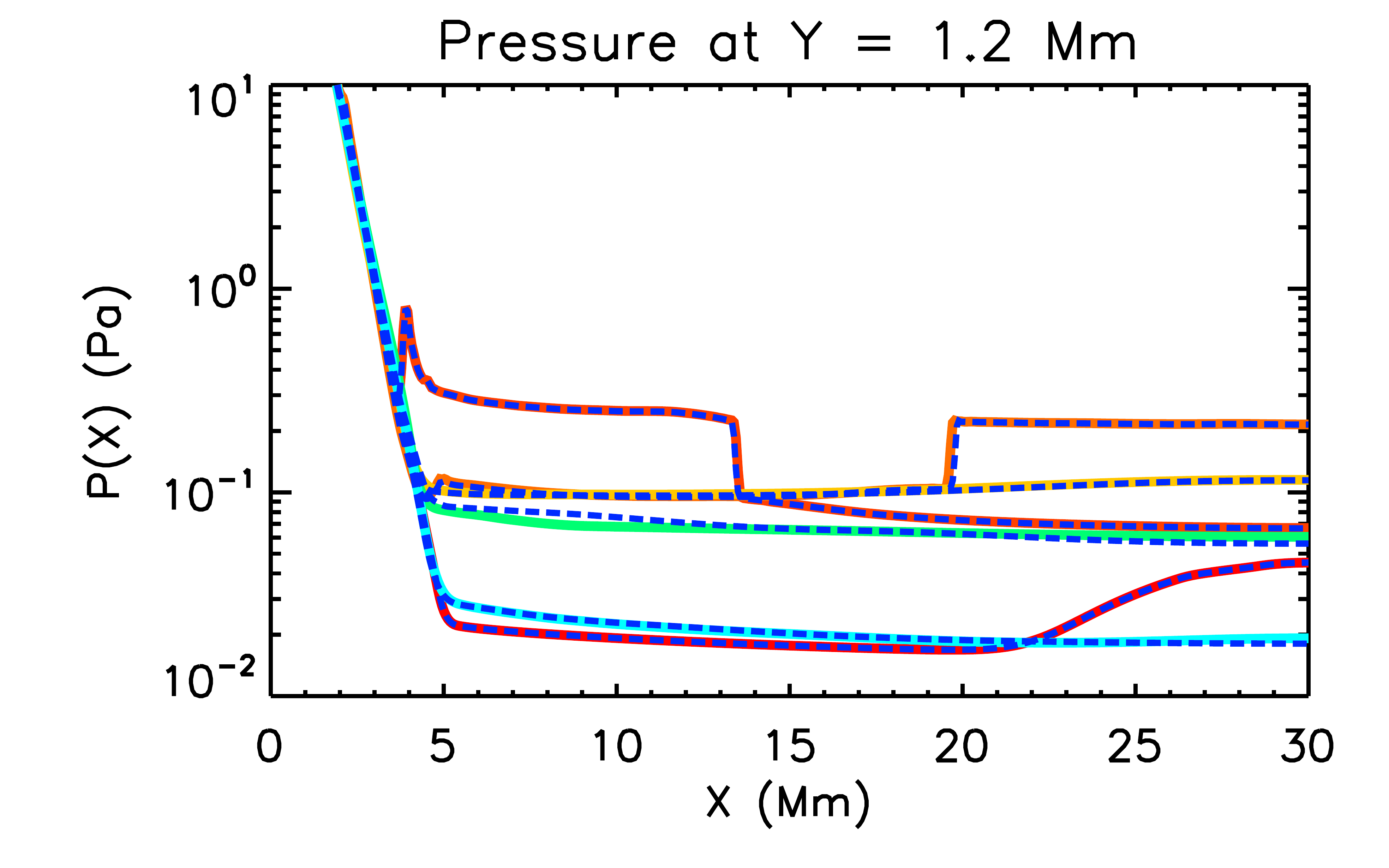}}
  \hspace*{-0.03\linewidth}
  \subfigure{\includegraphics[width=0.53\linewidth]
  {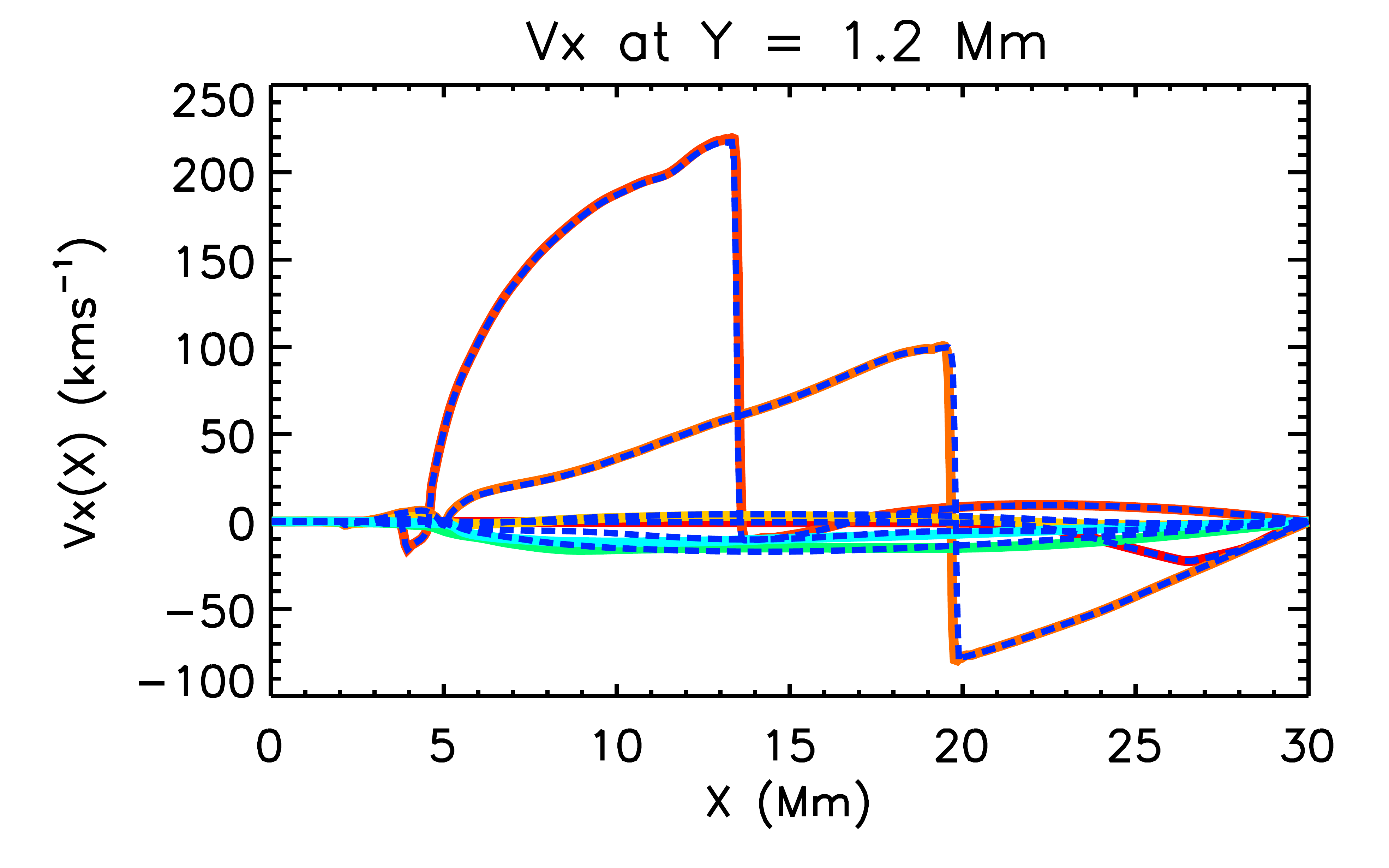}}
  \\[-5mm]
  \hspace*{-0.05\linewidth}
  \subfigure{\includegraphics[width=0.53\linewidth]
  {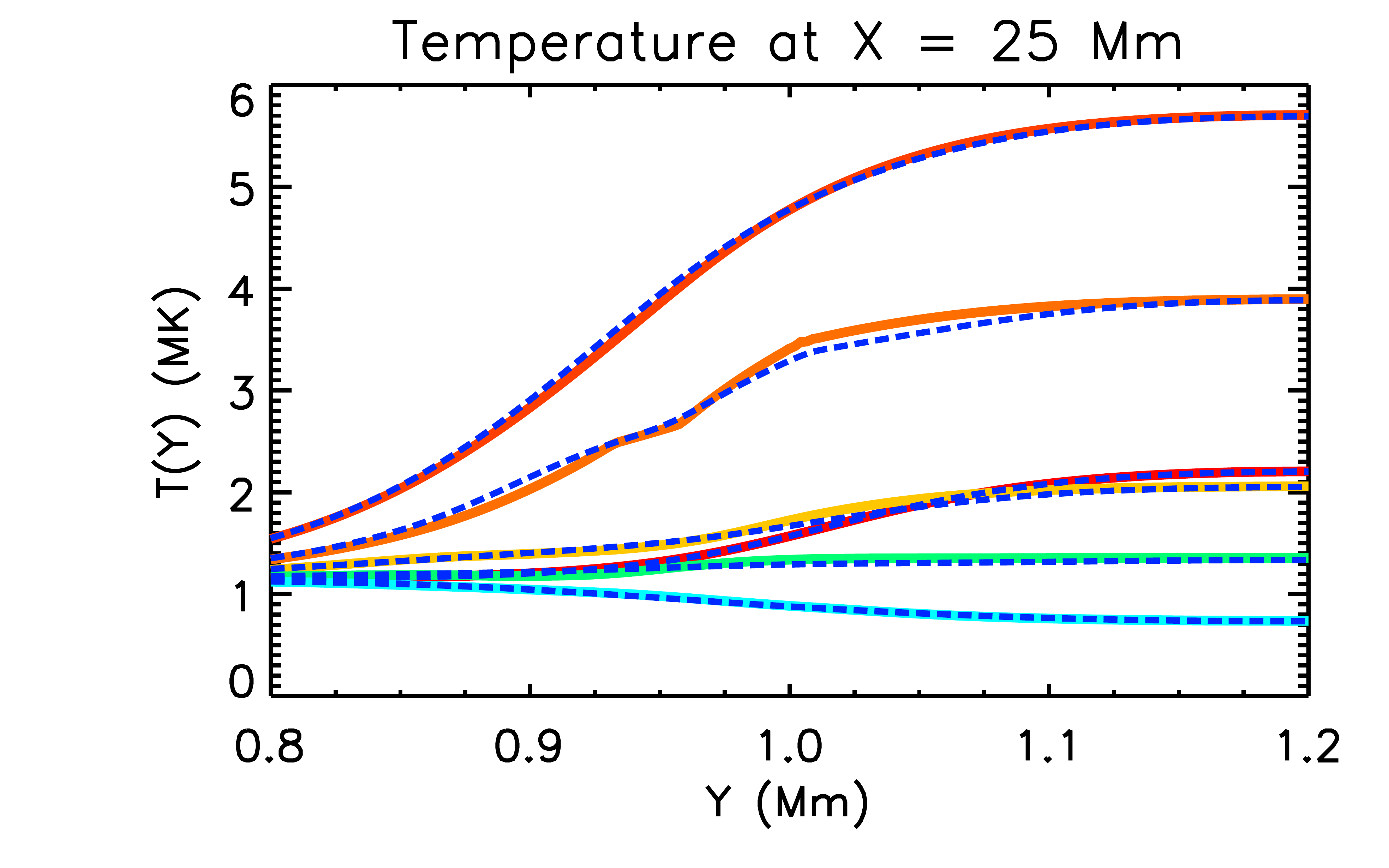}}
  \hspace*{-0.03\linewidth}
  \subfigure{\includegraphics[width=0.53\linewidth]
  {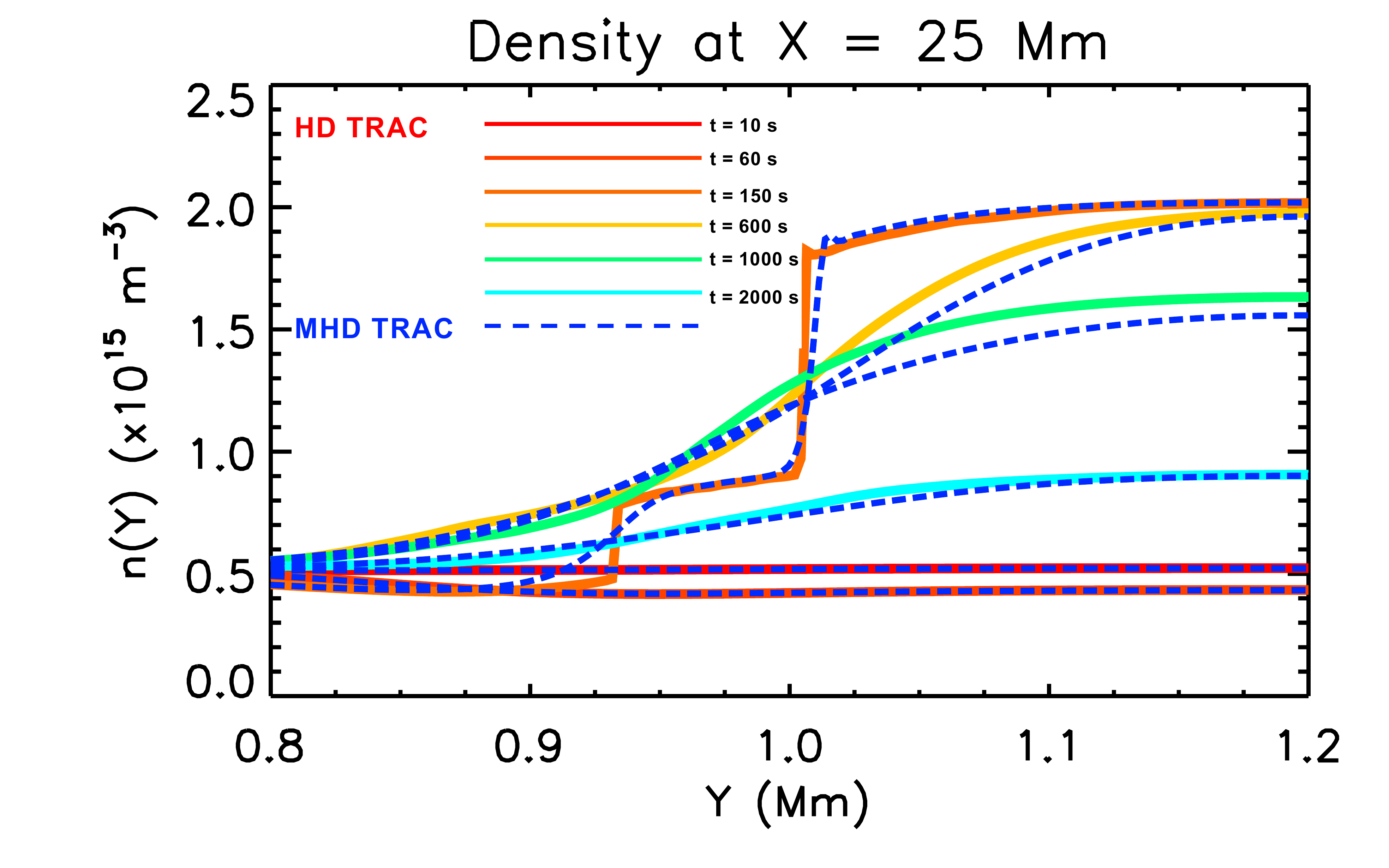}}
  \\[-5mm]
  \hspace*{-0.05\linewidth}
  \subfigure{\includegraphics[width=0.53\linewidth]
  {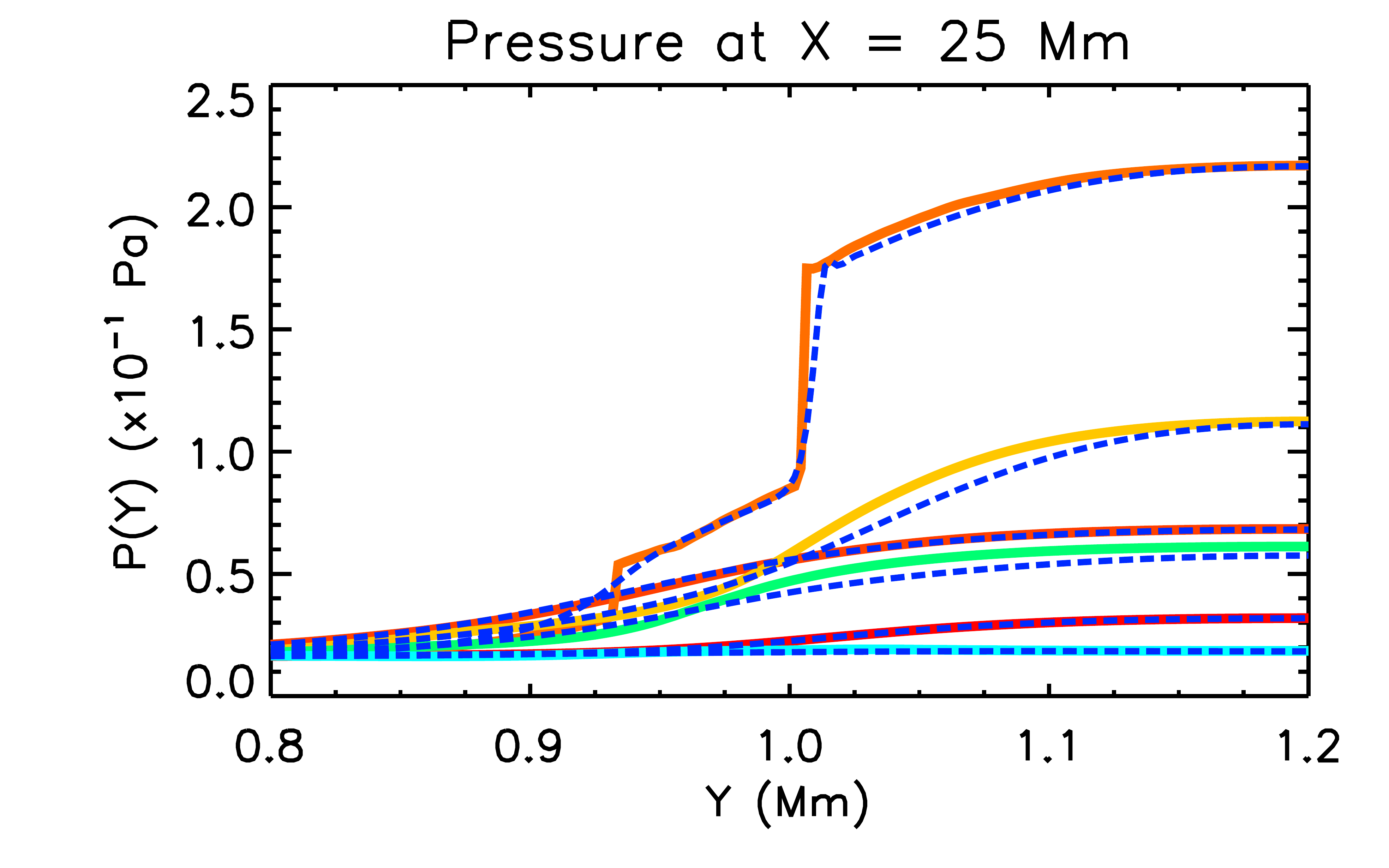}}
  \hspace*{-0.03\linewidth}
  \subfigure{\includegraphics[width=0.53\linewidth]
  {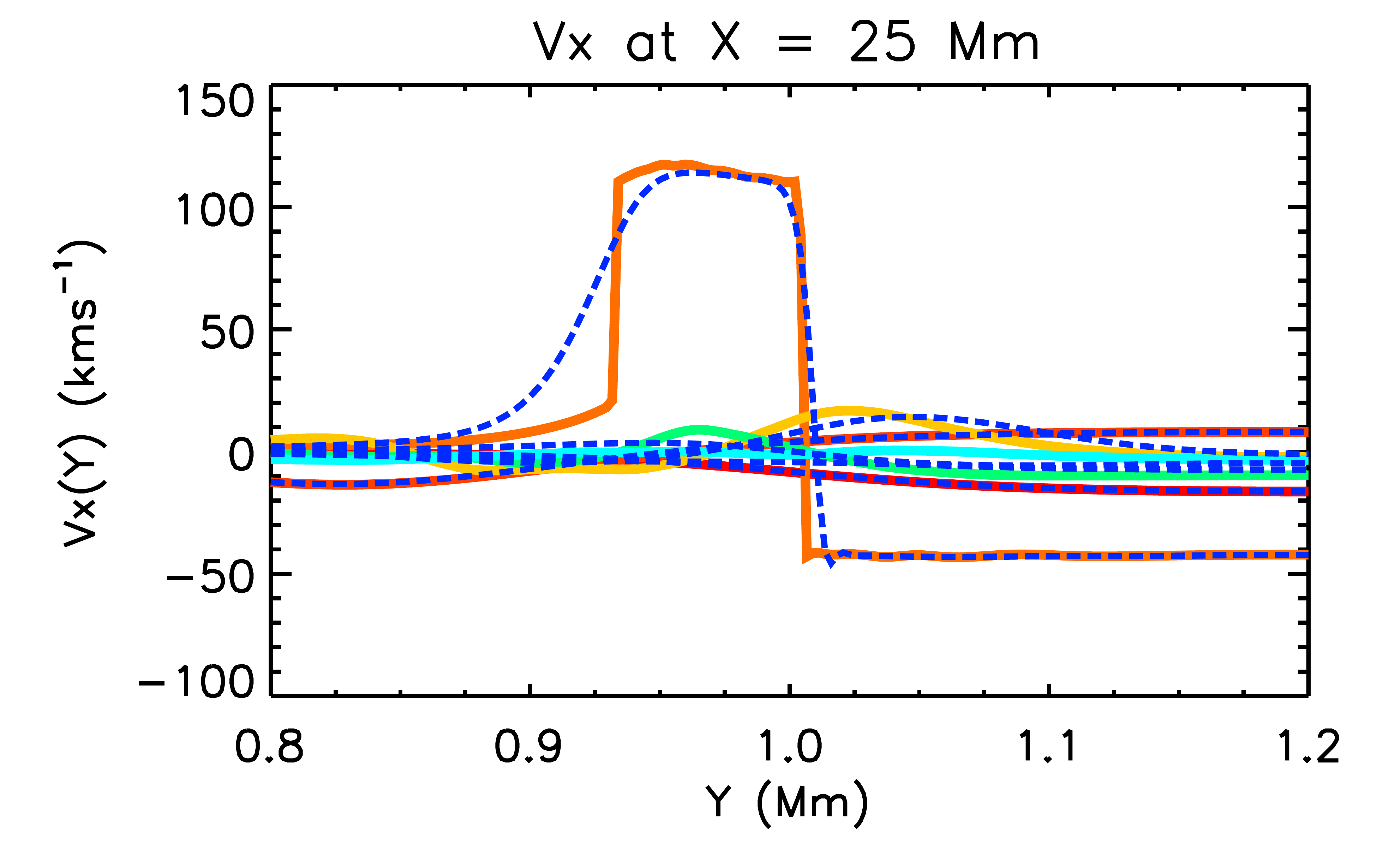}}
  \\[-8mm]
  \caption{
    Comparison of the field-aligned and transverse
    temporal evolution
    obtained from the two different models of the unsheared arcade
    (Sect. \ref{Sect:2D_Unsheared_arcade}), 
    in response to the 
    non-uniform coronal heating pulse.
    Upper (lower) four panels:     
    Time ordered snapshots of the temperature, density, pressure and
    field-aligned velocity as functions of position 
    along (across) the magnetic field
    at $y=1.2$~Mm ($x=25$~Mm).
    The various solid curves represent the HD TRAC solution at different 
    times
    (with the lines colour coded in a way 
    that reflects the temporal evolution)    
    and the dashed blue curves correspond to the 
    MHD TRAC solution at these times.
    \label{Fig:Unsheared_arcade_field_aligned_tranverse_evolution}
  }
\end{figure*}
  %
  %
  %
  %
  %
  %
  %
  %
\begin{figure*}
  \hspace*{-0.05\linewidth}
  \subfigure{\includegraphics[width=0.53\linewidth]
  {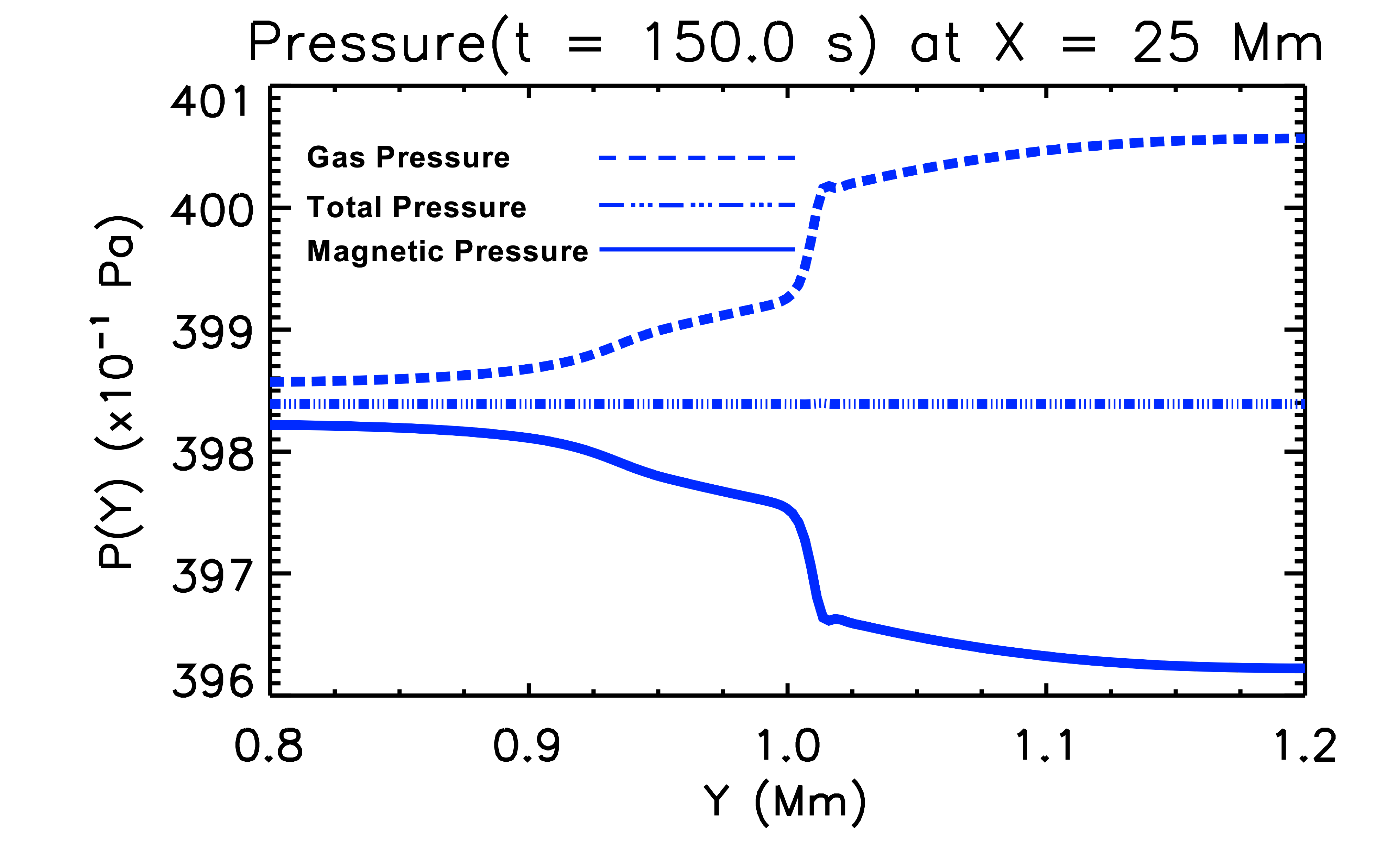}}
  \hspace*{-0.03\linewidth}
  \subfigure{\includegraphics[width=0.53\linewidth]
  {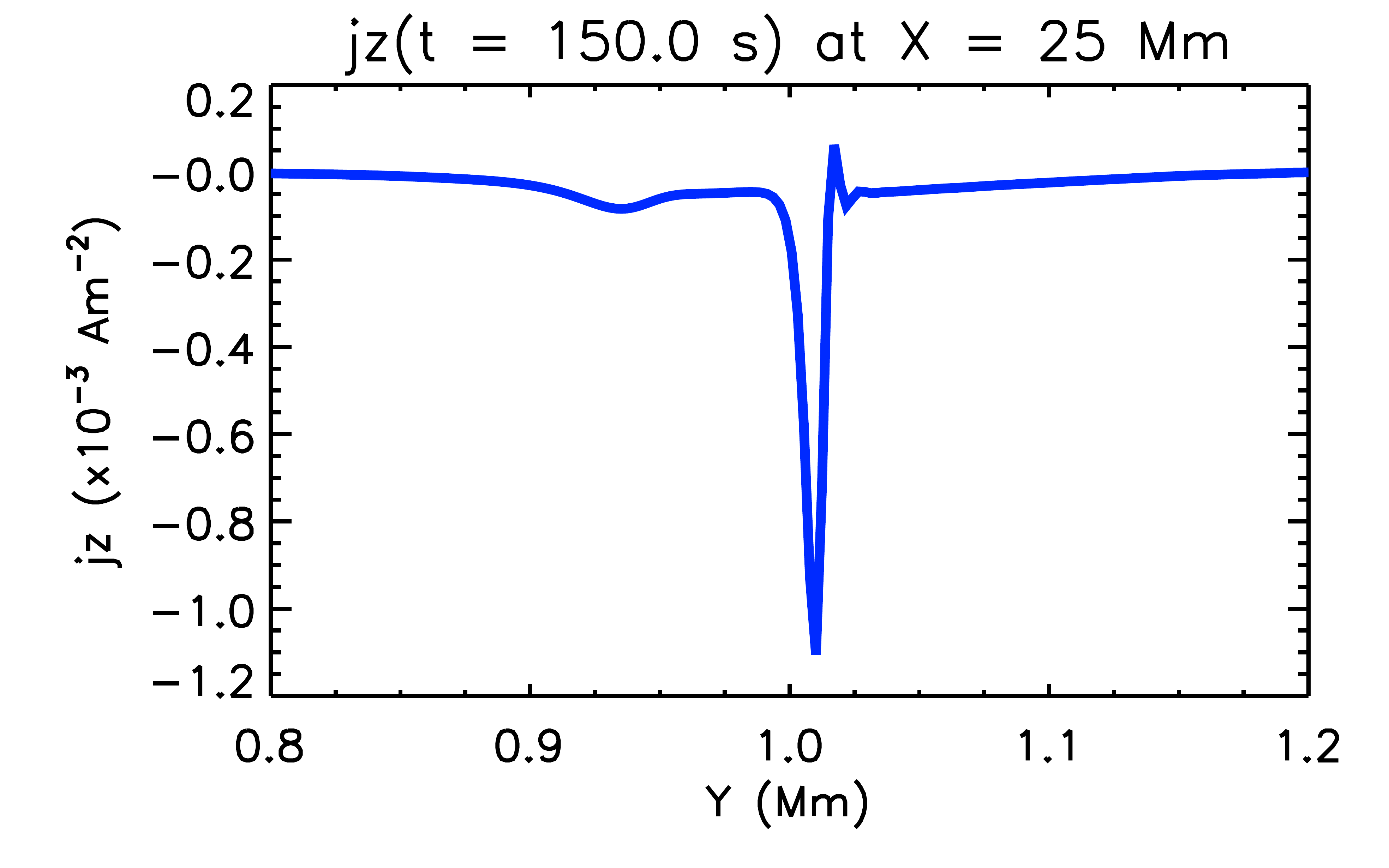}}
  \\[-8mm]
  \caption{
    Quantities from the 
    MHD model of the unsheared arcade
    (Sect. \ref{Sect:2D_Unsheared_arcade}),
    at a coronal height of $x=25$~Mm,
    at the time of the first density peak $(t=150$~s).
    Left-hand panel: 
    Gas pressure (dashed line),
    magnetic pressure (solid line),
    and total pressure (dash-dotted line)
    as functions of position across the arcade.
    We note that the gas pressure has been offset by 
    $398.4 \times 10^{-1}$~Pa in the pressure plot.
    Right-hand panel:
    $j_z$ component of the current current density.
    \label{Fig:2D_Simulation_UA_P_jz}
  }
\end{figure*}
  %
  %
  %
  %
  %
  %
  %
  %
\begin{sidewaysfigure*}
  \hspace*{0.02\linewidth}
  \subfigure{\includegraphics[width=0.28\linewidth]
  {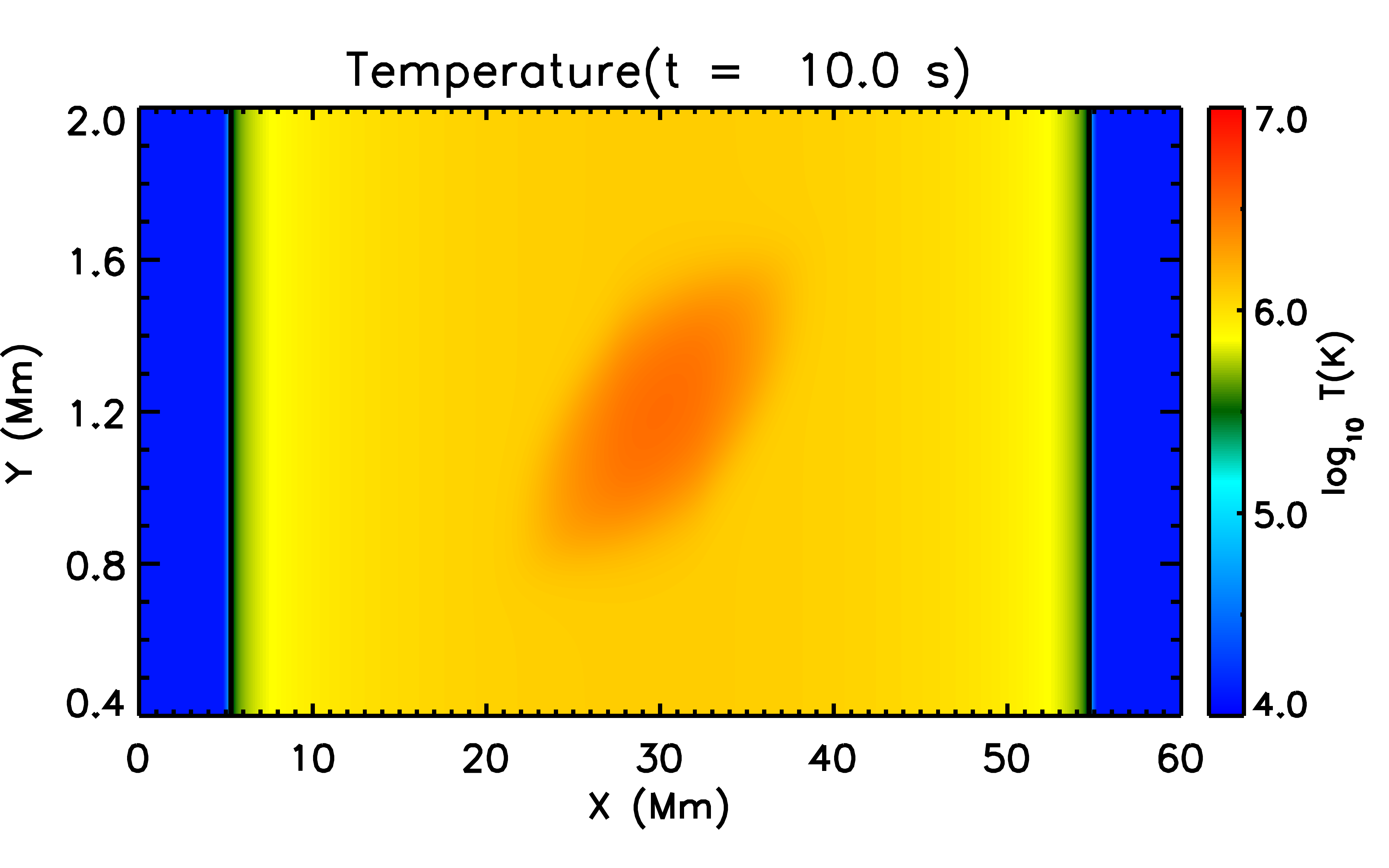}}
  \hspace*{0.04\linewidth}
  \subfigure{\includegraphics[width=0.28\linewidth]
  {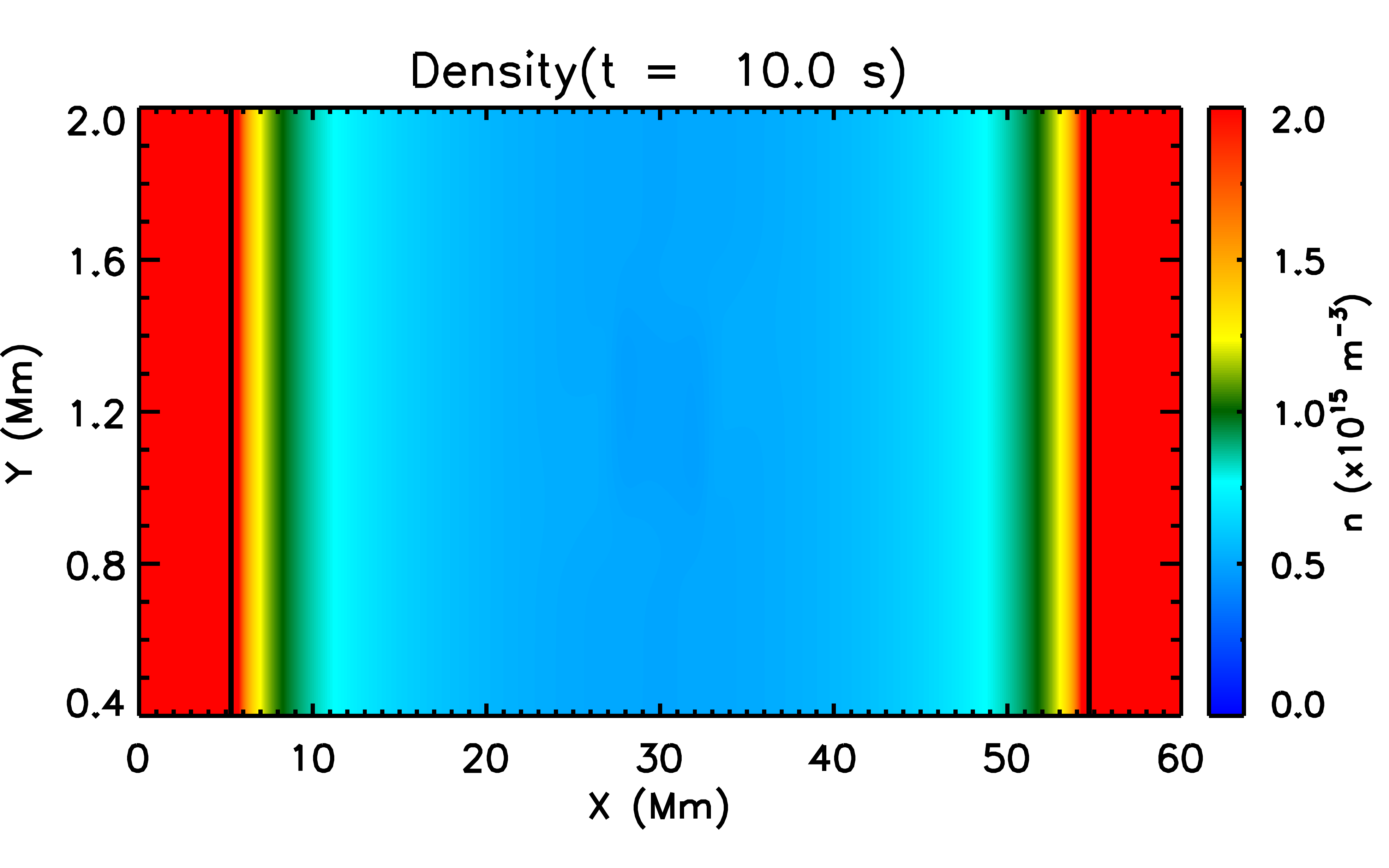}}
  \hspace*{0.04\linewidth}
  \subfigure{\includegraphics[width=0.28\linewidth]
  {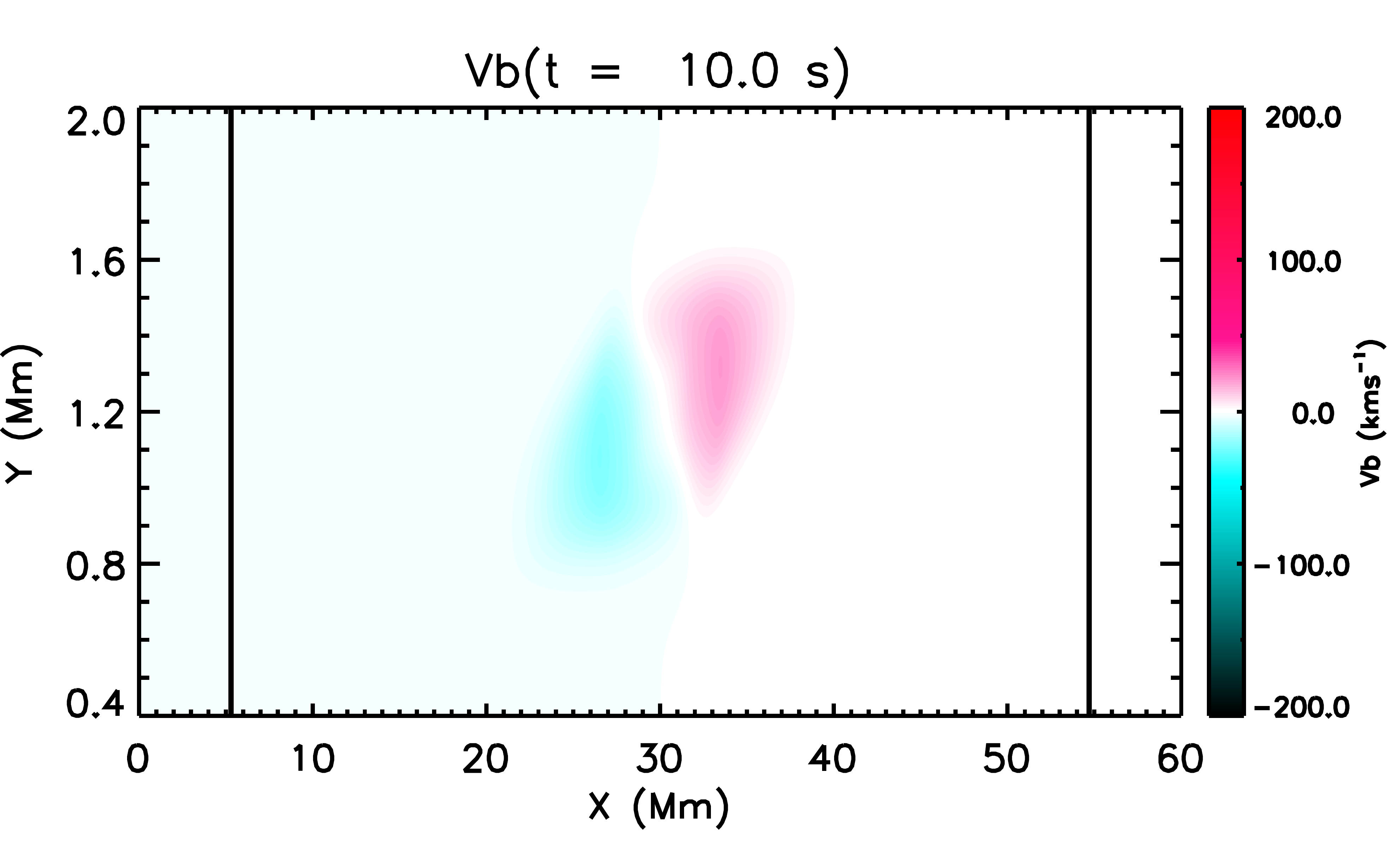}}
  \\[-2mm]
  \hspace*{0.02\linewidth}
  \subfigure{\includegraphics[width=0.28\linewidth]
  {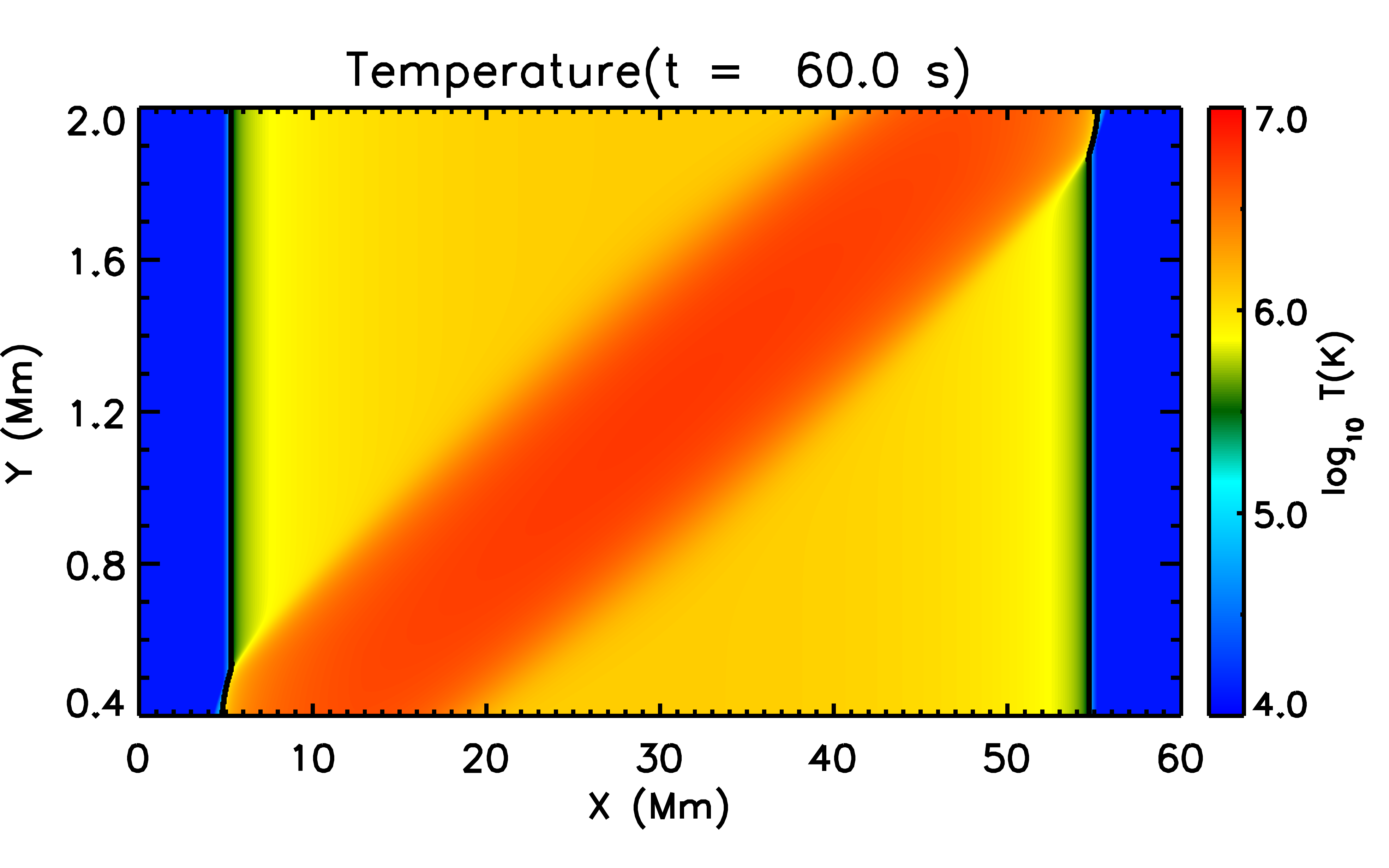}}
  \hspace*{0.04\linewidth}
  \subfigure{\includegraphics[width=0.28\linewidth]
  {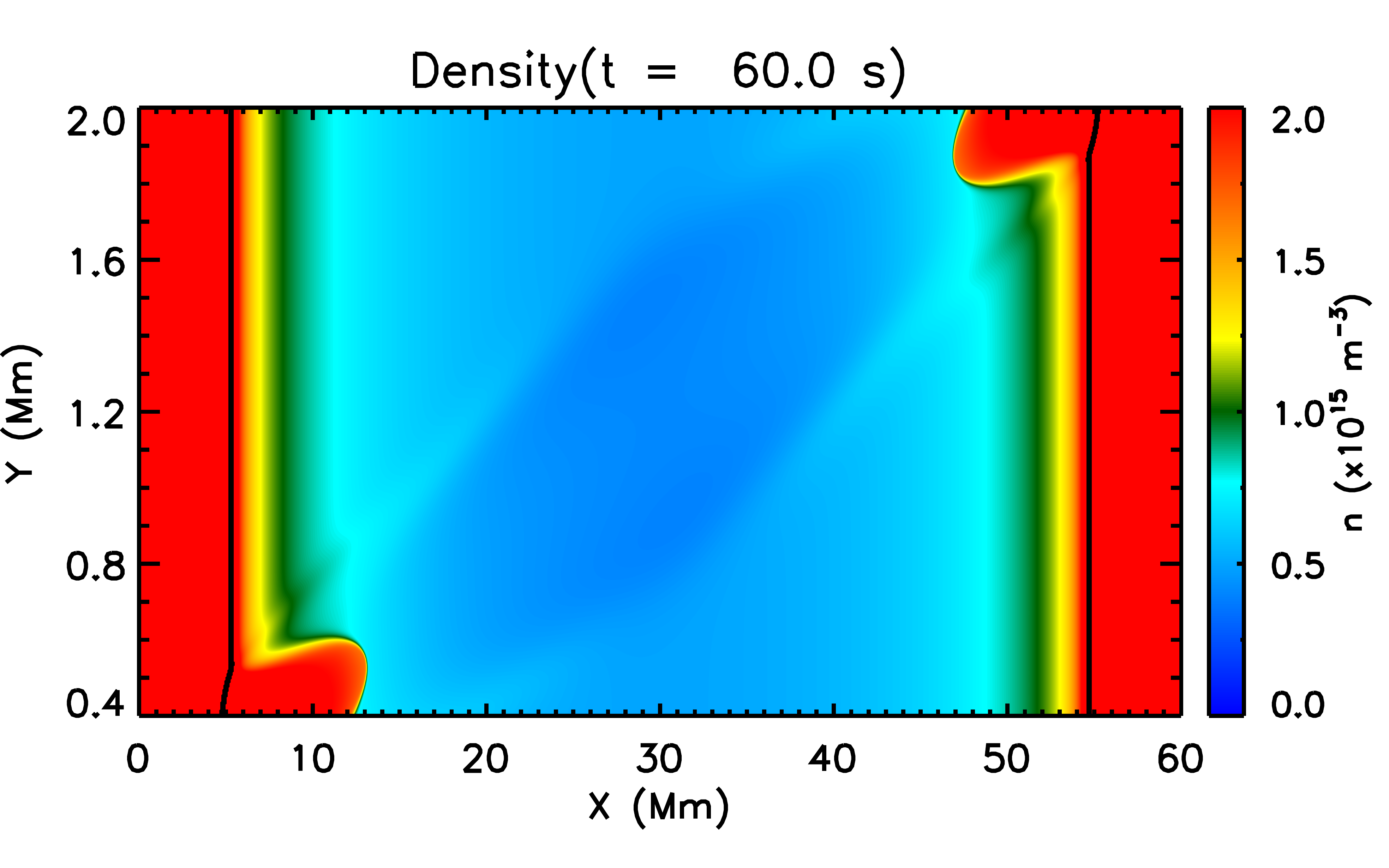}}
  \hspace*{0.04\linewidth}
  \subfigure{\includegraphics[width=0.28\linewidth]
  {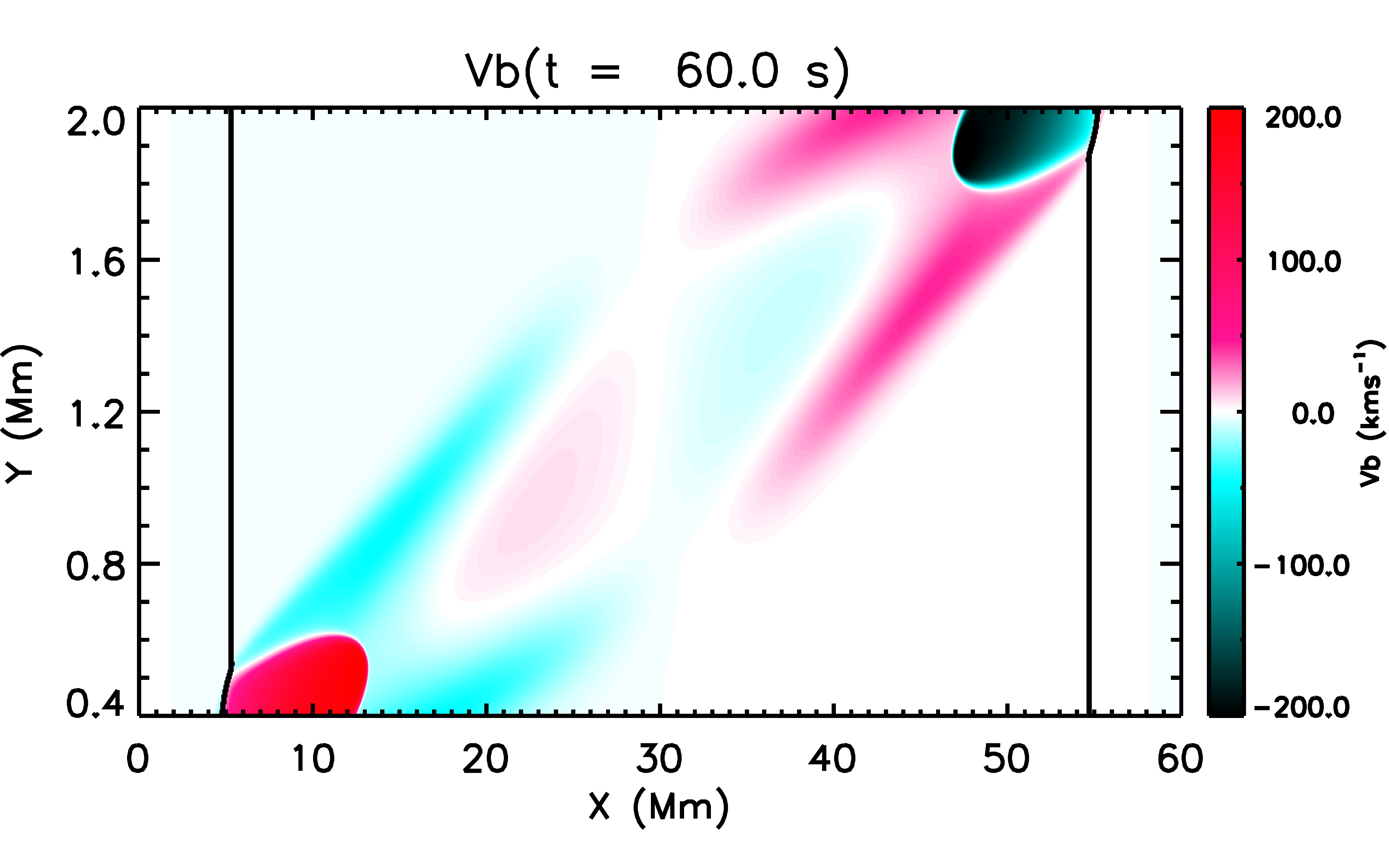}}
  \\[-2mm]
  \hspace*{0.02\linewidth}
  \subfigure{\includegraphics[width=0.28\linewidth]
  {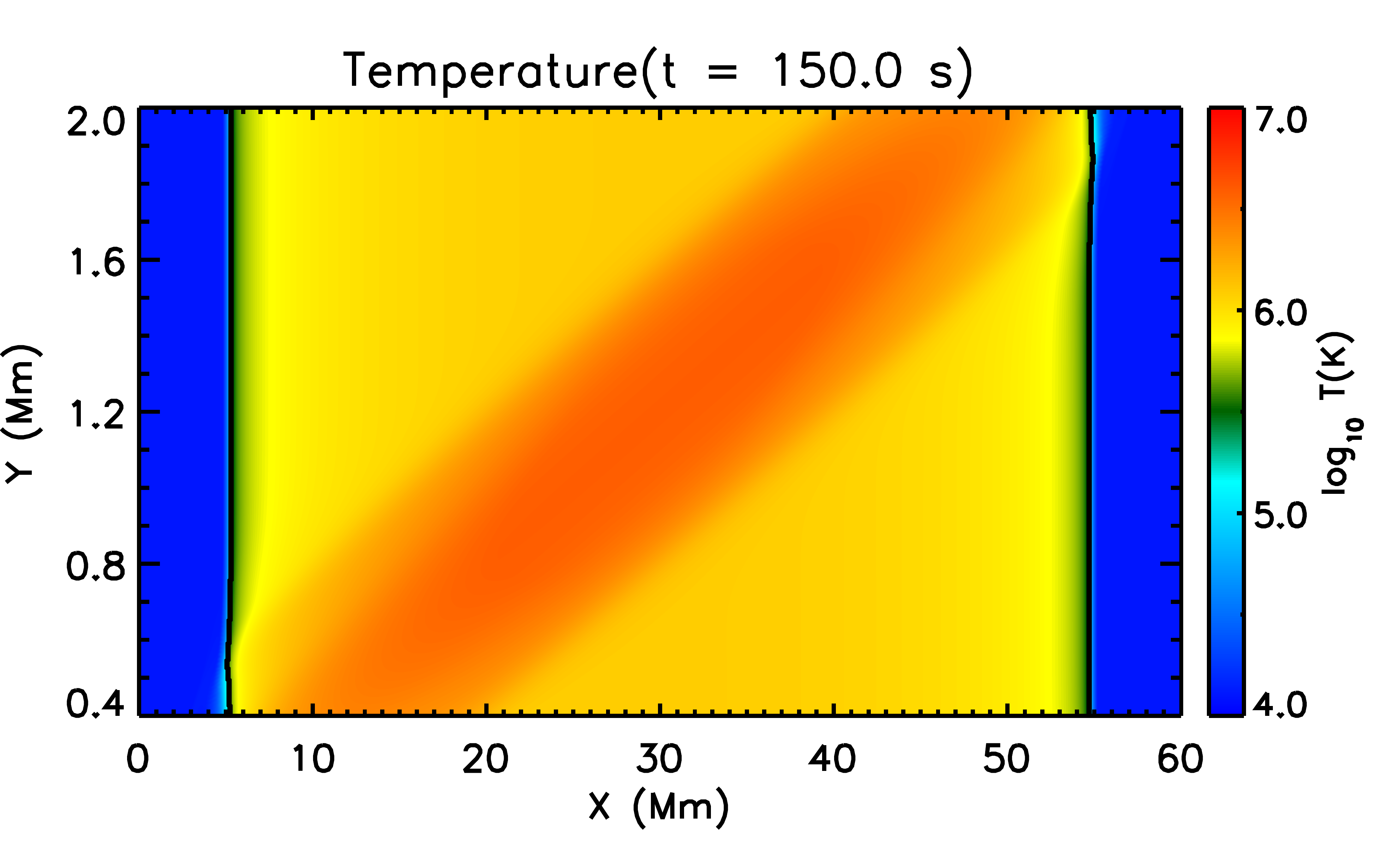}}
  \hspace*{0.04\linewidth}
  \subfigure{\includegraphics[width=0.28\linewidth]
  {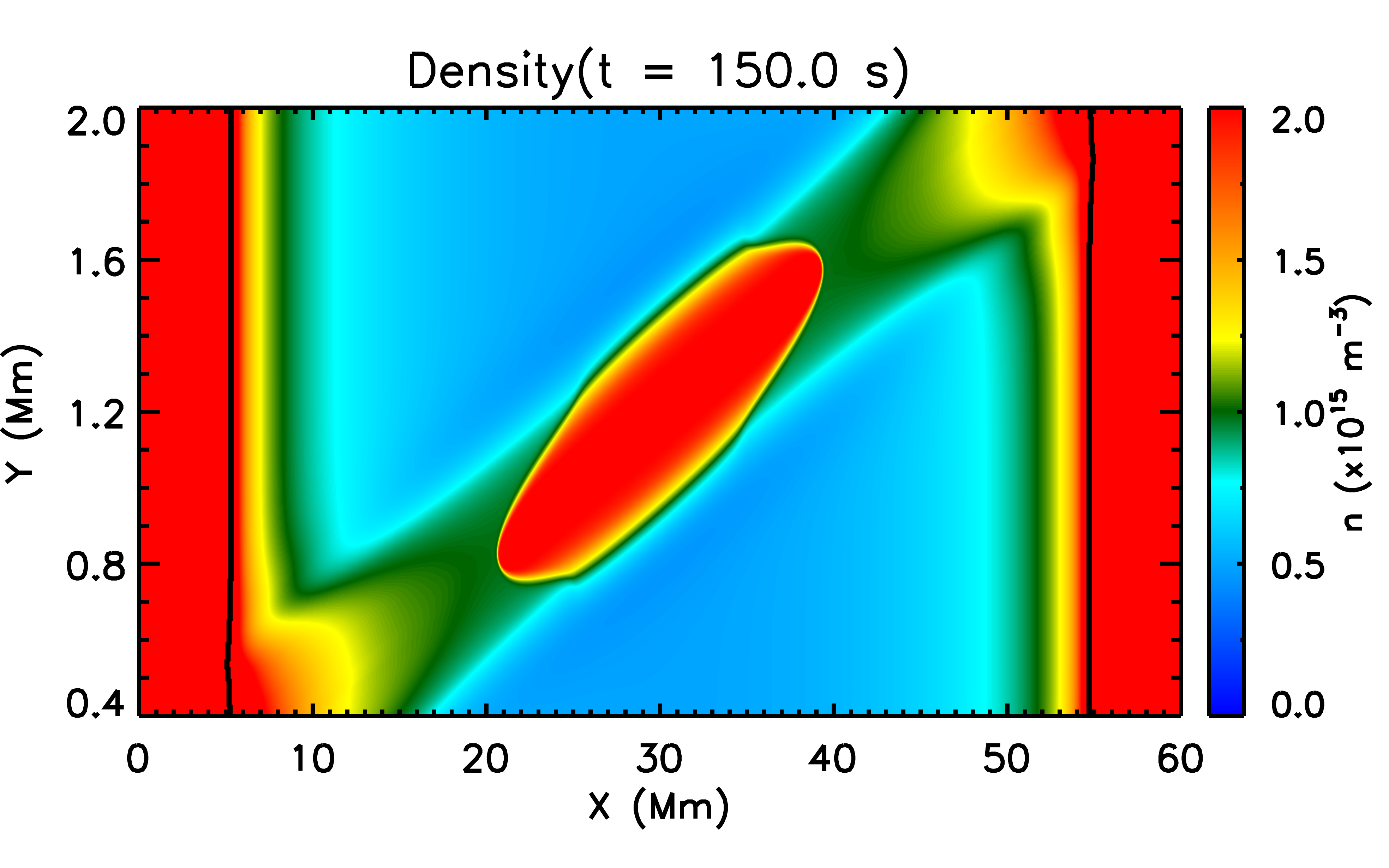}}
  \hspace*{0.04\linewidth}
  \subfigure{\includegraphics[width=0.28\linewidth]
  {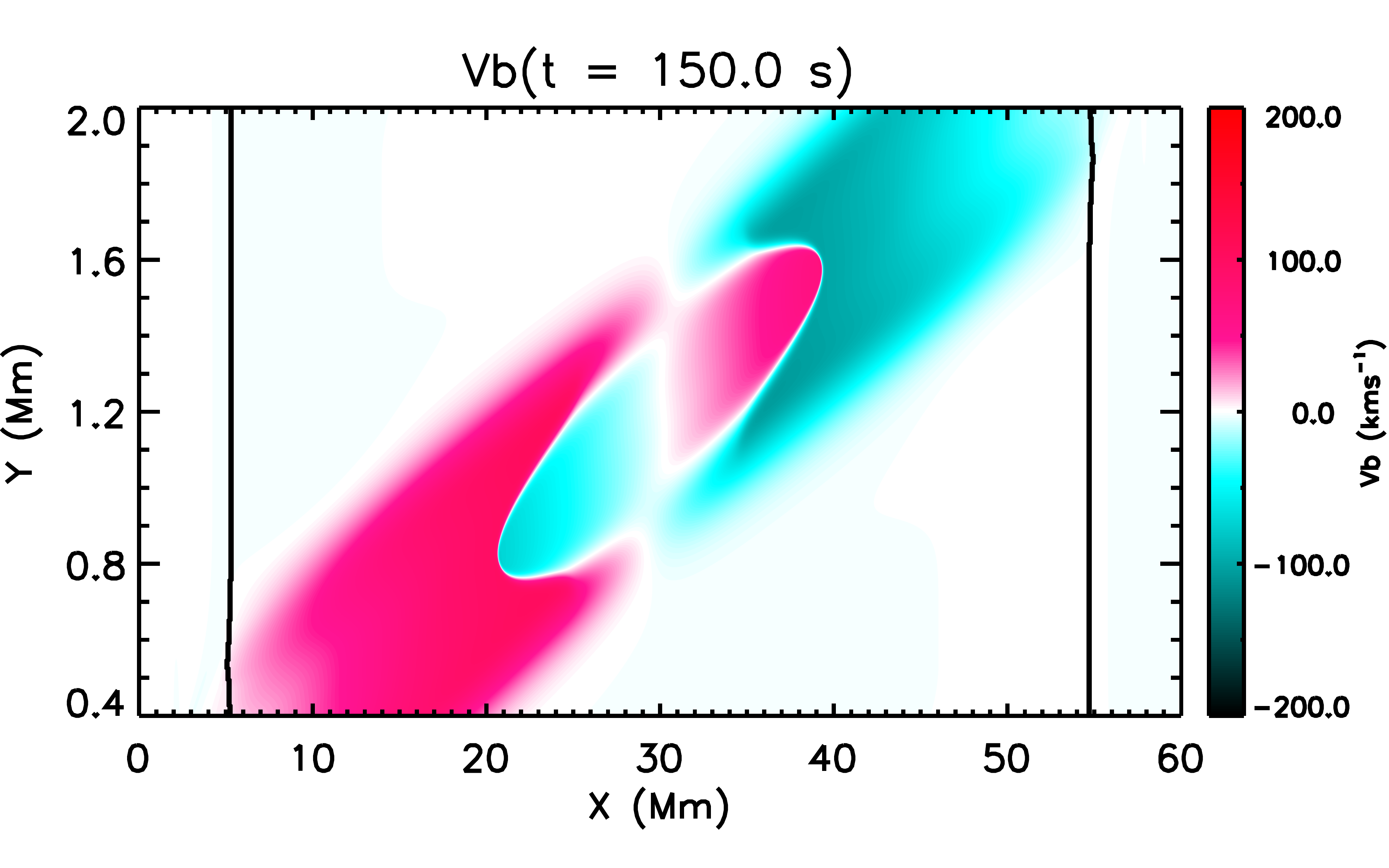}}
  \\[-2mm]
  \hspace*{0.02\linewidth}
  \subfigure{\includegraphics[width=0.28\linewidth]
  {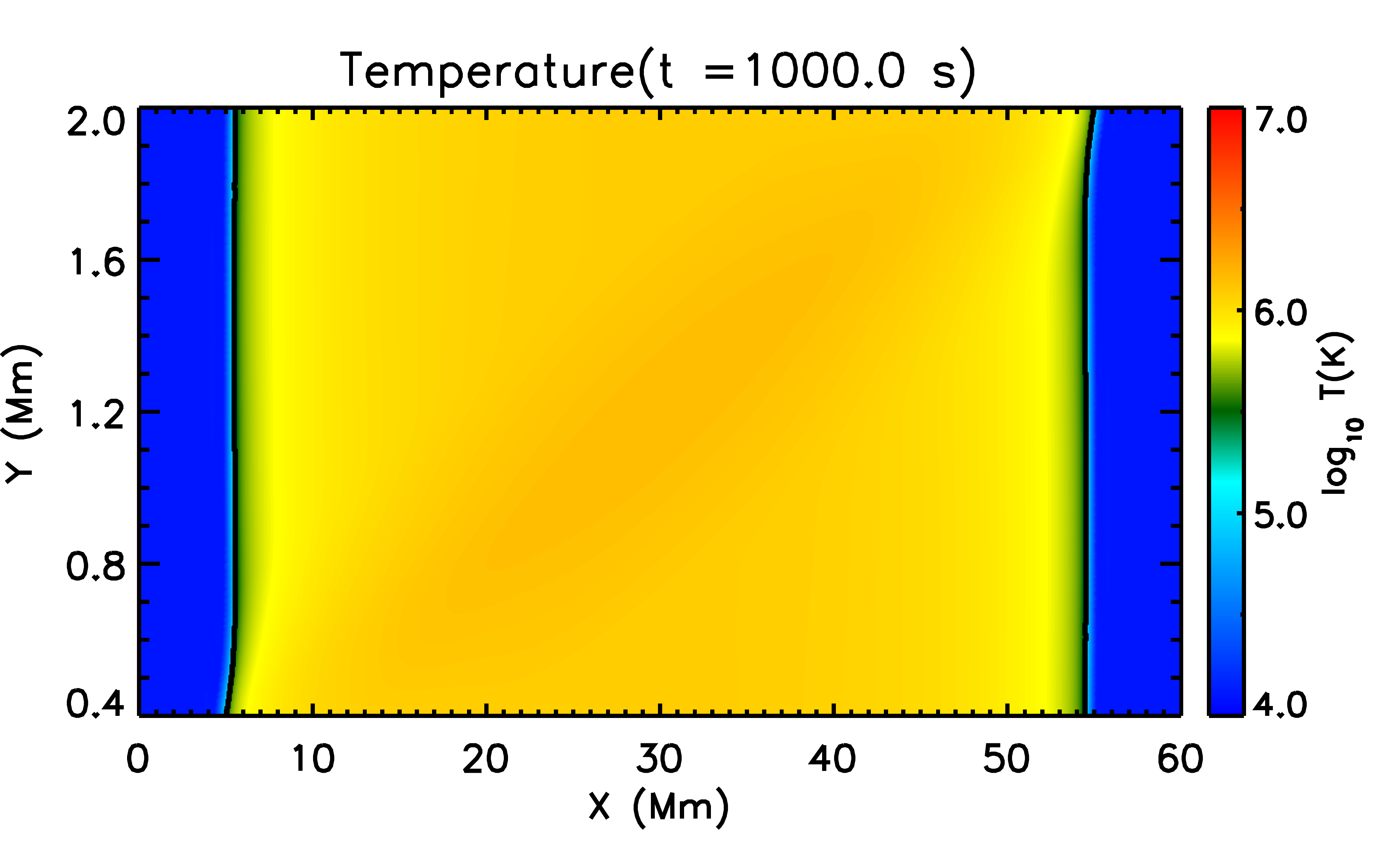}}
  \hspace*{0.04\linewidth}
  \subfigure{\includegraphics[width=0.28\linewidth]
  {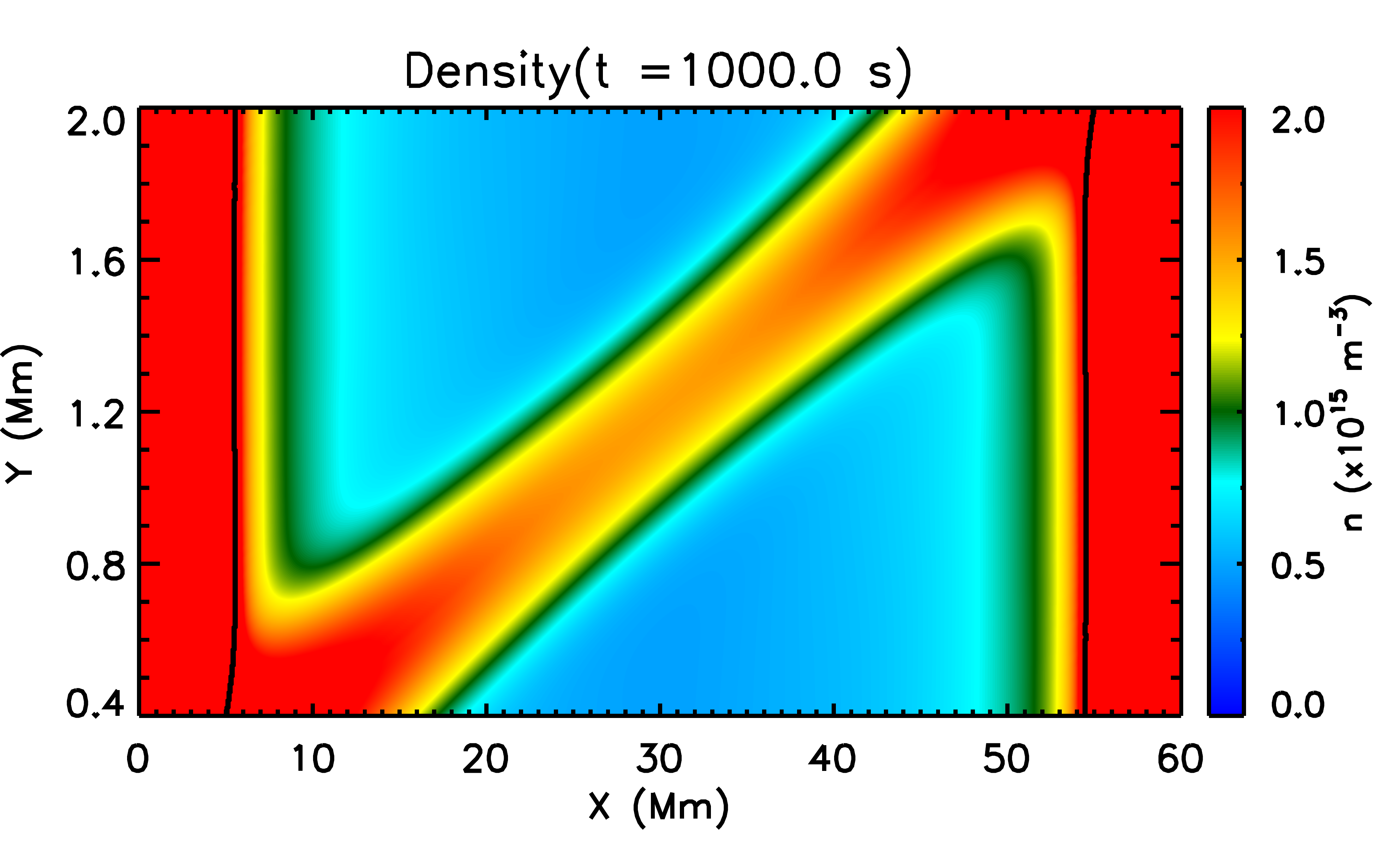}}
  \hspace*{0.04\linewidth}
  \subfigure{\includegraphics[width=0.28\linewidth]
  {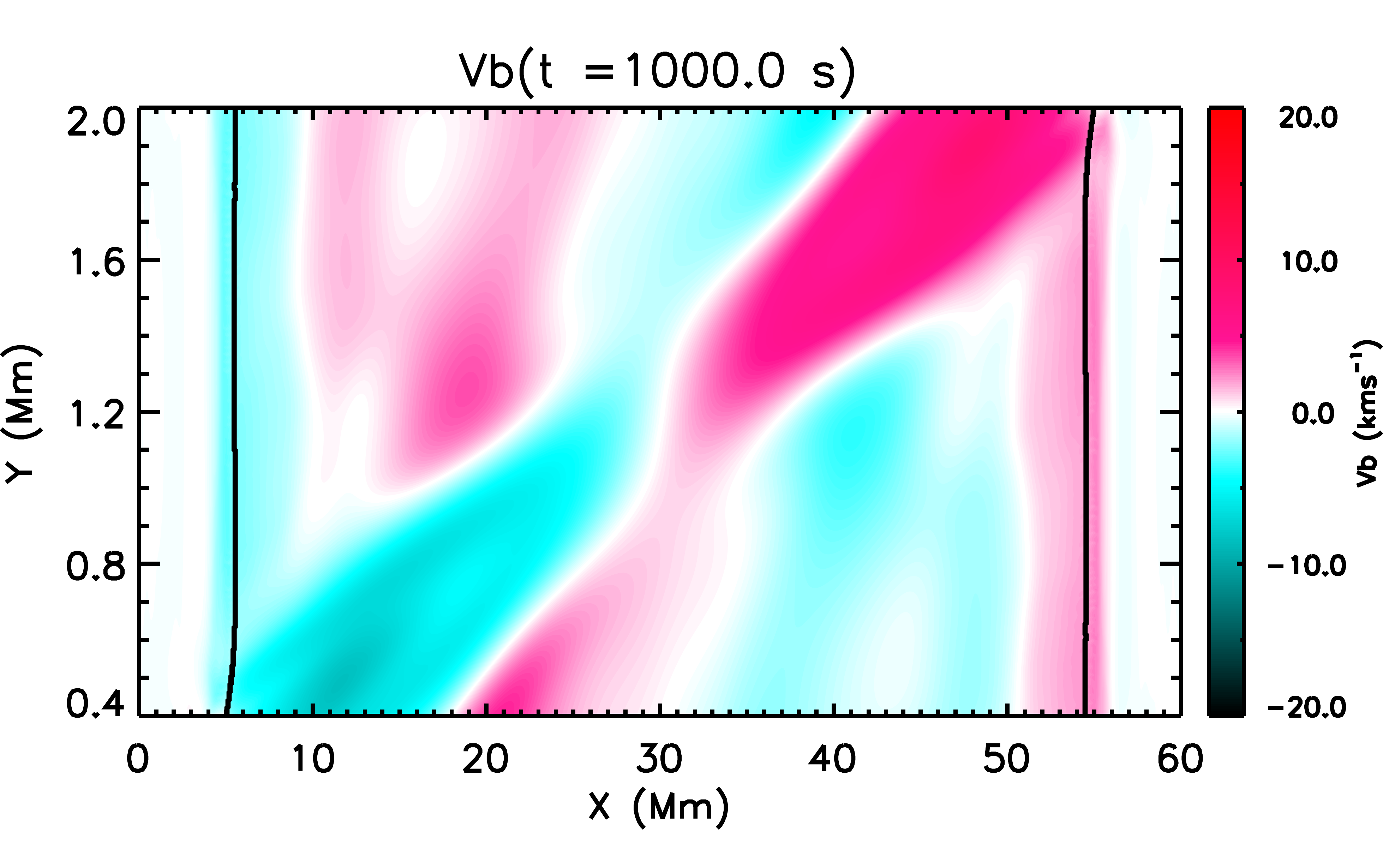}}
  \\[-6mm]
  \caption{
    Results for the 
    non-uniform coronal heating pulse using a two-dimensional
    MHD simulation of the sheared arcade
    (Sect. \ref{Sect:2D_Sheared_arcade}).
    Notation is the same as that in Fig. 
    \ref{Fig:2D_Simulation_UA_time_evolution_contours}.    
    Movies of the full time evolution of the $T$, $n$ and $v_b$
    contour plots can be viewed in the online version of this article.
    \label{Fig:2D_Simulation_SA_time_evolution_contours}
  }
\end{sidewaysfigure*}
  %
  %
  %
  %
  %
  %
  %
  %
\begin{figure*}
  \hspace*{-0.05\linewidth}
  \subfigure{\includegraphics[width=0.53\linewidth]
  {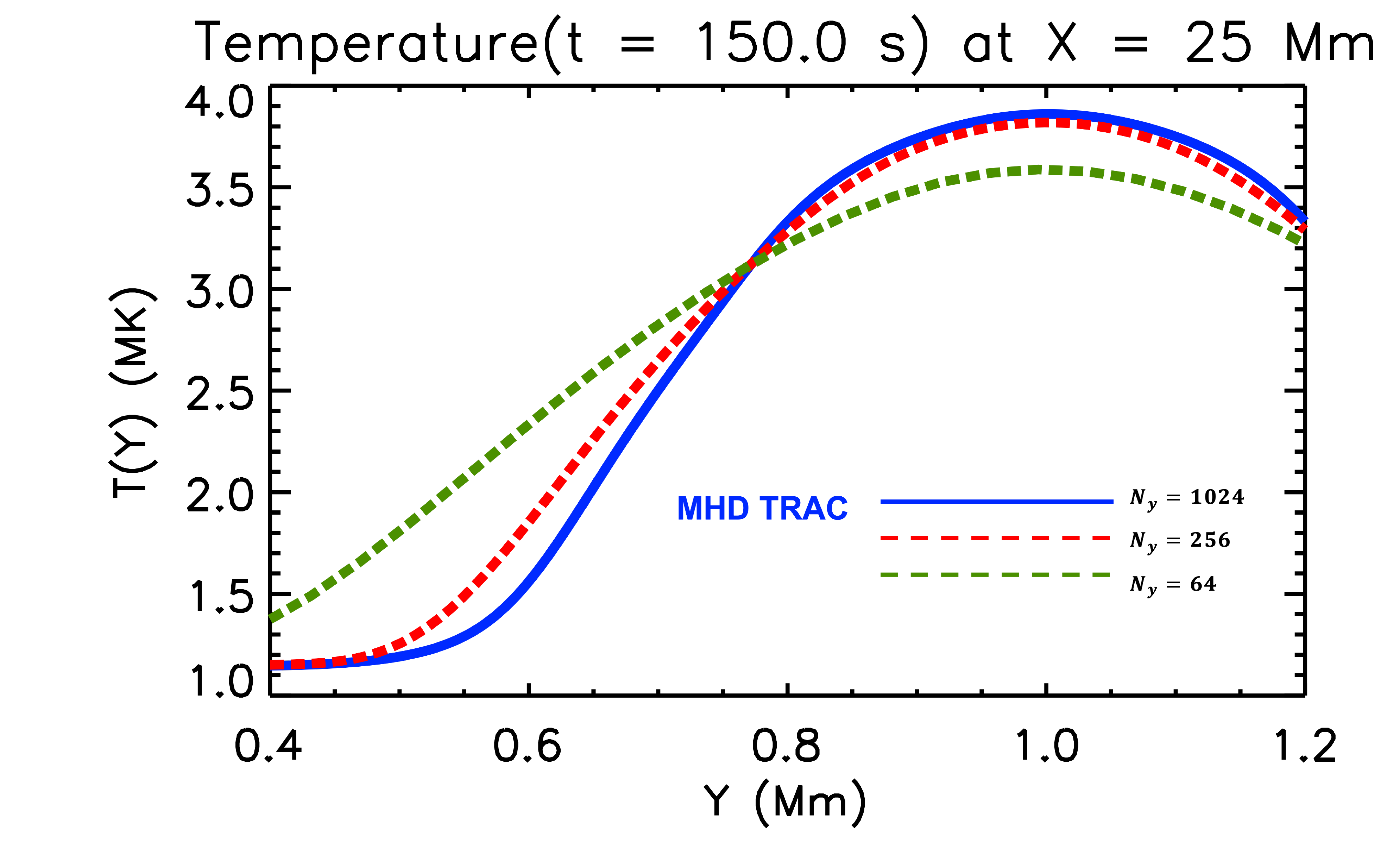}}
  \hspace*{-0.03\linewidth}
  \subfigure{\includegraphics[width=0.53\linewidth]
  {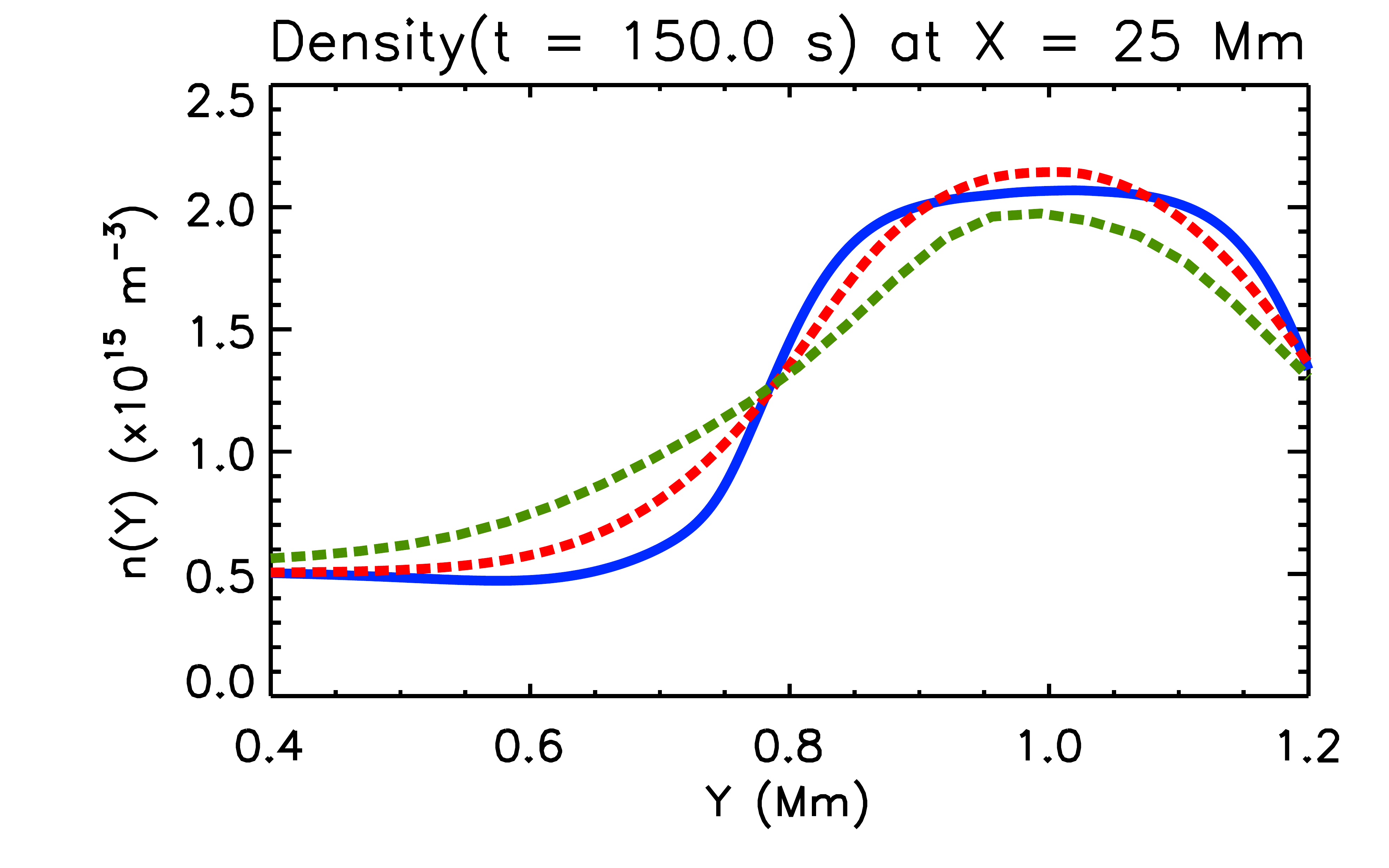}}
  \\[-5mm]
  \hspace*{-0.05\linewidth}
  \subfigure{\includegraphics[width=0.53\linewidth]
  {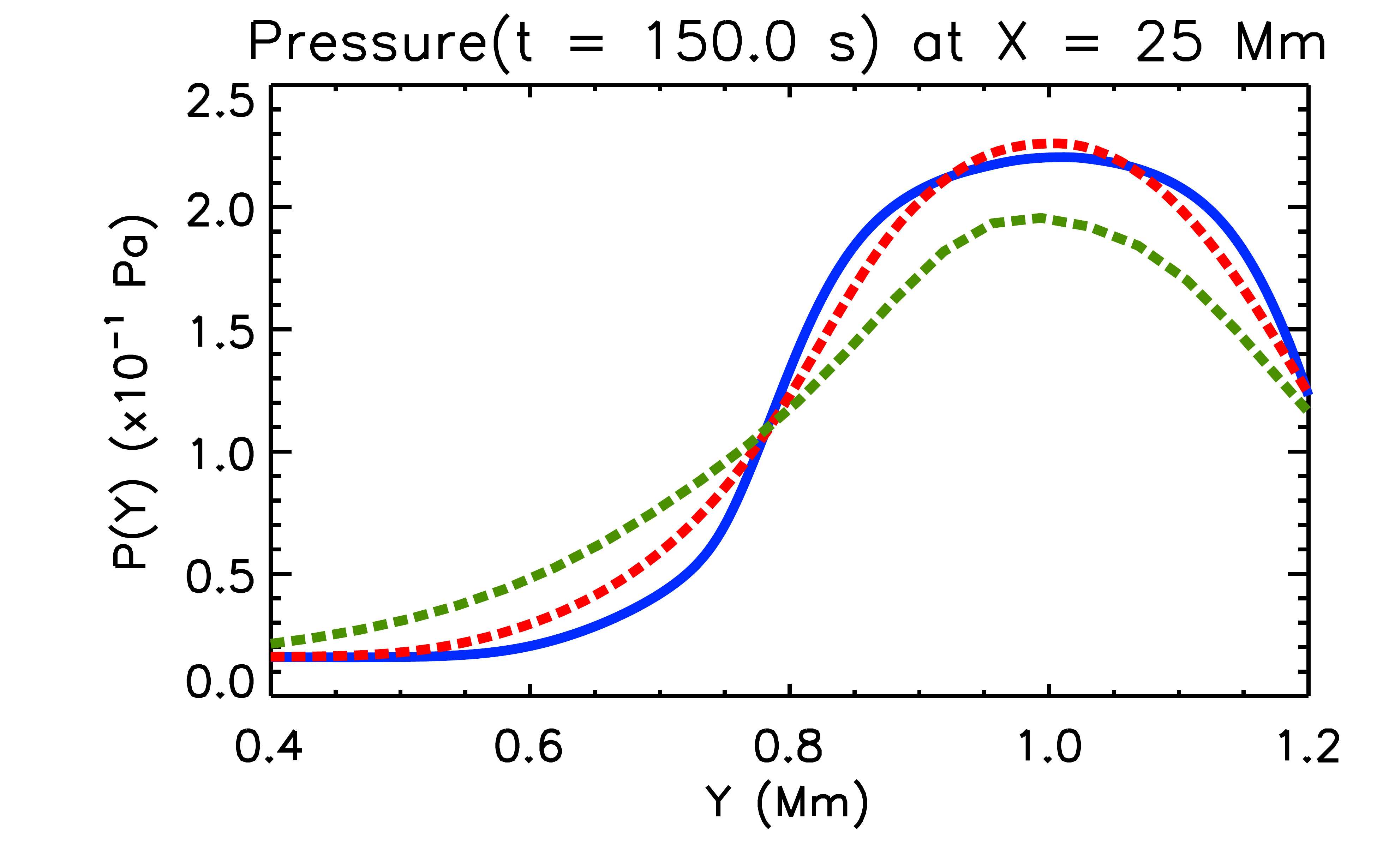}}
  \hspace*{-0.03\linewidth}
  \subfigure{\includegraphics[width=0.53\linewidth]
  {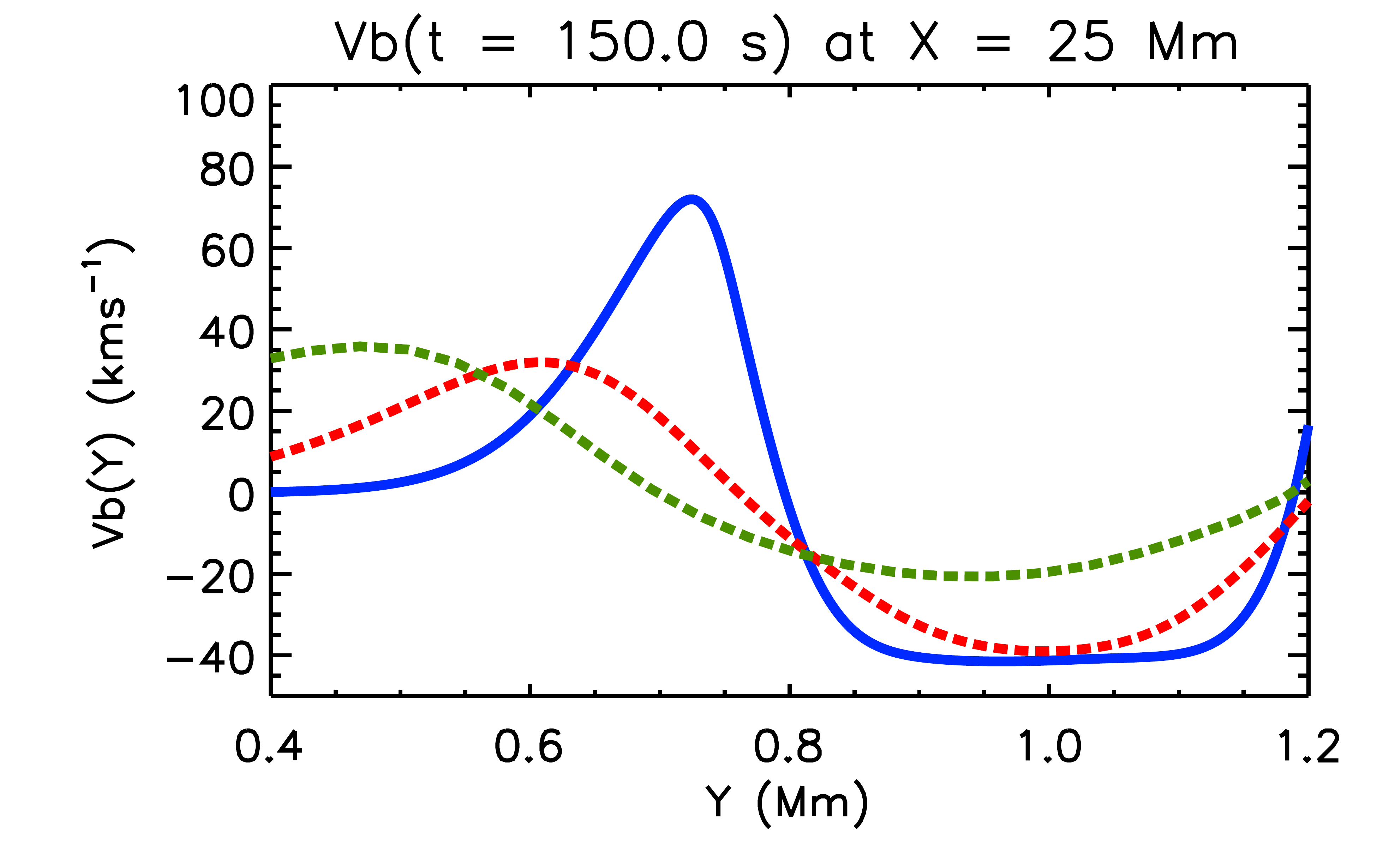}}
  \\[-8mm]
  \caption{
    Results for the
    non-uniform coronal heating pulse released in MHD
    simulations of the sheared arcade
    (Sect. \ref{Sect:2D_Sheared_arcade}),
    run with different levels of transverse resolution.
    The panels show the temperature, density, pressure and 
    velocity parallel to the magnetic field ($v_b$)
    as functions of position across the arcade,
    at a coronal height of $x=25$~Mm,
    at the time of the first density peak $(t=150$~s).
    The lines are colour coded in a way that reflects the
    transverse resolution used with solid blue
    representing the $N_y=1024$ solution
    and dashed red (dashed green) corresponding to
    $N_y=256$ ($N_y=64$).
    \label{Fig:2D_SA_transverse_profiles}
  }
\end{figure*}
  \indent
  First we solve the unsheared arcade model
  using a series of 1D field-aligned TRAC simulations
  that reconstruct the two-dimensional (2D) plasma response
  to the imposed non-uniform coronal heating pulse. 
  In particular, we treat each field line independently 
  by solving the field-aligned MHD equations
  with a heating function that is given by the
  transverse position of the field line in the 2D reconstruction. 
  \\
  \indent
  The justification for using such an approach is two-fold. 
  Firstly, 
  the 2D reconstruction obtained from the HD simulations 
  will be used as a benchmark solution for comparison 
  with the MHD implementation of TRAC,
  due to our previous demonstration of excellent
  agreement with fully resolved field-aligned models
  \citep[see e.g. JB19,][]{paper:Johnstonetal2020}. 
  Secondly,
  the results from the 1D simulations also give
  predictions for the transverse resolution that is required 
  in the MHD model
  in order to accurately capture 
  all of the features that form in the 2D plasma response.
  \\
  \indent
  Figure \ref{Fig:1D_Reconstruction_transverse_profiles} shows
  a snapshot at $t = 150$~s of
  a number of variables that are 
  reconstructed from the 1D simulations, as a function 
  of the transverse direction $(y)$, at
  a coronal height of $x=25$~Mm.
  This snapshot corresponds to the time of the first coronal density peak.
  In the four panels we focus on 0.4~Mm 
  around the centre of the arcade, 
  showing the temperature, density, pressure and 
  field-aligned velocity $(v_x)$.
  Each quantity is shown imposed on top of the transverse heating profile
  ($Q(y)$), which is displayed as the red curve.
  In these panels,
  the solid blue and dashed green lines are 
  2D reconstructions that are calculated from the 1D simulations using 
  $N_y=1024$ and $N_y=64$ grid points in the transverse direction,
  respectively.
  These transverse resolutions
  correspond to grid cell widths of approximately 2.3~km (solid blue curve)
  and 37.5~km (dashed green curve)
  in the 2D reconstructions, 
  enabling both grids to
  fully resolve the 
  minimum transverse heating length scale given by
  Eq. \eqref{Eqn:L_Q_y}.
  \\
  \indent
  Starting with the temperature structure across the arcade, 
  it is clear that the transverse variations in the temperature
  are consistent with the
  imposed heating function.
  This happens because the initial temperature evolution is being set
  by the direct in situ heating. 
  Consequently,
  the temperature profiles calculated using
  $N_y=1024$ and $N_y=64$ grid points show good agreement as
  both solutions adequately resolve the resultant
  transverse temperature length scale.
  \\
  \indent
  On the other hand, it is striking that the
  transverse variations in the density 
  have significantly shorter length scales than the  
  heating profile.
  The minimum transverse density length scale (defined in the same was as 
  the minimum transverse heating length scale) is of order 10~km,
  which is five times smaller than that of the heating function.
  Therefore, the density profiles show major differences
  between low ($N_y=64$) and high ($N_y=1024$)
  transverse resolution because the $N_y=64$ grid is unable to 
  resolve the narrow transverse variations that are 
  observed with the $N_y=1024$ solution.
  \\
  \indent
  The generation of these small length scales 
  can be attributed to the coronal density evolution relying on the 
  interplay in the TR
  between downward conduction and upward 
  enthalpy, entangling the scaling with the
  imposed heating function 
  (as discussed in detail in Sect. \ref{Sect:Shear_flow}).
  This is also the case for the pressure, which shows  
  small length scales that are of similar order 
  to those seen in the density, 
  with steep transverse gradients forming 
  around $y=0.93$ and $y=1.01$~Mm.
  \\
  \indent
  Furthermore, the field-aligned velocity across the arcade
  shows the formation of
  two shear flow layers that
  are co-spatial with the steep transverse 
  density and pressure gradients.
  However, the length scales that are associated with these 
  shear flows are of
  order 1~km, 
  making it more challenging to resolve the transverse variations
  in the field-aligned velocity than the density and pressure.
  The outcome is that there is considerable departure between the
  velocity profiles calculated using
  $N_y=1024$ and $N_y=64$ grid points, with the low resolution solution
  significantly under-resolving the shear flow layers.
  Therefore, based on the predictions of the
  length scales that result in the HD
  simulations, the MHD model will be run with a transverse resolution of
  2.3~km ($N_y=1024$), so that the steepness of the
  shear flow layers is captured
  reasonably well. 
%
%
\section{Results: Two-dimensional simulations
  \label{Sect:2D_Results}}
  %
  %
%
%
\subsection{Unsheared arcade
  \label{Sect:2D_Unsheared_arcade}}
  \indent
  Figure \ref{Fig:2D_Simulation_UA_time_evolution_contours}
  summarises the temporal evolution of the MHD simulation of the unsheared 
  arcade, in response to the non-uniform coronal heating pulse.
  The three columns show contour plots of the temperature, density and 
  field-aligned velocity ($v_x$).
  Each row shows a snapshot at a different time: $t=10$~s (row 1),
  $t=60$~s (row 2), $t=150$~s (row 3) and $t=1000$~s (row 4).
  These correspond to times during the heating phase, at the start of
  the evaporation phase, at the first coronal density peak,
  and during the arcade’s draining phase, respectively. 
  The black curves on the contour plots
  show the top of the TRAC region
  on each of the field lines in the unsheared arcade.
  \\
  \indent
  Individual 
  field lines located inside the heated region
  follow an evolution characteristic of an impulsively heated loop
  \citep[see e.g.][]{paper:Bradshaw&Cargill2006, paper:Klimchuk2006,
  paper:Klimchuketal2008,
  paper:Cargilletal2012a, paper:Cargilletal2012b, paper:Cargilletal2015,
  paper:Reale2016}. 
  For such an evolution,
  the field-aligned flows associated with the draining phase 
  are typically an order of magnitude smaller than during the evaporation
  phase \citep[JB19,][]{paper:Johnstonetal2020}.
  Thus, the scale used for the colour table of the
  field-aligned velocity
  in Fig. \ref{Fig:2D_Simulation_UA_time_evolution_contours},
  has been adjusted accordingly,
  to give an accessible range for the snapshot at $t=1000$~s.
  %
  %
%
%
\subsubsection{Shear flow formation
  \label{Sect:Shear_flow}}
  \indent
  The collective evolution of
  the individual field lines leads to the formation of a shear flow layer
  across the arcade, as shown
  in Fig.
  \ref{Fig:2D_Simulation_UA_time_evolution_contours}.
  As only a small part of the corona is heated,
  this initially produces a region with high temperature and pressure.
  The high pressure causes a downflow, which 	  
  pushes plasma out of the corona towards the footpoints, 
  while the high temperature causes a conduction front to
  propagate downwards along the magnetic field, driving a 
  heat flux into the TR.
  \\
  \indent 
  While the conduction front on a particular field line propagates
  ahead of the flows generated by the pressure gradient, the 
  speed of the conduction front depends 
  on the temperature reached, which depends on the transverse 
  heating profile.
  As the conduction front propagates downwards,
  the plasma lower in the corona is unable to radiate the
  excess conductive heating and so the gas pressure increases
  locally.
  The conduction front then slows down as it reaches the TR
  and the accompanying increased gas pressure
  piles up at the base of the TR. 
  This creates an upward pressure gradient that drives an 
  upflow of dense material from the TR to the corona, 
  increasing the coronal density. 
  \\
  \indent
  However, the timing of this upward pressure gradient 
  depends on the amount of heating in the corona,
  which depends on the transverse position of the field line. 
  Therefore, 
  a shear flow occurs between strongly and weakly heated regions
  when there are transverse 
  variations in the energy deposition.
  This shear flow formation is seen clearly at $t=60$~s in
  the field-aligned velocity plot shown in
  Fig. \ref{Fig:2D_Simulation_UA_time_evolution_contours}.
  Furthermore, row 3 of 
  Fig. \ref{Fig:2D_Simulation_UA_time_evolution_contours} demonstrates 
  that the shear flow remains present 
  at the time of the first coronal density peak ($t=150$~s).
  This corresponds to a time
  when the evaporation fronts on strongly heated field lines have
  rebounded at the apex and 
  subsequently reversed, transporting
  the evaporated plasma back 
  towards the footpoints, 
  while the flows on weakly heated field lines are still evaporating
  mass upwards into the corona.
  The flows associated with the directly heated plasma do not show
  such short length scales across the arcade.
  %
  %
%
%
\subsubsection{Comparison between the HD and MHD models}
  \indent
  Figures \ref{Fig:Unsheared_arcade_coronal_averages}  
  and \ref{Fig:Unsheared_arcade_field_aligned_tranverse_evolution}
  compare the response of the
  2D MHD model with the results from the 1D HD reconstruction of the
  unsheared arcade. 
  Hereafter, we refer to these simulations as the MHD and HD TRAC models,
  respectively. 
  Starting with the coronal response,
  the two panels in Fig. \ref{Fig:Unsheared_arcade_coronal_averages} 
  show the time evolution of the
  coronal averaged temperature and density, for both models,
  at four different transverse positions inside the heated region.
  The coronal averages are calculated by spatially averaging over the 
  uppermost 50\% of each field line. 
  Each solid curve represents the particular 
  coronal average on the selected field lines  
  from the HD TRAC reconstruction 
  and the dashed blue curves imposed on top are 
  the corresponding averages from the MHD TRAC simulation.
  \\
  \indent
  Both models show excellent agreement across each of the field lines, 
  with a rapid rise in temperature, followed by an increase in density 
  due to 
  evaporation, then, after the time of maximum density, a radiative cooling
  and draining phase
  \citep{paper:Bradshaw&Cargill2010a,paper:Bradshaw&Cargill2010b}.
  The density oscillations that are typical for the short 
  heating pulse imposed 
  \cite[e.g.][]{paper:Reale2016}, 
  are damped slightly faster in the MHD model,
  but the resulting differences are sufficiently small
  so that the correct draining rate is retained during the decay phase.
  Therefore,
  Fig. \ref{Fig:Unsheared_arcade_coronal_averages} demonstrates that
  the MHD code, with the multi-dimensional TRAC method, 
  accurately captures the enthalpy exchange between the corona and TR,
  through all phases of an impulsive heating event.
  \\
  \indent
  This conclusion is confirmed by the temporal
  comparisons that are  presented in
  Fig. \ref{Fig:Unsheared_arcade_field_aligned_tranverse_evolution}.
  In the upper four panels, we focus on the field-aligned evolution 
  of the unsheared arcade at $y=1.2$~Mm, 
  showing the temperature, density, pressure and 
  field-aligned velocity as functions of position 
  along the magnetic field.  
  In the lower four panels, we show the transverse evolution of the same 
  quantities at $x=25$~Mm, which is
  consistent with the coronal height
  considered previously in 
  Fig. \ref{Fig:1D_Reconstruction_transverse_profiles}.
  In these panels,
  using the same line styles as before,
  each solid curve represents a different snapshot from the 
  HD TRAC model and the dashed blue curves imposed on top are the 
  corresponding snapshots from the MHD TRAC simulation.
  \\
  \indent
  First we examine the field-aligned evolution. 
  The third panel of 
  Fig. \ref{Fig:Unsheared_arcade_field_aligned_tranverse_evolution}
  shows the high level of agreement between the HD and MHD TRAC models,
  for the evolution of the pressure 
  gradients that form along the field. 
  These pressure gradients subsequently drive the field-aligned flows.
  It therefore follows that the MHD TRAC solution correctly models the
  evolution of the field-aligned flows,
  throughout the
  evaporation and draining cycle, and
  this is confirmed in the fourth panel.
  The outcome is that the mass and energy exchange
  between chromosphere, TR, and corona is correctly captured
  by the MHD implementation of TRAC,
  which, in turn, ensures accuracy in 
  simulating the coronal temperature and density 
  evolution.
  \\
  \indent
  Next we look at the transverse evolution.
  As shown in the fifth panel of 
  Fig. \ref{Fig:Unsheared_arcade_field_aligned_tranverse_evolution},
  the temperature structure across the arcade shows excellent agreement
  between the HD and MHD TRAC models,  
  for each of the snapshots.
  Likewise, the evolution of the transverse variations in the
  density and pressure shows good agreement between the two models.
  However, there are some small differences.
  For example,
  while the steep transverse density and pressure 
  gradients that form around $y=1.01$~Mm,
  at the time of the first coronal density peak
  ($t=150$~s, orange curve),  
  are accurately captured by the MHD TRAC 
  solution, the accompanying gradients
  that form further out,
  around $y=0.93$~Mm, are
  slightly smoothed over by the MHD TRAC model.
  \\
  \indent
  Consequently, 
  the MHD TRAC solution correctly models the 
  shear flow layer in the field-aligned velocity that is 
  co-spatial with the transverse 
  gradients at $y=1.01$~Mm, 
  but slightly broadens the outer layer at $y=0.93$~Mm.
  Thereafter, minor differences remain for the evolution
  of the transverse density and pressure structure in the heated
  region, before both models reconcile later in the decay phase.
  Overall, 
  Fig. \ref{Fig:Unsheared_arcade_field_aligned_tranverse_evolution}
  shows that the TRAC
  method can be used to model the coronal plasma evolution 
  with confidence in multi-dimensional MHD simulations.
  %
  %
%
%
\subsubsection{MHD effects}
  \indent
  Finally, the first panel of
  Fig. \ref{Fig:2D_Simulation_UA_P_jz} shows that
  there is a modification to the 
  magnetic field in the MHD simulation in order to keep the total
  pressure constant in the transverse direction. 
  The second panel shows that this introduces a narrow current layer,
  which is co-spatial with the shear flow region that forms
  at $y=1.01$~Mm, for the snapshot shown at $t=150$~s.
  We note that this narrow current sheet is fundamentally different 
  from a tangential discontinuity as proposed by Parker (1972) 
  because it is driven by pressure differences
  instead of constant pressure across 
  a tangentially discontinuous flux surface.
  Hence, this demonstrates 
  for the first time
  that the thermodynamic response  
  to spatially non-uniform heating events 
  can generate small transverse length scales 
  in the form of pressure
  driven current sheets, which are 
  significantly shorter than those that are associated with
  the heating profile or mechanism.
  %
  %
  %
  %
  %
  %
\begin{figure*}
  \hspace*{-0.05\linewidth}
  \subfigure{\includegraphics[width=0.53\linewidth]
  {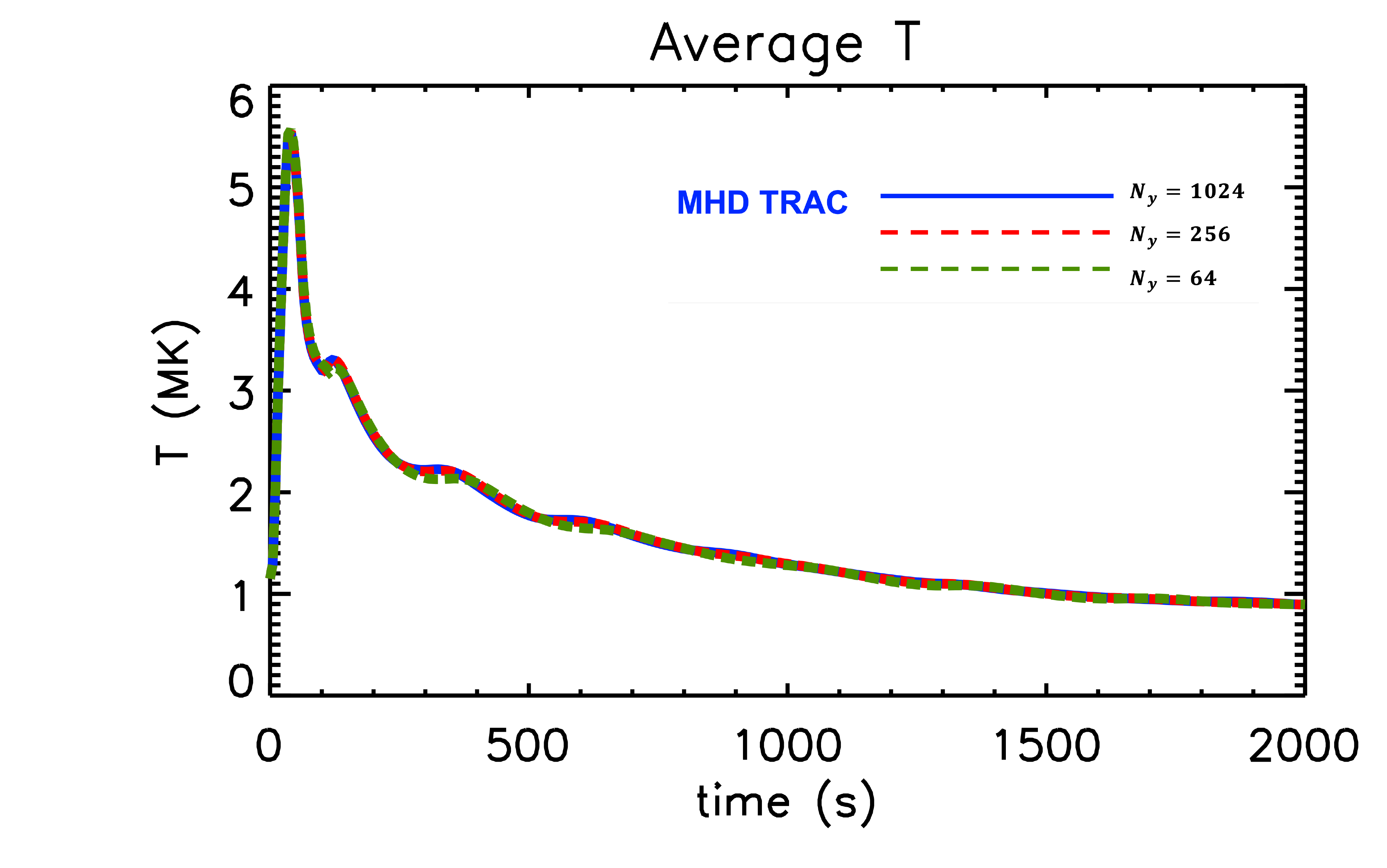}}
  \hspace*{-0.03\linewidth}
  \subfigure{\includegraphics[width=0.53\linewidth]
  {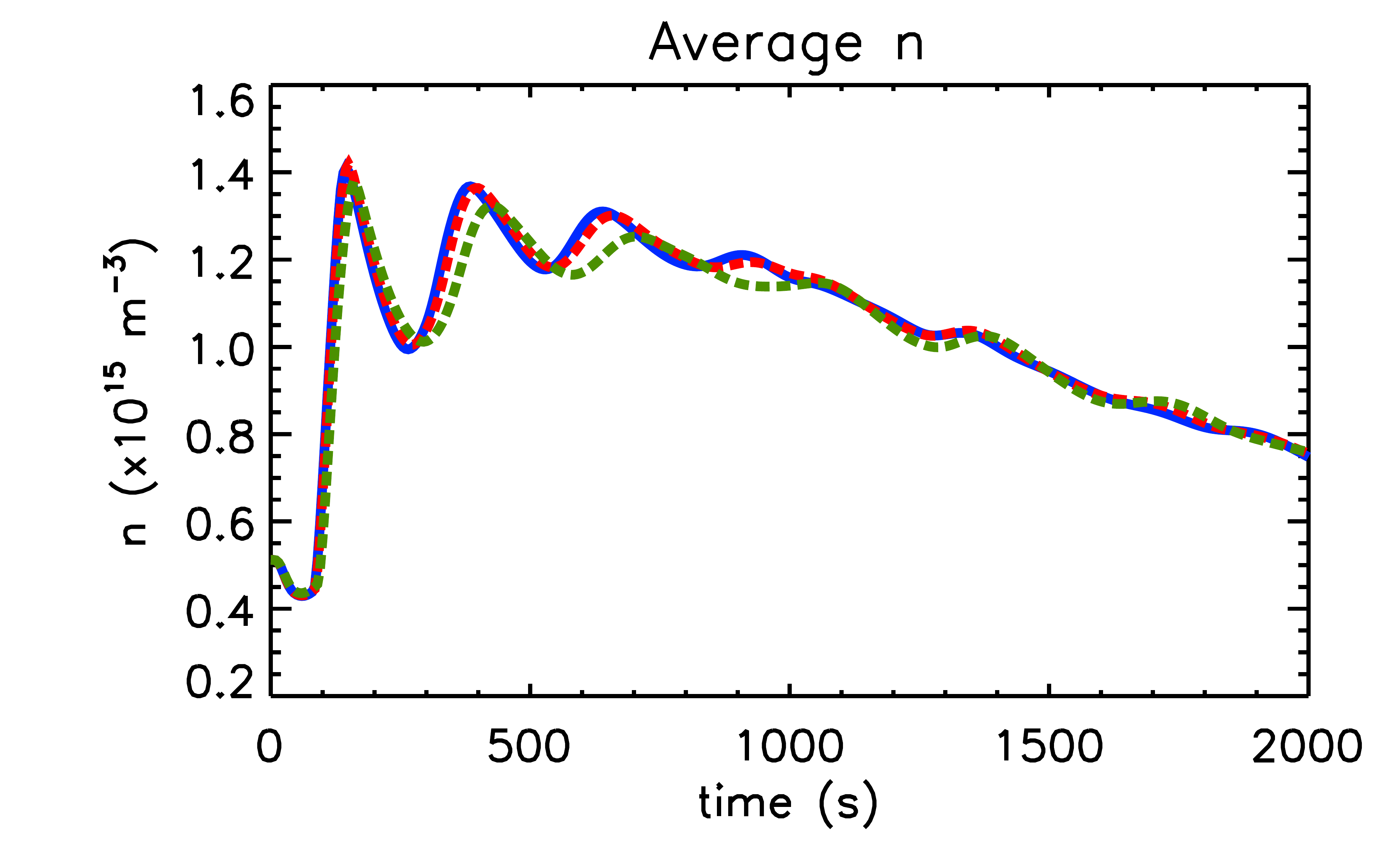}}
  \\[-8mm]
  \caption{
    Comparison of the coronal evolution
    obtained from the sheared arcade
    simultaions 
    run with different levels of transverse resolution
    (Sect. \ref{Sect:2D_Sheared_arcade}).
    The panels show the coronal averaged temperature and density as 
    functions of time,
    where the spatial average was calculated
    in both $x$ and $y$
    over the uppermost 25\% of the computational domain.
    The lines are colour coded in the same way as
    Fig. \ref{Fig:2D_SA_transverse_profiles}.
    \label{Fig:Sheared_arcade_coronal_averages}
  }
\end{figure*}
  %
  %
  %
  %
%
%
\subsection{Sheared arcade
  \label{Sect:2D_Sheared_arcade}}
  \indent
  Figure \ref{Fig:2D_Simulation_SA_time_evolution_contours}
  shows the outcome of imposing the non-uniform coronal heating pulse 
  in the sheared arcade model,
  in the same format as 
  Fig. \ref{Fig:2D_Simulation_UA_time_evolution_contours}.  
  The temporal evolution shows the same fundamental properties as the
  unsheared arcade, but with
  the TRAC region and thermodynamic response aligning
  with the tilted magnetic field accordingly.
  In particular, we see the formation of a shear flow layer
  between strongly and weakly heated field lines at 
  the start of the evaporation phase ($t=60$~s), 
  which remains prominent at the time of 
  the first coronal density peak ($t=150$~s).
  Therefore, the shear flow is a signature of the evaporated plasma, 
  which is not observed in the directly heated material.
  \\
  \indent
  We note that
  the simulation presented in 
  Fig. \ref{Fig:2D_Simulation_SA_time_evolution_contours} 
  used a transverse resolution of 2.3~km ($N_y=1024$) in
  order to resolve the resulting shear flow layer. 
  However, such high spatial resolution across the magnetic field 
  is not typically achieved in active region sized 3D MHD models.
  Thus, to study the influence of transverse resolution on the shear flow
  formation, we repeated the sheared arcade simulation
  using intermediate ($N_y=256$) and low ($N_y=64$)
  levels of transverse resolution.
  \\
  \indent
  Figure \ref{Fig:2D_SA_transverse_profiles} 
  contrasts the results with the shear flow formed
  in the high resolution simulation
  at $t=150$~s,
  showing the temperature, density, pressure and 
  velocity parallel to the magnetic field,  as functions of position 
  across the sheared arcade.  
  The low (dashed green) and intermediate (dashed red) transverse
  resolution simulations both show broadened 
  temperature, density and pressure profiles
  that smooth over the transverse gradients
  of the high resolution simulation (solid blue).
  Consistent with numerical diffusion 
  artificially influencing the evolution (due to the finite grid), 
  more significant broadening
  is associated with lower transverse resolution.
  This broadening makes it increasingly difficult to detect
  any local signatures of the shear flow when using lower
  transverse resolution.
  \\
  \indent
  However,
  Fig. \ref{Fig:Sheared_arcade_coronal_averages} demonstrates 
  that this has little effect on the
  global evolution of the corona
  because the evaporative response to heating
  is dominated by the field-aligned mass and energy exchange
  that takes place between the chromosphere, TR, and corona.
  A process which is 
  modelled accurately by the TRAC method  
  for each of the different levels of transverse resolution.
  The outcome is that the coronal
  averaged temperature and density
  show good agreement between
  the high ($N_y=1024$), intermediate ($N_y=256$) 
  and low ($N_y=64$) transverse
  resolution simulations of the sheared arcade.
  Therefore, lower transverse resolution
  does not lead to erroneous conclusions for the coronal plasma evolution
  when using the MHD implementation of TRAC. 
  %
  %
%
%
\section{Discussion and conclusions
  \label{Sect:Discussion}}
  \indent
  This paper extends the work of JB19 and \cite{paper:Johnstonetal2020},
  presenting a highly efficient formulation of the TRAC
  method for use in multi-dimensional 
  MHD simulations.
  Extending the TRAC method to MHD has required optimisation 
  in order to efficiently
  account for the magnetic field evolution,
  without the need to trace field lines
  at each time step.
  In particular, to move from one-dimensional 
  HD to multi-dimensional MHD, 
  we have modified the method from requiring
  the calculation of a global
  cutoff temperature that is associated with individual field lines, 
  to employing a local cutoff temperature that is calculated 
  using only local grid cell quantities.
  However, despite
  this change from using a global to a local cutoff 
  temperature for broadening the steep gradients in the TR, the total 
  radiative losses and heating remain conserved
  under the MHD formulation.
  The outcome is that multi-dimensional MHD simulations using the 
  MHD extension of the
  TRAC method can accurately model the coronal plasma evolution 
  through all phases of an impulsive heating event.
  \\
  \indent
  The advantages of using this novel extension of the TRAC
  method over 
  field line tracing approaches
  \citep[see e.g.][]{paper:Zhouetal2021} are multiple.  
  For multi-dimensional MHD models,
  the ability to side-step the need to trace magnetic field lines
  when applying the MHD TRAC method means that 
  (1) 
  the implementation of the method is substantially simpler,
  (2) 
  the cutoff temperatures are calculated 
  significantly faster at a fraction of the
  computational cost,
  (3) 
  it is fundamentally easier to account for
  changes in field line connectivity, 
  permitting the plasma response to be modelled accurately
  with relative ease
  in coronal heating simulations where the 
  energy release is generated self-consistently through 
  magnetic reconnection events
  \citep[e.g.][]{paper:Hoodetal2016,
  paper:Realeetal2016,
  paper:Reidetal2018,paper:Reidetal2020}
  and
  (4)
  the method is more readily employed in 
  large-scale 3D MHD simulations, which have more realistic and complex
  magnetic field configurations 
  \citep[e.g.][]{paper:Warneckeetal2017,
  paper:Mikicetal2018,
  paper:Sykoraetal2018,
  paper:Knizhniketal2019,
  paper:Howsonetal2019,
  paper:Howsonetal2020,
  paper:Kohutova2020,
  paper:Antolinetal2021}.
  Furthermore, the MHD TRAC method 
  only increases the thermal conductivity relative to the
  SH value 
  in under-resolved grid cells, while reducing to the SH model elsewhere. 
  Therefore, the method automatically switches off 
  in properly resolved parts of the atmosphere.
  \\
  \indent
  While the MHD TRAC method successfully removes the 
  influence of under-resolving the TR on the coronal density response to 
  heating,
  the broadening modifications act only in the field-aligned 
  direction. 
  This means that 
  full numerical resolution is still required in the transverse
  direction  in order
  to resolve the current sheets that are responsible for the heating
  \citep[see e.g.][]{paper:Leakeetal2020}.
  Moreover, in this paper, 
  we have demonstrated that the evaporative response  
  to impulsive heating events
  can generate transverse length scales that are much smaller
  than those
  associated with the heating mechanism.
  In particular, we presented the formation of a shear flow,
  which we identified
  as a unique signature of the 
  evaporated plasma because
  such shortened length scales are  
  not observed in the directly heated material,
  and associated pressure driven current sheets.
  \\
  \indent
  In summary, the MHD TRAC method efficiently 
  addresses the difficulty of obtaining the correct
  evaporative response to impulsive heating events
  in multi-dimensional MHD simulations, without the need for high 
  spatial resolution in the TR.
  Indeed, our results suggest that high levels 
  of accuracy
  can be obtained with grid cell widths of order 50 km
  in the field-aligned direction, which 
  is achievable in current three-dimensional MHD models.
  Therefore, the method helps to 
  free up computational resources to better 
  resolve the heating mechanism and subsequent shear flow dynamics.
  Furthermore, the MHD TRAC method is simple to implement, 
  fast to run and is easily employed 
  in MHD simulations of coronal heating
  that study the build-up of magnetic energy
  in complex field configurations and subsequent dissipation
  through magnetic reconnection. 
  %
  %
%
%
\begin{acknowledgements} 
  The research leading to these results has received funding from the UK
  Science and Technology Facilities Council 
  (consolidated grant ST/N000609/1), 
  the European Union Horizon 2020 research and innovation programme 
  (grant agreement No. 647214). 
  IDM received funding from the Research Council of Norway through its 
  Centres of Excellence scheme, project number 262622. 
  CDJ acknowledges support from the International Space Science 
  Institute (ISSI), Bern, Switzerland to the International Teams on
  \lq\lq
  Observed Multi-Scale Variability of Coronal Loops as a Probe of Coronal 
  Heating\rq\rq\ 
  and 
  \lq\lq
  Interrogating Field-Aligned Solar Flare Models: 
  Comparing, Contrasting and Improving\rq\rq .   
  We also acknowledge very helpful 
  comments from the referee, Dr. Philip Judge.
  This work used the DiRAC Data Analytic system at the University of 
  Cambridge, operated by the University of Cambridge High Performance 
  Computing Service on behalf of the STFC DiRAC HPC Facility 
  (www.dirac.ac.uk). 
  This equipment was funded by BIS National E-infrastructure capital grant 
  (ST/K001590/1), STFC capital grants ST/H008861/1 and ST/H00887X/1, and 
  STFC DiRAC Operations grant ST/K00333X/1. 
  DiRAC is part of the National e-Infrastructure.
\end{acknowledgements}
  %
  %
%
%
\bibliographystyle{aa}
\bibliography{MHD_TRAC_Paper}
  %
  %
%
%
\begin{appendix}
  %
  %
%
%
  \section{Influence of numerical resolution on coronal response to 
  heating \label{App:IoNR}}
\begin{figure*}
  \hspace*{-0.05\linewidth}
  \subfigure{\includegraphics[width=0.53\linewidth]
  {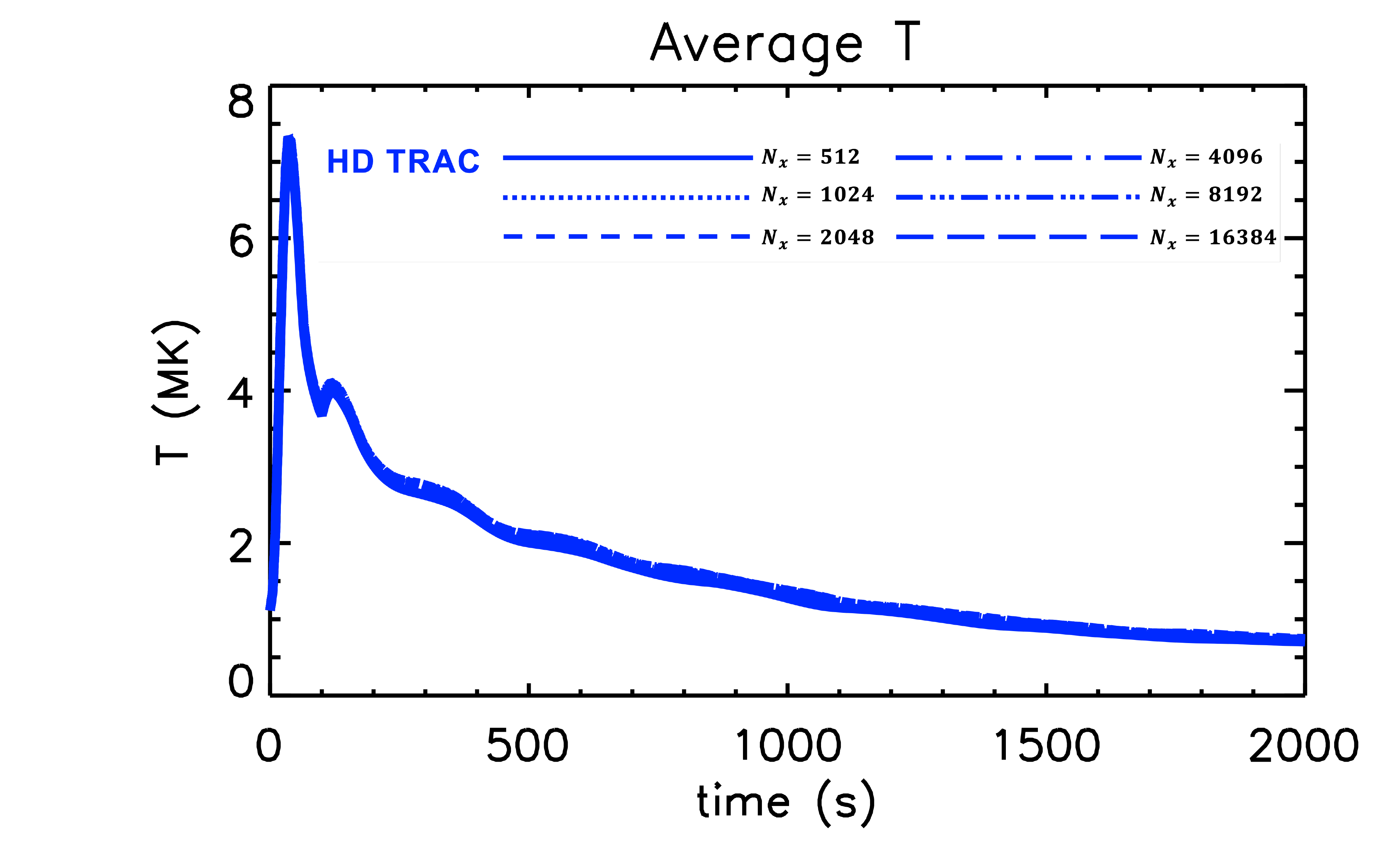}}
  \hspace*{-0.03\linewidth}
  \subfigure{\includegraphics[width=0.53\linewidth]
  {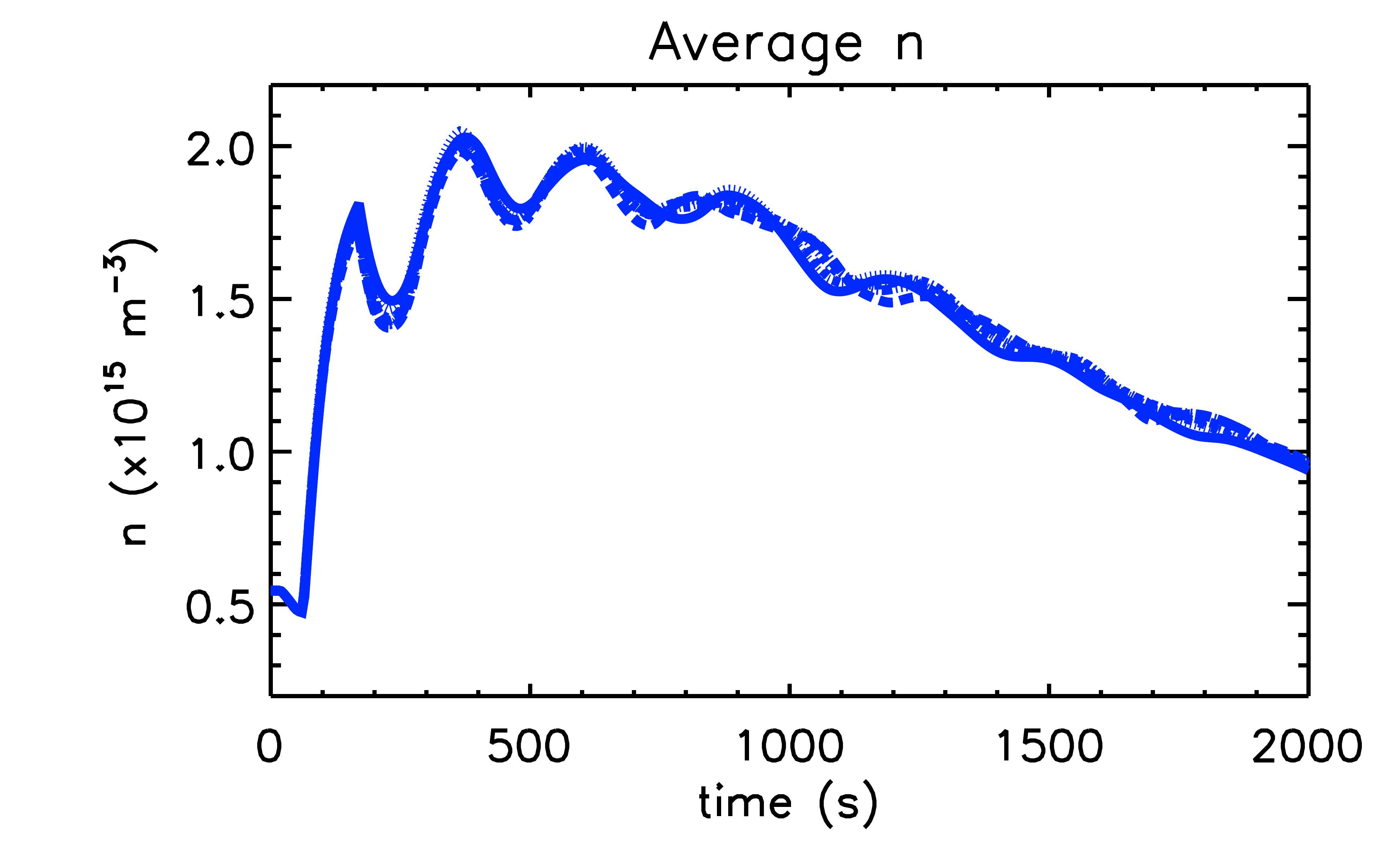}}
  \\[-5mm]
  \hspace*{-0.05\linewidth}
  \subfigure{\includegraphics[width=0.53\linewidth]
  {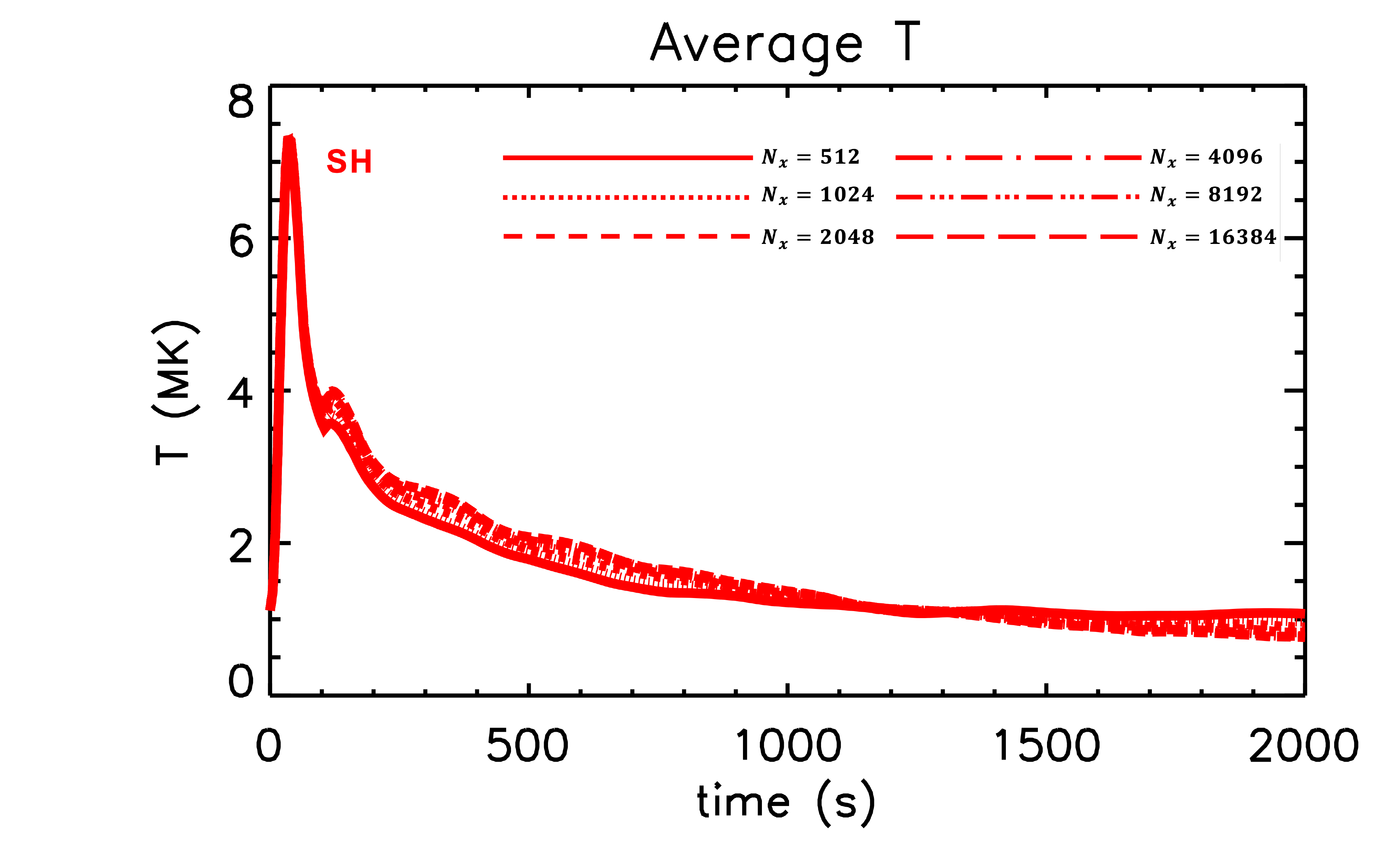}}
  \hspace*{-0.03\linewidth}
  \subfigure{\includegraphics[width=0.53\linewidth]
  {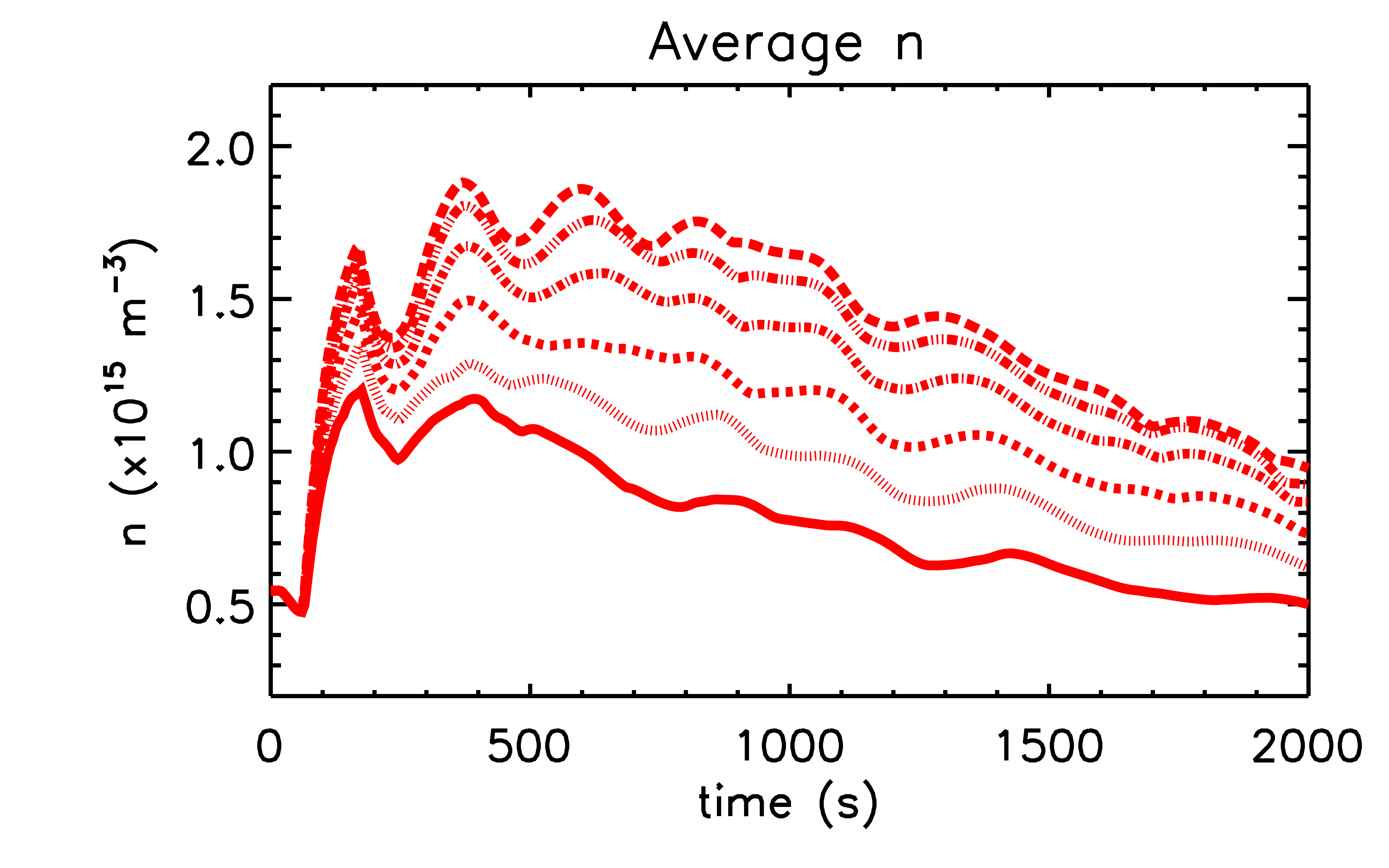}}
  \\[-8mm]
  \caption{
    Results for the non-uniform coronal heating pulse using one-dimensional 
    HD simulations of the unsheared arcade 
    (Sect. \ref{Sect:1D_Results}).    
    The panels show the
    coronal averaged temperature (left-hand column) and density (right-hand 
    column) at $y=1.2$~Mm, 
    as functions of time. 
    The various curves represent different values of $N_x$, 
    which converge as $N_x$ increases (higher spatial resolution 
    in the field-aligned direction
    is associated with larger $N_x$).
    Rows 1 and 2 correspond to simulations run 
    with the HD
    implementation of TRAC (Sect. \ref{Sect:TRAC_HD})
    and the
    SH conduction method, respectively. 
    The lines are colour-coded in a way that reflects 
    the conduction method used.
    \label{Fig:Unsheared_arcade_coronal_averages_IoNR}
  }
\end{figure*}

  \indent
  Figure \ref{Fig:Unsheared_arcade_coronal_averages_IoNR}
  shows the temporal evolution of the coronal averaged
  temperature and density at $y=1.2$~Mm,
  for the non-uniform coronal heating pulse considered in 
  Sect. \ref{Sect:1D_Results}.
  The blue curves correspond to the TRAC method (first row) and the
  red curves represent the SH conduction method (second row).
  In the panels of each method, each curve corresponds to a simulation run 
  with a different number of grid points
  that are uniformly spaced 
  along the length of the loop.
  Simulations run 
  with $N_x=[512, 1024, 2048, 4096, 8192, 16384]$ are identified
  with different line styles, as shown in the figure legend on the
  temperature plot.
  \\
  \indent
  Consistent with JB19, the coronal density evolution
  in the TRAC simulations is only
  weakly dependent on the spatial resolution.
  Grid cell widths  of approximately 60~km ($N_x=1024$) 
  are sufficient to observe convergence
  between the TRAC results. 
  Therefore, 
  TRAC solutions that are calculated using
  local cutoff temperatures show the same
  fundamental properties
  as those employing global cutoff temperatures
  \citep{paper:Johnstonetal2020},
  accurately capturing the interaction between
  the corona and chromosphere
  through all phases of an impulsive heating event.
  \\
  \indent
  On the other hand, 
  the SH solutions are
  strongly dependent on the spatial resolution. 
  In agreement with the detailed investigation of 
  \cite{paper:Bradshaw&Cargill2013},
  even when using $N_x=16384$,
  the grid cells widths 
  remain too large
  to observe convergence in the SH runs. 
  Furthermore, we note that we
  have had to limit the most refined resolution used here
  because of the increased computation time that is required 
  every time the number of grid points is doubled 
  \citep{paper:Johnstonetal2017a}. 
  Thus, it is not computationally feasible to obtain
  a fully resolved SH solution 
  when using a uniform grid.
  Therefore, in this paper,
  we benchmark the MHD implementation 
  of TRAC using 
  1D field-aligned TRAC simulations.
  %
  %
  %
  %
  %
  %
  \end{appendix}
\end{document}